\documentclass[aps,amsmath,amssymb,nofootinbib,twocolumn,a4paper,prb,showpacs]{revtex4-1}
\usepackage{verbatim}
\usepackage{graphicx}
\usepackage{color}
\usepackage[vcentermath]{youngtab}
\usepackage{multirow}
\usepackage{caption}
\usepackage{subfig}
\usepackage{float}
\usepackage{eufrak}
\usepackage{verbatim}

\newcommand{\fref}[1]{Fig.~\ref{#1}}
\newcommand{\tref}[1]{Table~\ref{#1}}
\newcommand{\eref}[1]{Eq.~(\ref{#1})}
\newcommand{\cref}[1]{Chapter~\ref{#1}}
\newcommand{\sref}[1]{Sec.~\ref{#1}}
\newcommand{\aref}[1]{Appendix~\ref{#1}}
\newcommand{\generalizedspin}[1]{generalized spin wave function#1}
\newcommand{\gssubscript}[1]{gsw}
\newcommand{\sboson}{S_{\mbox{\small boson}}}
\newcommand{\sfermion}{S_{\mbox{\small fermion}}}
\newcommand{\note}[1]{$\color{red}{\bullet}$}

\usepackage{color}
\definecolor{darkblue}{rgb}{0.,0.,0.7}
\definecolor{darkred}{rgb}{0.7,0.,0.}

\usepackage[	 	pdfpagelabels, 
					bookmarks = true,
					plainpages=false,
                       	bookmarksopen = true,
                      	bookmarksnumbered = true,
                	     	breaklinks = true,
                     	     colorlinks = true,
                      	linkcolor = darkred,
                      	urlcolor  = darkblue,
                	     	citecolor = darkred,
                      	anchorcolor = green,
		            	unicode,
                	     hyperindex = true,
                      	hyperfigures
                  ]{hyperref}

\begin{document}

\title{\textbf{Multiparticle pseudopotentials for multicomponent quantum Hall systems}}
\author{Simon C. Davenport and Steven H. Simon}
\affiliation{Rudolf Peierls Centre for Theoretical Physics, 1 Keble Road, Oxford OX1 3NP, United Kingdom}
\date{\today}

\begin{abstract}
The Haldane pseudopotential construction has been an extremely powerful concept in quantum Hall physics---it not only gives a minimal description of the space of Hamiltonians but also suggests special model Hamiltonians (those where certain pseudopotential are set to zero) that may have exactly solvable ground states with interesting properties.  The purpose of this paper is to generalize the pseudopotential construction to situations where interactions are $N$-body and where the particles may have internal degrees of freedom such as spin or valley index.  Assuming a (spatially) rotationally invariant Hamiltonian, the essence of the problem is to obtain a full basis of wave functions for $N$ particles with fixed relative angular momentum $L$.  This basis decomposes into representations of $SU(n)$ with $n$ the number of internal degrees of freedom. We give special attention to the case where the internal degree of freedom has $n=2$ states, which encompasses the important cases of spin-1/2 particles and quantum Hall bilayers.   We also discuss in some detail the cases of spin-1 particles ($n=3$) and graphene ($n=4$, including two spin and two valley degrees of freedom). 
\end{abstract}


\maketitle
 
\section{Introduction}

In the absence of interactions or disorder, the impact of a magnetic field on the band structure of two dimensional electrons is profound: the spectrum of single particle eigenstates breaks into degenerate bands called Landau levels (LLs).  Fractional quantum Hall (FQH) physics \cite{prangebook} occurs when interactions between electrons break the degeneracy of partially filled Landau bands leading to an incompressible fluid ground state.

In most discussions of fractional quantum Hall effect (FQHE), one assumes that the interaction between particles is a pair-wise potential $V(|{\bf r_1 - r_2}|)$ and, further,  that the electrons have no additional degrees of freedom such as a spin or valley index.  These two assumptions are ones we would like relax in the current work. In this paper, we study so-called multicomponent systems (particles with an additional degree of freedom, such as spin) interacting via very general $N$-body potentials  $V({\bf r_1}, s_1;{\bf r_2}, s_2; \ldots {\bf r}_N, s_N)$ where here $s_j$ represents the additional quantum number of the $j^{th}$ particle. Motivation for expanding our attention to this more complicated case is discussed below in \sref{subMultiQuantumHall}.

In analyzing FQH physics a very important tool is the Haldane pseudopotential. \cite{haldane1983} In the simple case where the interaction is two-body and there  is a single species of spin-polarized electrons (i.e., without the generalizations discussed in this paper), the Haldane pseudopotential, $V_L$, is defined as the energy cost for two electrons to have a given relative angular momentum $L$. These pseudopotentials give a complete (and minimal) description of any rotationally and translationally invariant two-body interaction within a single Landau level. The language of pseudopotentials not only provides a convenient parametrization of the problem but also makes it easy to write down certain model Hamiltonians that are solvable, thus providing a key piece of our understanding of the FQHE.  The purpose of the current paper is to generalize the idea of the pseudopotential to the multicomponent case where particles have additional degrees of freedom such as spin or valley index and where interactions between particles are of a general $N$-body form.

\subsection{Pseudopotentials and Quantum Hall States}

Let us elaborate for a moment on  the connection between pseudopotentials and exactly solvable model interactions in the case where interactions are of the conventional two-body form. Let us choose a model interaction potential such that the value of $V_1$ is positive, but $V_L =0$ for $L>1$. In this case, we are imposing an energy cost for any two particles to have relative angular momentum of $L=1$: The exact highest density zero energy state  (the highest density state where no two electrons have relative angular momentum of 1) is precisely the $\nu=1/3$ Laughlin wave function\cite{laughlin1983a,haldane1983,trugman1985} (as always, we define filling fraction $\nu$ to be the fraction of single particle states within a LL which are occupied or, equivalently, $\nu$ is the ratio of number of electrons to number of flux quanta).    Because of this exact solvability, much can be established in detail about this type of model Hamiltonian. While the model Hamiltonian may not be too similar to any particular physical Hamiltonian (such as the Coulomb Hamiltonian), nonetheless, the ground state may be very close to the physical one.  More importantly, the ground state of the model gives an easily studied representative of an entire phase of matter.

In recent years, due to a large extent to interest in more exotic non-Abelian FQH states, \cite{nayak2008} interest has turned from two body interactions (such as Coulomb) to $N$-body interactions with $N>2$. While, in principle, such $N$-body interactions occur as a result of integrating out higher LLs, \cite{bishara2009} and could, in principle, be engineered to exist in certain cold-atom systems, \cite{cooper2004} the main interest in multi-body interactions is, again, due to the fact that certain interesting many-body wave functions are the exact highest density zero energy state of certain $N$-body interactions. For example, the celebrated Moore--Read wave function \cite{moore1991} is the exact ground state of a model three-body interaction, \cite{greiter1991} and, more generally, the $Z_k$ Read--Rezayi wave function \cite{read1999} is the ground state of a $(k+1)$-body interaction. Other interesting examples include the Gaffnian \cite{simon2007b} and Haffnian \cite{green2002} wave functions, which are also ground states of special three-body interactions.

As with the simple case of two-body interactions, when studying FQH Hamiltonians with $N$-body interactions, it is quite useful to work with a generalization of Haldane's pseudopotential formalism first proposed in Ref.~\onlinecite{simon2007c}. While this formalism is similar to that discussed by Haldane in that it decomposes the interaction into angular momentum components, it is somewhat more complicated because specifying the total relative angular momentum of $N$ particles does not completely specify the relative wave function of the $N$-particles as it does in the case of two particles. This complication is explained in detail in \sref{secPseudoMultiPseudoReview} below where we review the multiparticle pseudopotential formalism. Accounting for this complication, the multiparticle pseudopotentials can simply describe all of the special model Hamiltonians pointed out above (Moore--Read, Read--Rezayi etc.). In fact, in each of the above discussed cases, the Hamiltonian can always be described as simply forbidding any $N$ particles from having relative angular less than some $L$.

All of the wave functions discussed above turn out to be special cases of a very broad set of so-called Jack polynomial wave functions. \cite{bernevig2008b} While there is a general belief that local $N$-body Hamiltonians may exist for all of the Jacks, only limited further cases have actually been explored. \cite{jackson2013, simon2010}  Such Hamiltonians, should they be constructed, can be phrased in the general language of pseudopotentials.  In these more general cases, the necessary Hamiltonians  cannot be of a simple form that simply prohibits clusters of particles from having certain angular momenta \cite{simon2007c} but, rather, will require a particular balance of certain pseudopotential coefficients.\cite{jackson2013}

\subsection{Multicomponent Quantum Hall}
\label{subMultiQuantumHall}

All of the FQH states discussed so far describe spin-polarized or ``spinless" system, i.e., the only degree of freedom for each electron is its orbital position. However, there are many cases where we may want to consider a more complex model where electrons are endowed with additional degrees of freedom, such as spin, and the FQH ground state may not be fully polarized.  In these situations we say that the ground state is ``multicomponent", meaning that it involves an important contribution from electrons having different values of these additional internal degrees of freedom.

The simplest example of a multicomponent quantum Hall system is one with spin.  Naively, one might expect that the high magnetic field characteristic of the fractional quantum Hall regime might remove any spin degree of freedom entirely.   However, in many quantum Hall systems this is not the case, and we must consider the spin degree of freedom as well. For example, in conventional GaAs systems, due to the small $g$ factor, even in fairly high fields, electron spin may not be polarized. \cite{du1995,cho1998} Furthermore, there are experimental methods to engineer an even smaller $g$ factor, making both spin states even more relevant. \cite{leadley1997}

Valley degrees of freedom are another way which quantum Hall systems may be multicomponent.  For example, in both AlAs quantum wells, \cite{bishop2007,padmanabhan2010b} and graphene, \cite{dean2011,bolotin2009,du2009} the semiconductor band structure is such that each electron has a spin and valley degree of freedom.  Analogous to the electron spin, in these cases, the valley index may take one of two possible values, so we think of the valley as a ``pseudospin" or ``isospin". Silicon MOSFETs may be even more complicated: Depending on the crystal orientation with respect to the 2D electron layer, the electrons will have a valley index that takes one of two, four, or even six values. \cite{lai2004,arovas1999,eng2007} Yet another important case where electrons have an additional degree of freedom is in quantum Hall bilayers, where the layer index plays the role of a pseudospin. \cite{eisensteinbook,eisenstein2004a,eisenstein2004b}  All of these multicomponent quantum Hall systems have been the subjects of intense theoretical and experimental study.

A related topic of recent theoretical interest is the study of the quantum Hall effect of cold bosons.  While such a quantum Hall effect has not yet been observed, it seems potentially feasible in the not-so-far future. \cite{cooper2008}.   Bosons, as compared to electrons, must have integer spin, but one could easily imagine a quantum Hall effect of spin-1 bosons that would have three internal states rather than the two of an electron. Other possibilities for multicomponent Bose systems exploit multiple hyperfine states of an atom, or multiple subbands that occur for bosons in a magnetic field on an optical lattice. \cite{palmer2006,hormozi2012} Another application would be to systems where multiple Landau levels can be occupied where the internal degree of freedom would be the Landau level index. 

The many possibilities of experimentally realizing multicomponent quantum Hall states have driven a large number of theoretical studies. \cite{eisensteinbook} These studies began with very early work by Halperin \cite{halperin1983} which generalized Laughlin's wave function \cite{laughlin1983a} to the multicomponent case. Quite naturally, the concept of Haldane's pseudopotentials were quickly generalized to the multicomponent case as well.\cite{yoshioka1989,macdonald1989,haldane1988,rezayi1987a,rezayi1987b} Analogous to the Laughlin case, the Halperin wave functions can also be described as the exact highest density zero energy states of special (two-body) interactions, and these interactions can in turn be described by particular multicomponent pseudopotential coefficients.

The added richness of multicomponent systems has made them a prime place to search for new and exciting physics.  In the search for novel non-Abelian quantum Hall systems, \cite{nayak2008} several multicomponent candidates have been proposed, \cite{ardonne1999,read2000,ardonne2001,ardonne2002,reijnders2002,reijnders2004, barkeshli2010} including the so-called non-Abelian spin singlet (NASS) states \cite{ardonne1999} and the spin-charge separated states\cite{ardonne2002} for the two component case, as well as generalizations of these constructions to higher numbers of components. \cite{reijnders2002,reijnders2004}  Recently the idea of Jack polynomials was generalized to certain multicomponent systems as well. \cite{estienne2012,ardonne2011}

Analogous to the situation with single component wave functions, many of these novel multicomponent wave functions are exact ground states of special $N$-body interactions.  Here, however, interactions may be more complicated, depending on the ``spin-state" as well as the position of the particles.  As such, it seems natural to try to generalize the pseudopotential formalism to the case of multicomponent $N$ body interactions.  This is the aim of the current paper.

As in the single component case, the motivation for developing the pseudopotential formalism for multicomponent many-body interactions is severalfold.  On the one hand, pseudopotentials provide a complete parametrization of the problem for such systems.   The usefulness of this is evidenced by recent works that have introduced pseudopotentials for multicomponent many-body interactions for important special cases. \cite{yang2008,bishara2009} More importantly, the pseudopotential structure hints at what sorts of simple Hamiltonians may be written such that interesting quantum Hall states might be found as the highest density zero energy state of a particular set of pseudopotential coefficients. While in the current paper we do not yet undertake to identify new wave functions in this way, the research program is nonetheless clear and will be a topic of future research (see also Refs.~\onlinecite{ardonne2011,estienne2012}).

\subsection{Structure of this Paper}

In \sref{secPseudoMultiPseudoReview} we review the idea of (multiparticle) pseudopotentials and we define our problem in more detail.  Following this necessary background, our main results are summarized in \sref{secPseudoMultiResults}.  While our methods are generally applicable to any number of internal degrees of freedom per particle, we emphasize in particular the most experimentally relevant multicomponent cases that are the cases where particles (fermions or bosons) have two, three, or four possible internal states for interactions between small numbers (2--5) of particles. We present tables for these simple cases indicating the number $p$ of linearly independent wave functions that exist for $N$ particles having total relative angular momentum $L$. This specifies a $p \times p$ Hermitian matrix of pseudopotential parameters that can be defined for that value of angular momentum. In this same section we, further, give the explicit form of these linearly independent wave functions.  For many readers who are interested in applications of our work, these tables in the results section should provide most of the relevant information. Within this section, \sref{subPseudoTwoComponentResults} addresses the case of two-component wave functions (applicable to spin-1/2 systems, bilayers, etc).  This section is the most straightforward of our results and is also probably of the most wide interest.  \sref{subPseudoSymmetryTypes} gives a brief primer on symmetry types and Young tableaux that is necessary for the description of the extension of this work to higher number of components, which we give in \sref{subMultiComponentResults}. Following the results section we give a brief discussion in \sref{secDiscussion}  of applications of our results as well as some simple examples and we direct our readers to this section for a more general conclusion and discussion of results. 

The main mathematical formalism that derives the results presented here is relegated to the appendices.  For the interested reader, these appendices have been made fairly extensive and pedagogical.  However, for most readers who are interested in the application to quantum Hall physics, the group-theoretical details of the derivation will not be necessary.  Although this presentation may be a departure from typical structure of most publications, we hope that it provides the clearest approach. 

%%%%%%%%%%%%%%%%%%%%%%%%%%%%%%%%%%
%%%%%%%%%%%%%%%%%%%%%%%%%%%%%%%%%%

\section{Review of Multiparticle Pseudopotentials}
\label{secPseudoMultiPseudoReview}

To pedagogically introduce the concept of the pseudopotential, we start by examining the single-component situation where the particles have no internal degree of freedom.  We will further simplify to two-body interaction then generalize to multi-body interaction.  Much of this exposition follows that of Ref.~\onlinecite{simon2007a} and we refer the reader to that reference for more detail.  Finally, we will introduce multicomponent wave functions in \sref{secPseudoMultiResults} below.

Before studying interactions between particles, we must first discuss wave functions for noninteracting particles in a magnetic field.   In the lowest Landau level (LLL) with a planar geometry, the noninteracting Hamiltonian is \[\frac{1}{{2{m_e}}}{\left| {{\bf{p}} + \frac{e}{c}{\bf{A}}} \right|^2},\] where in the symmetric gauge
 ${\bf{A}} =   - {\textstyle{1 \over 2}}{\bf{r}} \times {\bf{B}}$, with the magnetic flux density given by ${\bf{B}}= B \hat{\bf{z}} $.  The solutions of the Schr\"{o}dinger equation in the LLL are given by
\[{\varphi _m} = \frac{z^m {e^{ - \frac{1}{4}{{\left| z \right|}^2}}}}{{\sqrt {2\pi {2^m}m!} }} ,\]
where $z = {\left( {x + iy} \right)}$ is a complex number representing the position of the particle in the plane, and we have set the magnetic length $(\hbar c/eB)^{1/2}$ to unity.  Here, the angular momentum around the origin is $\hbar m$ and  $m \ge 0$ is an integer.  On other geometries such as the sphere, solutions still will take the form $z^m$ only changing the normalization and the Gaussian measure. \cite{simon2007a,read1999}

More generally, a solution to the single-particle Schr\"{o}dinger equation is any linear combination of the above basis states, thus taking the form
\[
\psi  = f\left( z \right){e^{ - \frac{1}{4}{{\left| z \right|}^2}}},
\]
where $f(z)$ is any analytic function of $z$. Note that due to an isomorphism between Landau levels it will be sufficient to study our problem in the lowest Landau level (LLL) only.  Interacting systems within higher Landau levels may be treated with appropriate transformations of the lowest Landau level (see Ref.~\onlinecite{simon2007a} for a detailed discussion).   We also comment that if the particle has an additional degree of freedom, such as spin, which we discuss below, we will need to specify the state of this spin as well.  For example, we might write
\[
\psi  = f\left( z \right){e^{ - \frac{1}{4}{{\left| z \right|}^2}}} \,  \left|\uparrow\right\rangle .
\]

For the many-body version of the problem we may construct linear combinations of products of the single-particle wave functions in the coordinates $z_i$ and impose an appropriate symmetry (a wave function describing fermions must be overall fully antisymmetric, and a wave function describing bosons must be fully symmetric).  In the case of a single component wave function (no internal degree of freedom) one can simply write
\[
\psi = \phi(z_1, \ldots, z_N) e^{-\frac{1}{4} \sum_{j=1}^N |z_i|^2},
\]
where $\phi$ is an analytic function of all of its arguments and is overall symmetric for bosons and antisymmetric for fermions.   In the multicomponent case, the symmetry condition is much more complicated as we will see below.  For now, we will continue to focus on the simpler single-component case.

\subsection{Single-Component Two-Body Interactions}

We begin our discussion of interactions with the case of simple two-body interaction. Very generally a wave function for two particles can be decomposed into relative and center-of-mass components.  For two-body wave functions within the LLL, we can write a complete basis
\[
|\Psi_{L,l}\rangle = |l \rangle \otimes |L \rangle,
\]
where $l$ is the center-of-mass angular momentum and $L$ is the relative angular momentum (i.e the relative angular momentum about the common center of mass of the $N$ particles).  Explicitly we mean the center of mass wave function is given by
\[
  |l \rangle \propto \left( {{z_1} + {z_2}} \right)^l  {e^{ - {\textstyle{1 \over {8}}} {{{\left| {{z_1+z_2}} \right|}^2}} }}
\]
and the relative wave function is
\begin{equation}
|L \rangle \propto {\left( {{z_1} - {z_2}} \right)^L}{e^{ - {\textstyle{1 \over {8}}} {{{\left| {{z_1-z_2}} \right|}^2}} }},
\label{eqRelativeWaveFunction}
\end{equation}
It is important that these wave functions form a complete set. Physically we deal with particles that are either fermions or bosons and so in fact the space of two particle eigenstates exists only for odd $L$ for fermions (or even $L$ for bosons) in order to obey the correct symmetry.

After projecting to a single Landau level (which can be justified by a large gap between Landau levels), the Hamiltonian is simply the interaction between the particles. We can, thus, write
\begin{equation}
H =   \sum\limits_{j < k} V(\left| {{\bf r_j} - {\bf r_k}} \right|).
\label{eqPhysicalHamiltonian}
\end{equation}
where $V$ is some interparticle interaction such as the Coulomb interaction. To decompose this interaction into pseudopotentials, we define
\begin{equation}
{V_L} = \left\langle L \right|\, V(\left| {\bf r_1} - {\bf r_2}\right|)\, \left| L \right\rangle. 
\label{eqTwoBodyPseudopotential}
 \end{equation}  
 Since the interaction is translationally invariant, it is independent of the center of mass degree of freedom of the two particles.

We can then rewrite the Hamiltonian given in \eref{eqPhysicalHamiltonian} as
\begin{equation}
H = \sum\limits_{L,l} {\sum\limits_{i < j} {\left| L \right\rangle \left| l \right\rangle {V_L}} } \left\langle l \right| \left\langle L \right|, 
\label{eqTwoBodyHamiltonian}
\end{equation}  
where it is implied that in each term of the sum, the two-particle ket $|L\rangle$ involves particles $i$ and $j$.  Again, since the interaction is translationally invariant, interaction between particles $i$ and $j$ never changes their common center of mass, so we may work with only the relative wave functions. Further, the interaction,
\eref{eqTwoBodyPseudopotential}, and, hence, the Hamiltonian \eref{eqTwoBodyHamiltonian} is diagonal in $L$ on account of the fact that the interaction potential is rotationally invariant and, therefore, conserves relative angular momentum. The intuition behind this rewriting of the Hamiltonian is that any two particles with relative angular momentum $L$ incur an energy cost $V_L$.

\subsection{Single-Component Multiparticle Interactions}

In order to extend this discussion to multiparticle pseudopotentials, we consider a general many body interaction potential ${V\left( {{\bf{r}}_1 ,\ldots,{\bf{r}}_N } \right)}$, and we restrict ourselves to a rotationally and translationally invariant system.  Analogously to \eref{eqPhysicalHamiltonian} the LLL Hamiltonian is written in terms of the $N$-particle potential as
\[
H = \sum\limits_{i_1  < i_2  < \ldots < i_N } {V\left( {{\bf{r}}_1 ,\ldots,{\bf{r}}_N } \right).}
\]
By analogy with Haldane's original pseudopotential construction, we decompose the wave function of the $N$ particles into a center of mass and a relative motion.   Further, we would like to write a complete basis for the possible relative wave functions of the $N$ particles which we will use for our pseudopotential construction.

Determining this complete basis turns out to be the tricky part of the $N$-body pseudopotential construction.  As in the two-particle case we can use the total relative angular momentum $L$ of the cluster of $N$ particles as a useful parameter. Again by rotational invariance of the interaction, $L$ will not be changed by the interaction $V$ between the particles of the cluster.  However, here the parameter $L$ is not sufficient to fully describe the $N$-particle wave function as it is in the two-body case [see \eref{eqRelativeWaveFunction}]. We must, therefore, instead write an orthonormal set of possible relative wave functions $\left| {L,q} \right\rangle$ all having the same total relative angular momentum between the $N$-particles.  Here the index $q$ runs from 1 to the number of states $p (N,L)$ in the basis for that given $L$ and $N$.  We will elaborate more on the structure of these wave functions below.

Given this basis of relative wave functions, we define $N$-particle pseudopotentials by:
\begin{equation}
V_{L,N}^{q,q'}  = \left\langle {L,q} \right|V\left( {{\bf{r}}_1 , \ldots ,{\bf{r}}_N } \right)\left| {L,q'} \right\rangle.
\label{eqDefinitionMultiParticlePseudopotentials}
\end{equation}
As in the two-body case, by translational invariance of the interaction, we need not specify the center-of-mass wave function in the definition of the pseudopotential.  Again, by  rotational invariance of the interaction, the relative angular momentum $L$ must be conserved (the matrix element is diagonal in this variable).  However, there is no need for the matrix element to be diagonal among the states $q$ with the same angular momentum.  Hence for each $N$ and $L$ we define a $p (N,L)$-dimensional Hermitian pseudopotential matrix with indices $q$ and $q'$.  

The Hamiltonian can be written in terms of these new pseudopotentials via a resolution of the identity (and making use of the fact that the interaction potential is rotationally invariant by construction):
\begin{equation}
H = \sum\limits_{ i_1  <  \ldots < i_N } {\sum\limits_{L,l,q,q'} {\left| {L,q} \right\rangle \left| l \right\rangle V_{L,N}^{q,q'} }  \left\langle {l} \right| \left\langle {L,q'} \right|},
\label{eqMutliParticleProjectionHamiltonian}
\end{equation}
where the sum over $i_1,\ldots, i_N$ indicates which $N$ particles are considered in a particular term of the sum and it is implied that $|L,q\rangle$ is the relative wave function for that given set of particles.

We now turn to the issue of determining the basis $\left| {L,q} \right\rangle$. These wave functions must be made of LLL variables and must be overall symmetric for boson wave functions and antisymmetric for fermion wave functions. Further, the basis states represent relative motion so they should be translationally invariant (i.e., the center-of-mass coordinate should not appear).

To be more specific, let us factor out the Gaussian exponential factors throughout the discussion (indeed, these factors are geometry dependent anyway \cite{simon2007a,read1999}). The remaining wave function must be a homogenous analytic polynomial of degree $L$ in the variables $z_i$ analogous to \eref{eqRelativeWaveFunction}. Translational invariance of the wave function implies that the polynomial must be invariant under any global shift in all of the coordinates $z_i \rightarrow z_i + a$ for any complex number $a$. Finally, the wave function must be overall symmetric or antisymmetric for bosons or fermions respectively.

The enumeration of such polynomials is a task that turns out to be fairly straightforward. \cite{simon2007a,liptrap2010} First, we note that the problem of enumerating the antisymmetric wave functions is essentially equivalent to that of enumerating the symmetric wave functions. To see this we note that any antisymmetric polynomial in $N$ variables can be written as a symmetric polynomial in $N$ variables times a Jastrow factor (or Vandermonde determinant) as follows:
\begin{equation}
J_{1\ldots N}
=\prod\limits_{\scriptstyle  i,j = 1 ; i < j }^N {\left( { z_i  -  z_j } \right)} ,
\label{eqJastrowFactor}
\end{equation}
Thus, there is a precise isomorphism between homogeneous symmetric polynomials of overall degree $L$ and homogeneous antisymmetric polynomials of degree $L+N(N-1)/2$.

To establish a complete basis of translationally invariant symmetric polynomials we use the basis of the elementary symmetric polynomials, which are defined in the following way:
\begin{eqnarray}
\label{eqElementarySymmetricPolyDefinition}
& & e_{m,N} \left( {z_1 ,\ldots,z_N } \right)  \\ & & ~~~~~~~=\begin{cases}
 \sum\limits_{0 < i_1  < i_2  < \ldots < i_m  \le N} {z_{i_1 }z_{i_m } } & m \le N \\
\,\,\,\,\,\,\,\,\,\,\,\,\,\,\,\,\,\,0 & {\rm{otherwise}} \nonumber
 \end{cases} 
\end{eqnarray}
All possible symmetric polynomials in $N$ variables can be written as sums and products of these generators (i.e., these generate the \emph{ring} of symmetric polynomials).

To impose the condition that the polynomials are translationally invariant we shift each variable by the overall center-of-mass coordinate to give
\begin{equation}
\tilde z_i  = z_i  - \frac{1}{N}\sum\limits_{j = 1}^N {z_j }.
\label{eqTildeCoordinates}
\end{equation}
By writing elementary symmetric polynomials of the relative coordinates ${\tilde z}_i$, we then obtain generators for the ring of translationally invariant symmetric polynomials. \cite{simon2007a,liptrap2010} It is easy to check that
\begin{equation}
e_{1,N } \left( {\tilde z_{1} ,\tilde z_{2} \ldots \tilde z_{N}} \right) = 0.
\label{eqTranslationallyInvariantSymmPoly}
\end{equation}
therefore there is one generator fewer once we impose translational invariance. It can be shown that the remaining generators $e_{m,N } \left( {\tilde z_{1} ,\tilde z_{2} \ldots \tilde z_{N}} \right)$ for $1 < m \leq N$ do not vanish and are still linearly independent. \cite{simon2007a,liptrap2010}

Given that we know the generators, with some combinatorics, we can calculate the dimension $d_{{\rm{sym}}} \left( {L,N} \right)$ of the space of translationally invariant symmetric polynomials in $N$ variables and of degree $L$. A table of the values of $d_{{\rm{sym}}} \left( {L,N} \right)$ are listed in Ref.~\onlinecite{simon2007a}. The analytic formula for $d_{{\rm{sym}}} \left( {L,N} \right)$ is reproduced in \eref{eqDimensionFullSymmetricPoly} in \aref{appendixSpatialWaveFunctions}, and these values are identical to the spin polarized cases presented in \tref{tableDimensionsOfPolynomialSpaces} (see rows with the maximum $\sboson$ value, i.e., the top row,  for each value of $N$). 

As an example, consider a translationally invariant symmetric polynomial of degree $L=4$ in $N=4$ variables.  From our allowed generators, the only basis states we can construct of degree $L=4$ are given by $[e_{2,4 } \left( {\tilde z_{1} ,\tilde z_{2}, \tilde z_{3}, \tilde z_{4}} \right) ]^2$ and $e_{4,4 } \left( {\tilde z_{1} ,\tilde z_{2}, \tilde z_{3}, \tilde z_{4}} \right)$.  Hence, $d_{{\rm{sym}}} \left( {4,4} \right) = 2$ (cf. \tref{tableDimensionsOfPolynomialSpaces}).

Our objective was to determine a basis for the states $\left| {L,q} \right\rangle$.  For bosonic wave functions, using combinations of these generators, we have found an appropriate basis of translationally invariant symmetric polynomial wave functions. The basis states given here are not orthonormal, but can easily be orthonormalized by hand (indeed, the concept of orthonormality depends on the integration measure.  For example, on the sphere it differs from the usual Gaussian we are familiar with on the plane).  For fermionic wave functions, we simply multiply these symmetric functions of degree $L$ by a Jastrow factor $J_{1\ldots N}$ to give a basis for the space of translationally invariant antisymmetric polynomials of degree $L+[N(N-1)]/2$ in $N$ variables.

Having determined our basis of states $\left| {L,q} \right\rangle$ it is then a straightforward matter to construct our pseudopotential representation of any given Hamiltonian using Eqs.~\ref{eqDefinitionMultiParticlePseudopotentials} and \ref{eqMutliParticleProjectionHamiltonian}.

%%%%%%%%%%%%%%%%%%%%%%%%%%%%%%%%%%
%%%%%%%%%%%%%%%%%%%%%%%%%%%%%%%%%%

\section{Results for multicomponent case}
\label{secPseudoMultiResults}

In this section, we present our main results for the multicomponent case. The essence of our objective in this section is, analogously to the single component case, to determine how many different linearly independent $N$-particle wave functions might exist with a fixed angular momentum $L$, and to form a complete basis for these states.  For the multicomponent case, we may be able to classify these states by some additional quantum numbers (such as overall spin in the case where we are considering the multiple components to be multiple spin states). While some amount of formalism is necessary in this section, it will be minimized.  The more detailed, and more formal, derivations are left to the appendices.

Very generally, we will consider an $N$-body interaction Hamiltonian that may depend on an internal degree of freedom $s_i$ (such as spin) of each particle,
\begin{equation}
H = \sum\limits_{i_1  < i_2  < \ldots < i_N } {V\left( {{\bf{r}}_1, s_1; \ldots ;{\bf{r}}_N, s_N } \right).} 
\label{eqGenericManyBodyHamiltonian}
\end{equation}
We will assume that the interaction $V$ is translationally invariant, and rotationally invariant in positional space (i.e., under rotation of the $\bf r$ variables), but we do not necessarily assume the interaction is invariant under any particular symmetry of the internal degree of freedom $s_i$ (for example, if we are considering particles with spin, we do not assume rotational symmetry in spin space).

Analogously to the approach in the spinless case, to decompose this interaction into pseudopotentials, we need to construct a complete set of states for an $N$-particle cluster. Again factoring out the center-of-mass degree of freedom of the cluster, let us write a complete set of relative wave functions as 
\begin{equation}
|L, {\mathfrak{q}}\rangle,
\label{eqMulticomponentRelativeWaveFunctionBasis}
\end{equation} 
where $L$ is the relative orbital angular momentum of the cluster and $\mathfrak{q}$ enumerates all basis states with this value of $L$ (note that the index $\mathfrak{q}$ indicates not only different spatial wave functions but also the different possible configurations of the internal degree of freedom, e.g., spin). Given such a complete basis, we can always define the corresponding pseudopotentials as the matrix elements of the form:
\begin{equation} 
V_{L,N}^{\mathfrak{q},\mathfrak{q}'} = \left\langle L, \mathfrak{q} \right| V\left( {{\bf{r}}_1, s_1; \ldots ;{\bf{r}}_N, s_N } \right) \left| L, \mathfrak{q}' \right\rangle.
\label{eqDefinitionGeneralizaedPseudopotential}
\end{equation}
Note that due to the rotational invariance of the potential, pseudopotentials are always diagonal in $L$ and, due to the translational invariance of the potential, the center-of-mass degree of freedom does not appear. This expression is analogous to \eref{eqDefinitionMultiParticlePseudopotentials} above.

We can now use a resolution of the identity to rewrite the Hamiltonian, \eref{eqGenericManyBodyHamiltonian}, in the following general form:
\begin{equation}
H = \sum\limits_{ i_1  < \ldots < i_N } {\sum\limits_{L,l \mathfrak{q},\mathfrak{q}'} {\left|L, {\mathfrak{q}} \right\rangle \left|l \right\rangle V_{L,N}^{\mathfrak{q},\mathfrak{q}'} }  \left\langle l \right| \left\langle L, {\mathfrak{q}'} \right|},
\label{eqGeneralizedProjectionHamiltonian}
\end{equation}
analogously to \eref{eqMutliParticleProjectionHamiltonian}.   

Thus, our task in this section is simply to determine the complete linearly independent basis $|L, {\mathfrak{q}}\rangle$  when the particles have an internal degree of freedom.

\subsection{Two Component Case:  Spin 1/2, Bilayers, etc.}
\label{subPseudoTwoComponentResults}

We begin with the simplest and most important multicomponent case: the two-component case.  This case applies, for example, to spin-1/2 fermions, such as (unpolarized) electrons where each fermion has two possible internal states (spin-up and spin-down).   This case also applies to (spin-polarized) bilayers, \cite{eisensteinbook,eisenstein2004a,eisenstein2004b} where the layer index (or iso-spin) corresponds to the two-state system.  We may also consider bosons with two internal states that are frequently called ``spin-1/2 bosons"  (although this nomenclature is not strictly correct).  These two internal states could be two available hyperfine states or potentially two possible layers or any other orbital index. \cite{hormozi2012}  Whatever the origin of the two possible states,  we will use the nomenclature of spin for simplicity.  (Note that in the case of more than two components, which we discuss in \sref{subMultiComponentResults} below, the language of spins becomes somewhat less useful.)

In order to describe wave functions that depend on a spin degree of freedom, it is convenient to work with a basis of states that are eigenstates of spin angular momentum.   These eigenstates are characterized by the spin quantum numbers $S$ and $S_z$, which are the eigenvalues of the combined total spin angular-momentum operator for $N$ particles, $S^2$, and of the combined $z$-component of spin angular-momentum operator for $N$ particles, $S_z$, respectively.

Thus, we propose to write a complete basis of states for $N$ spin-1/2 particles with total orbital relative angular momentum $L$, total spin angular momentum $S^2$, and $z$-component of spin angular momentum $S_z$. We denote this basis as
\begin{equation} 
\left| {L,S,S_z ,q} \right\rangle,
\label{eqSpinRelativeWaveFunctionBasis}
\end{equation}
where $q$ runs from 1 to the number of states in the basis (i.e., the total number of states of $N$ particles having $L$, $S^2$, and $S_z$). In the language of \eref{eqMulticomponentRelativeWaveFunctionBasis}, the index $\mathfrak{q}$ here represents $\{S,S_z,q\}$. 

Our first goal will be to determine the dimension of the space of the wave function basis (the number of $q$ values for a given set of spin eigenvalues and a given $N$, $L$, $S$, and $S_z$), as has been documented for the spinless case [see \eref{eqDimensionFullSymmetricPoly} below or Ref.~\onlinecite{simon2007a}].  The results of our calculation of these parameters for fermionic and bosonic cases are shown in \tref{tableDimensionsOfPolynomialSpaces}. For each $S$, there are always $2S + 1$ different possible values of $S_z$.  The table presents the number of states for all possible values of $S_z$. (Note that some rows of the table are not labeled with a spin quantum number $\sboson$ or $\sfermion$, for the bosonic or fermionic case, which means that the corresponding states cannot occur for a two-component system.)

Our second goal is to describe the forms of the basis wave functions. We shall now summarize our results, leaving the details of the derivation of the forms these wave functions take to Appendices \ref{appendixConstructionOfMulticomponentWaveFunctions}, \ref{appendixGeneralizedSpinWaveFunctions}, and \ref{appendixSpatialWaveFunctions}. As in the spinless case, the basis wave functions, of the form $\left| {L,S,S_z ,q} \right\rangle$,  are equivalent to wave functions describing a small number of spin-1/2 particles. Compared to the spin-polarized case, these basis wave functions are now composed of both a spin part and a spatial part. 

Once we have our complete basis, the set of spin-pseudopotentials can be defined as in \eref{eqDefinitionGeneralizaedPseudopotential}:
\begin{eqnarray}
& & V_{L,N}^{S,S_z ,q;S',S'_z ,q'}  =  \\
& & ~~~~~~~~\left\langle {L,S,S_z ,q} \right|V\left( {{\bf{r}}_1 ,s_1; \ldots; {\bf{r}}_N, s_N} \right)\left| {L,S',S'_z ,q'} \right\rangle \nonumber.
\end{eqnarray}
Using these spin pseudopotentials we can now write down an expression for the Hamiltonian \eref{eqGenericManyBodyHamiltonian} in the general form described by \eref{eqGeneralizedProjectionHamiltonian}.

As always, rotational invariance in the plane ensures that the Hamiltonian is diagonal in $L$. If the system is spin-rotationally invariant then the Hamiltonian is also diagonal in the eigenvalues $S$ and $S_z$. For a more general interaction however, the Hamiltonian might not be spin-rotationally invariant, but, nonetheless, it is still convenient to decompose the interaction using the spin basis.

We now turn to the explicit construction of our complete set $|L,S,S_z,q\rangle$. A wave function for fermions must be overall antisymmetric, whereas a wave function for bosons must be overall symmetric.  However, the wave functions we consider are a combination of both a spin and a spatial part, and only the combination of the two parts needs to have the overall fermionic or bosonic symmetry. The spatial and spin parts of the wave function can have more complicated symmetry as long as the two parts are appropriately sewn together and the combination has the correct overall symmetry. In fact, there is a direct correspondence between the type of symmetry and the spin quantum number $S$. The mathematical structure of forming this combination is discussed in detail in \aref{appendixConstructionOfMulticomponentWaveFunctions} below.  Here, however, we shall simply present the results of this procedure.

The presentation of our results is divided into two parts: first we shall introduce \emph{primitive polynomials}, which are the lowest-degree polynomials corresponding to a particular type of spatial symmetry type for a set of $N$ particles; second we shall describe how to use these primitive polynomials to construct a spatial function of arbitrary degree that still corresponds to a particular symmetry type. Each allowed spatial symmetry corresponds to a particular spin eigenvalue for the $N$ particles. Merging a spatial wave function of a given symmetry type (a given spin quantum number) with a corresponding spin wave function will give an overall wave function with the appropriate fermionic or bosonic symmetry. In a moment we shall discuss how the spatial and spin wave functions are merged.

In Tables~\ref{tablePrimitivePolynomialsFermions} and \ref{tablePrimitivePolynomialsBosons} we list what we have termed primitive polynomials, $\beta_L$, of degree $L$. These polynomials are written in terms of symmetric polynomials of the form of \eref{eqElementarySymmetricPolyDefinition} in terms of the relative coordinates as in \eref{eqTildeCoordinates}. Now, however, we frequently need to write symmetric polynomials in fewer than all $N$ of the $\tilde z$ variables. For compactness, we shall use the following short-hand notation:
\begin{equation}
e_{m ,i_1 i_2 i_3\ldots i_p}\equiv e_{m,p} \left( {\tilde z_{i_1} ,\tilde z_{i_2} ,\tilde z_{i_3} ,\ldots, \tilde z_{i_p}} \right).
\label{eqShortHandSymmetricPolynomial}
\end{equation}
Note that we will always take $\tilde z_i$ to have the center of mass of all $N$ particles subtracted off [as in \eref{eqTildeCoordinates}] independent of the value of $p$.  So, for example,
\[
  e_{2,245} \equiv  e_{2,3}(\tilde z_2,\tilde z_4,\tilde z_5) = \tilde z_2 \tilde z_4 + \tilde z_2 \tilde z_5 + \tilde z_4 \tilde z_5.
\]
We also use the short-hand notation for Jastow factors,
\begin{equation}
J_{i_1 \ldots i_p} = \prod_{k , l \in \{i_1 \ldots i_p\}; k < l }  (\tilde z_k - \tilde z_l),
\label{eqShortHandJastrow}
\end{equation}
so, for example,
\[
 J_{134}  = (\tilde z_1 - \tilde z_3)(\tilde z_1 - \tilde z_4)(\tilde z_3 - \tilde z_4).
\]

Multiplying an $N$-particle spatial wave function by any fully symmetric translationally invariant polynomial does not change its symmetry type. One can, thus, construct spatial wave functions with a given symmetry type and relative angular momentum $L$ by multiplying a primitive polynomial by any fully symmetric (in all $N$ variables) translationally invariant polynomial such that the combined polynomial degree of the resultant product is $L$. Thus, the most general spatial wave function of a particular symmetry and a particular relative angular momentum $L$ is a linear combination of these products of translationally invariant polynomials times primitive polynomials giving a homogeneous polynomial of degree $L$.

We shall now demonstrate how to use the primitive polynomials via a simple example. Consider a three-particle electron wave function with total spin eigenvalue $S=1/2$. For relative angular-momentum eigenvalue $L=1$ we construct from the relevant entry in \tref{tablePrimitivePolynomialsFermions} the only possible degree 1 polynomial conforming to this type of symmetry, namely
\begin{equation}
\beta_1 = J_{12} \equiv \left( {\tilde z_1  - \tilde z_2 } \right).
\label{eqFirstPrimitivePolynomial}
\end{equation}
For $L=2$ we can use the information in \tref{tablePrimitivePolynomialsFermions} to construct the valid degree 2 polynomials conforming to this type of symmetry. Note that, due to \eref{eqTranslationallyInvariantSymmPoly}, $e_{1,123} = 0$, and so we cannot construct an degree 2 polynomial from $J_{12}$ (which is degree 1) multiplied by any degree 1 translationally invariant fully symmetric polynomial. The only possibility according to \tref{tablePrimitivePolynomialsFermions} is the second primitive polynomial:
\[
\beta_2 = J_{12} e_{1,12} \equiv J_{12} \left( { \tilde z_1  + \tilde z_2   } \right).
\]
For $L=3$ we can have only the polynomial given by multiplying the $L=1$ result by a fully symmetric translationally invariant polynomial of degree 2
\[
\left( { \tilde z_1  - \tilde z_2 } \right)e_{2,123}.
\]

For $L=4$ we can have either the $L=1$ primitive polynomial multiplied by a translationally invariant fully symmetric polynomial of degree 3 or the $L=2$ primitive polynomial multiplied by a translationally invariant fully symmetric polynomial of degree 2. The most general result is a linear combination,
\[
\left( {\tilde z_1  - \tilde z_2 } \right)\left[ {A_1 e_{3,123} + A_2 \left( {\tilde z_1  +  \tilde z_2 } \right)e_{2,123} } \right],
\]
with two arbitrary coefficients, $A_1$ and $A_2$. Equivalently, we have two linearly independent basis states in the space of spatial wave functions.   At each polynomial degree the number of linearly independent basis vectors appearing in these polynomials is precisely the dimension appearing in \tref{tableDimensionsOfPolynomialSpaces}: in this case $0,1,1,1,2, \ldots,$ for polynomial degrees $L= 0,1,2,3,4,\ldots,$ and so on.   Since all fully symmetric translationally invariant polynomials are generated by $e_{2,123}$ and $e_{3,123}$, the most general spatial wave function with relative angular momentum $L$ is, thus,
\[
\begin{array}{l}
 \left( { \tilde z_1  - \tilde z_2 } \right)\left\{ {\sum\limits_{\scriptstyle 2l + 3m \hfill \atop
  \scriptstyle  = L - 1 \hfill} {A_{lm} e_{2,123}^l e_{3,123}^m }  + } \right. \\
 \left. {\left( {\tilde z_1  + \tilde z_2   } \right)\sum\limits_{\scriptstyle 2l' + 3m' \hfill \atop
  \scriptstyle  = L - 2 \hfill} {A'_{l'm'} e_{2,123}^{l'} e_{3,123}^{m'} } } \right\}, \\
 \end{array}
\]
where the total number of coefficients $A_{lm}$ and $A'_{l'm'}$ appearing in the wave function takes the value given in \tref{tableDimensionsOfPolynomialSpaces}.

To generate the complete basis wave functions we must combine the spatial wave function with a spin wave function. In this procedure, we follow Ref.~\onlinecite{paunczbook}. We shall define a \emph{primitive spin wave function} $\vartheta_i$ of a many-particle system to be an eigenfunction of the $S_z$ operator of every particle in the system. For example, $\left|\uparrow \downarrow \uparrow \uparrow \downarrow\right\rangle$ is a primitive spin wave function for five particles. We imagine ordering all of the primitive spin wave functions in lexicographical order (an alphabetical, or last letter sequence ordering scheme), so the \emph{first primitive spin wave function} (first in the sense of this ordering scheme) is given by
\begin{equation}
\vartheta_1 \left(N,S_z \right)= \left| { \uparrow  \uparrow  \uparrow \ldots  \downarrow  \downarrow  \downarrow } \right\rangle ,
\label{eqDefinitionFirstPrimitiveSpin}
\end{equation}
where the number of spin-up particles minus the number of spin-down particles times $1/2$ is equal to the total $S_z$ eigenvalue:
\[
{S_z} = \frac{1}{2}\left( {{N_ \uparrow } - {N_ \downarrow }} \right).
\]

To combine our spatial wave function with a spin wave function, we take a spatial wave function corresponding to the spin eigenvalue $S$ (i.e., a primitive polynomial in \tref{tablePrimitivePolynomialsFermions} or \tref{tablePrimitivePolynomialsBosons} times any translationally invariant symmetric polynomial) and then multiply this by the first primitive spin wave function with any valid $S_z$ eigenvalue for that particular $S$ eigenvalue. To recover the complete basis wave function we simply antisymmetrize (for fermions) or symmetrize (for bosons) the resulting combination in all of the coordinates.

Continuing with our example, let us consider the case of $N=3$ where the spin eigenvalue is $S_z = 1/2$.  At degree 1, for example, we multiply the spatial wave function \eref{eqFirstPrimitivePolynomial} with the first primitive spin wave function $\left|\uparrow \uparrow \downarrow\right\rangle$ to construct
\begin{equation}
\left| { \uparrow  \uparrow  \downarrow } \right\rangle \left( { \tilde z_1  -  \tilde z_2 } \right).
\end{equation}
By antisymmetrizing in all the particle coordinates, we then find the three-body wave function,
\begin{equation}
\begin{array}{l}
\left| {L=1,S=1/2,S_z=1/2} \right\rangle  = 2\left| { \uparrow  \uparrow  \downarrow } \right\rangle \left( { \tilde z_1  - \tilde z_2 } \right) \\ - 2\left| { \uparrow  \downarrow  \uparrow } \right\rangle \left( {\tilde z_1  - \tilde z_3 } \right) + 2\left| { \downarrow  \uparrow  \uparrow } \right\rangle \left( { \tilde z_2  - \tilde z_3 } \right). \\
\end{array}
\label{eqUpUpDownExpandedForm}
\end{equation}
Note that we do not write an additional $q$ index here [as in \eref{eqSpinRelativeWaveFunctionBasis}] since there is a unique wave function with this value of $N,L,S,S_z$. Note also that this wave function is not normalized (and normalization depends on an inner product that depends on whether the geometry is on the sphere, plane, or torus ).  We can obtain a similar wave function with $S_z =-1/2$ by simply following the same procedure with the primitive spin wave function $\left|\uparrow \downarrow \downarrow\right\rangle$. An identical result is obtained by instead applying the spin lowering operator $\hat S^- = \sum_i \hat S^-_i$ to \eref{eqUpUpDownExpandedForm}. 

To give another example briefly, to create a bosonic wave function of four particles with angular momentum $S=S_z = 1$ we look at \tref{tablePrimitivePolynomialsBosons} under $N=4, S=1$.  The listed spatial polynomials (such as $e_{1,123} = \tilde z_1 + \tilde z_2 + \tilde z_3$) can be multiplied by any overall symmetric polynomial (which does not change their symmetry).  The spatial wave function must be then multiplied by the appropriate first primitive spin wave function $\left| { \uparrow  \uparrow  \uparrow \downarrow  } \right\rangle$, and, finally, the entire wave function should be fully symmetrized over all particles.

The prescription laid out here is sufficient to construct a complete basis of states of a two-component system of $N$particles having overall relative (orbital) angular momentum $L$ and spin quantum numbers $S$ and $S_z$. For many readers this is the most important result of the current paper.

\subsection{Interlude: Primer on Symmetry Types}
\label{subPseudoSymmetryTypes}

In the above section we explained briefly how we take spatial and spin wave functions of certain symmetry types and sew them together to make a wave function with overall bosonic or fermionic symmetry. In this section we briefly explain in more detail precisely what we mean by a symmetry type.  While it is not essential to understand this material in order to follow the construction of the previous section, in the more complicated cases of the next two sections where there are more than two components to the wave function (e.g., spin-1, spin-3/2, graphene with spin and valley, etc.) it becomes more crucial to understand the concept of symmetry type. The information contained in this section is standard mathematics, reviewed here briefly for convenience. See, for example, Ref.~\onlinecite{hamermeshbook} for a more detailed discussion.

A key intellection we use to describe our basis wave functions is the concept of a symmetry type or antisymmetry type of a wave function (the original method was introduced in \mbox{Ref.~\onlinecite{hund1927})}. These symmetry types are intimately related to irreducible representations of the symmetric group (the group of permutations of $N$ objects). Such irreducible representations are visualized by \emph{Young frames} (closely related to \emph{Young tableaux}),  a series of $N$ boxes arranged in left-justified rows and columns (a more detailed description of the representation theory of the symmetric group and of Young tableaux is provided here in \aref{appendixMathPreliminaries}). We shall denote the shape of a frame containing $N_i$ boxes in the $i$th row by $\lambda = [N_1,N_2,\ldots]$ with $N_i \geq N_j \geq 0$ if $i < j$ and $N =\sum_i N_i$ . Thus, the shape of each frame represents a unique partition of the integer $N$  into pieces $[N_1,N_2,\ldots]$.

We will use the term \emph{conjugate} to refer to Young frames related by a reflection along the diagonal, for example, the following frame of shape $\lambda=[3,1]$,
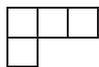
\[
\yng(3,1)
\]
is conjugate to the following frame of shape $\tilde \lambda=[2,1,1]$,
\[
\yng(2,1,1)
\]
Representations of the symmetric group associated with conjugate Young frames are called \emph{contragradient representations}.  We will use the notation $\tilde \lambda$ to denote the conjugate Young frames (or integer partition) to $\lambda$.

We now explain how Young frames can be used to describe the permutation symmetry of a function. Given a Young frame $\lambda = [N_1, N_2, \ldots]$, being an integer partition of $N$, we say a function has \emph{normal form} with symmetry $\lambda$ if (a) it is symmetric in the first $N_1$ variables, symmetric in the next $N_2$ variables, symmetric in the next $N_3$ variables, and so forth, and (b) we cannot use permutation and linear combinations to construct a nonvanishing function that is symmetric in greater than $N_1$ variables, we cannot use permutation of the variables $N_1 + 1, \ldots, N$  and linear combination to construct a nonvanishing function which is symmetric in greater than $N_2$ of these variables, we cannot use permutation of the variables $N_1 + N_2  +1, \ldots, N$ to construct a nonvanishing function symmetric in greater than $N_3$ of these variables, and so forth.  In other words, permutations and linear combinations cannot make the function any more symmetric than it already is. By permutation and linear combination any function can be brought to normal form to clearly show its symmetry type (a detailed method of reduction to normal form is given in Ref.~\onlinecite{hamermeshbook}). Antisymmetry type is entirely analogous except that one should replace the word \emph{symmetric} with \emph{antisymmetric} in the above definition. A nontrivial theorem is that any function that has symmetry type $\lambda$ also has the conjugate antisymmetry type, denoted by $\tilde \lambda$. \cite{hamermeshbook}

One can check that the polynomials given in \tref{tablePrimitivePolynomialsFermions} and \tref{tablePrimitivePolynomialsBosons} are listed by symmetry type. For example, in \tref{tablePrimitivePolynomialsBosons}, the $N=4$ primitive polynomials of symmetry type $[2,2]$ are clearly symmetric in the first two variables [which are $z_1$ and $z_2$, bearing in mind the definitions given in \eref{eqTildeCoordinates} and \eref{eqTranslationallyInvariantSymmPoly}] and then also symmetric in second two variables ($z_3$ and $z_4$). Further, any attempt to symmetrize in four variables ($z_1,\ldots,z_4$) will vanish.  Thus these entries are in normal form.  For  \tref{tablePrimitivePolynomialsFermions} the symmetry type is sometimes less obvious (it is not in symmetric normal form), but the antisymmetry type is evident (the polynomials are listed in antisymmetric normal form).   For example, symmetry type $\lambda = [2,1,1]$ must be equivalent to  antisymmetry type $\tilde \lambda =[3,1]$.  The polynomials labeled by $\lambda =[2,1,1]$ in \tref{tablePrimitivePolynomialsFermions} are indeed antisymmetric in the first three variables ($z_1,\ldots,z_3$) and attempts to antisymmetrize in four variables ($z_1,\ldots,z_4$) will fail.  

Having defined symmetry type, the general procedure to create a wave function with bosonic symmetry involves sewing together a spin and a spatial wave function of the same symmetry type. Similarly to create wave functions of fermionic symmetry one must sew together spin and spatial wave functions with conjugate symmetries.  The details of this procedure are presented in \aref{appendixConstructionOfMulticomponentWaveFunctions} (see also Ref.~\onlinecite{paunczbook}). Here we shall simply state a few key results derived in that appendix. The first result is that, for a two-component system, the spin wave functions of a particular symmetry type are in one-to-one correspondence with spin eigenfunctions, that is, such functions are eigenfunctions of the $\hat S^2$ operator. For example, the following function is an eigenfunction of $\hat S^2$ with eigenvalue $S=S_z = 1/2$:
\[
{\textstyle{1 \over {\sqrt 6 }}}\left( {2\left| { \uparrow  \uparrow  \downarrow } \right\rangle  - \left| { \uparrow  \downarrow  \uparrow } \right\rangle  - \left| { \downarrow  \uparrow  \uparrow } \right\rangle } \right).
\]
This can be shown by explicitly applying an $\hat S^2$ operator. One can also check that this function has symmetry type $[2,1]$, since it is symmetric in the first two arguments. In order to generate a fully antisymmetric combined spin and spatial wave function, the procedure is to combine the above spin wave function with a spatial wave function of conjugate symmetry type and then antisymmerize the resulting construction. To generate a symmetric combined function, we associate that spin eigenfunction with a spatial function of the same symmetry type and then symmetrize that construction.  If, for example, we wanted to generate a three-electron wave function then the corresponding spatial part would be a polynomial of conjugate symmetry type $[2,1]$ (note that $[2,1]$ is self-conjugate).

The second key result shown in \aref{appendixGeneralizedSpinWaveFunctions} and in Ref.~\onlinecite{paunczbook} is that, for a two-component system, the result of the construction procedure described in the previous paragraph can be obtained by using just the first primitive spin wave function [defined in \eref{eqDefinitionFirstPrimitiveSpin}] instead of the full spin eigenfunction. The simplest construction procedure is then to associate a spatial function of a given symmetry type with a first primitive spin wave function only, and then to (anti-) symmetrize or symmetrize that result to obtain an overall (anti-) symmetric combined spin and spatial wave function. The reason for this simplification lies with the (anti-) symmetrization operation: as long as the spatial part corresponds to a particular symmetry type, the (anti-) symmetrization procedure will automatically impose the correct symmetry type on the spin part.

Once we have more than two components in the wave function (higher spin, or multiple spins and valley degrees of freedom) this simplification no longer fully holds, nevertheless our guiding principle is to try to specify a (generalized) spin wave function of the simplest possible form, which can be associated with a spatial wave function of a particular symmetry such that the result of (anti-) symmetrizing that construction is an appropriate basis wave function.

A system built from $N$ particles with $n$ multiple components is described mathematically by the irreducible representations of the Lie algebra of the group SU($n$). For example, we refer to spin-1/2 particles as being described by representations of the Lie algebra of SU(2). Irreducible representations of SU($n$) are in correspondence with a subset of irreducible representations of the symmetric group of $N$ objects \cite{hamermeshbook} and, hence, the symmetry types. The simplest way to visualize this correspondence is via Young frames associated with these irreducible representations. The set of Young frames describing the irreducible representations of SU($n$) corresponds precisely to the subset of symmetric group Young frames containing no more than $n$ rows; the corresponding set of conjugate Young frames contain no more than $n$ columns. There is a corresponding restriction on the possible symmetry types that can be used, for example, in the case of a two-component system the possible symmetries of the spin and spatial wave function are restricted to being symmetric or antisymmetric in two subsets of particle indices.  Hence, in \tref{tablePrimitivePolynomialsBosons} where bosons are considered the only allowed symmetry type for the two-component case (spin-1/2 case) are those where there are only two columns of the frame $\lambda$, whereas in \tref{tablePrimitivePolynomialsFermions} where fermions are considered the only allowed symmetry type for the two-component case (spin-1/2 case) are those where there are only two rows of the frame $\lambda$ (or two columns of the conjugate frames).

\subsection{Extension to Systems with $n$ Components}
\label{subMultiComponentResults}

Having described the general properties of symmetry types, we shall now turn our attention to systems containing $n$ components.  In this case the possible symmetries of the spin and spatial wave functions are symmetric or antisymmetric in up to $n$ subsets of particle indices. The corresponding symmetry types are $\lambda = [N_1, N_2, \ldots ,N_n]$ (for bosons) or conjugate symmetry types $\tilde \lambda$ (for fermions). In order to generate a complete basis for interactions with multicomponent symmetry, we have chosen to decompose the interparticle interaction into a basis of all the possible combinations of symmetry types that the spatial and spin wave functions can have.  The forms of the spatial wave functions remain translationally invariant analytic polynomials.

In \tref{tableDimensionsOfPolynomialSpaces} we have listed the dimensions of the space of polynomials for an $n$-component system ($n$ possible values of the internal degree of freedom) for $N \le 5$ particles. The dimensions are labeled by the symmetry type $\lambda$ of the corresponding pseudopotential basis functions. Additionally, primitive polynomials for all symmetry types with $N \le 5$  are listed in \tref{tablePrimitivePolynomialsFermions} and \tref{tablePrimitivePolynomialsBosons}. We stress that not all symmetry types are available to a given $n$-component system---such a system can only be classified into symmetry types of the form $\lambda = [N_1, N_2, \ldots ,N_n]$ (for bosons) or conjugate symmetry types  $\tilde \lambda$ (for fermions). In particular, for the two-component case, one can see that only certain symmetry types are allowed depending on whether the system is bosonic or fermionic (e.g., the fully symmetric [3] type is not allowed for three fermions).  

In order to construct complete basis wave functions, we also require a modified spin eigenfunction to describe the multiple internal degrees of freedom. We shall refer to this wave function as a \emph{\generalizedspin{}}.  When a particle has an internal degree of freedom $s$, let us notate the different possible values of this degree of freedom by $\alpha$, $\beta$, $\gamma$, and so on. The notation can be interpreted, for example, in the two-component case of spins as, $\alpha \equiv \left| { \uparrow  } \right\rangle$ and $\beta \equiv \left| { \downarrow } \right\rangle$; but now there $n$ possible values that our parameter can take.  For a multiparticle system with multiple internal degrees of freedom, we similarly define primitive \generalizedspin{s} such as
\begin{equation}
\vartheta = \left| { \alpha  \zeta \beta \beta \alpha } \right\rangle.
\label{eqPrimitiveGeneralizedSpinWaveFunction}
\end{equation}
It is useful to define the index $\lambda_z$ which represents the list $\left\{ N_{\alpha,} N_{\beta},\ldots ,N_{\zeta}\right\} $ of how many times each type each component ($\alpha, \beta, \ldots$)  occurs in the primitive \generalizedspin{}.  So, for example, in the primitive  \generalizedspin{} $|\alpha \alpha \gamma\rangle$ for three particles we have the index  $\lambda_z = \left\{ 2,0,1\right\}$. 

A consequence of the increased number of component degrees of freedom is that the complete wave function can no longer be generated solely by using a so-called first primitive \generalizedspin{} [first in the sense of a lexicographic ordering scheme, in analogy with \eref{eqDefinitionFirstPrimitiveSpin} for the two-component case].  Instead, we consult a list of the primitive components that occur for each symmetry type in order to generate a complete basis. These results are presented for the case of three-component and four-component systems in \tref{tableGeneralizedSpinWaveFunctions}.

As an example, consider the three-component $N=3$, $\lambda=[2,1]$, and $\lambda_z=\left\{ 1,1,1 \right\}$ state.  From \tref{tableGeneralizedSpinWaveFunctions} we see that there is a choice of two possible primitive \generalizedspin{s}, $| \alpha \beta \gamma \rangle$ or $|\alpha \gamma \beta\rangle$. Using these functions to construct a complete basis wave function will generate two orthogonal wave functions. (An interesting aside is that the three-component, $N=3$, \generalizedspin{s} are identical to the colour wave functions describing baryons in the SU(3) quark model; see, for example, Ref.~\onlinecite{halzenbook}.)

The problem of constructing $n$-component $N$-particle wave functions with prescribed symmetry is equivalent to the problem of constructing tensor products of $N$ fundamental irreducible representations of the Lie algebra of SU($n$). We shall explain the mathematical details of that procedure in \aref{appendixMathPreliminaries}. Our general aim is to decompose our parameter space of wave functions into a basis of wave functions labeled by irreducible representations of the Lie algebra of SU($n$). We shall describe the general method for selecting the appropriate primitive  \generalizedspin{s} in \aref{appendixGeneralizedSpinWaveFunctions}. Due to symmetry arguments, note that it is not possible to construct linear combinations of certain types of primitive \generalizedspin{} corresponding to certain symmetry types. For example, it is clearly not possible to generate a fully antisymmetric  \generalizedspin{} from $\left| {\alpha \ldots  \alpha}\right\rangle $.

The method to construct a full basis wave function is similar to the two-component case: first we choose a primitive  \generalizedspin{} from \tref{tableGeneralizedSpinWaveFunctions}, corresponding to a given symmetry type $\lambda$ and having a given $\lambda_z$. For a boson wave function we construct a polynomial corresponding to a symmetry $\lambda$ and at a given degree $L$ using the primitive polynomials in \tref{tablePrimitivePolynomialsBosons}.  The full basis wave function is generated by putting these two parts together and then fully symmetrizing. To obtain a fermion wave function, we would instead construct a polynomial corresponding to a symmetry type $\tilde \lambda$ using the primitive polynomials given in \tref{tablePrimitivePolynomialsFermions}, and the complete basis function is given by fully antisymmetrizing. We shall explain how these results were derived in Appendices \ref{appendixConstructionOfMulticomponentWaveFunctions}, \ref{appendixGeneralizedSpinWaveFunctions}, and \ref{appendixSpatialWaveFunctions}.

To obtain the total size $p(N,L)$ of the space of pseudopotentials, $|L, {\mathfrak{q}} \rangle$, for a given $L$, we need to include both the number of polynomials $p(N,L,\lambda)$ with a given symmetry type $\lambda$ and the number of \generalizedspin{s} for that same $\lambda$ but with different values of $\lambda_z$ (for example, in the two-component case the spin wave functions have $2S+1$ values of $S_z$). For clarity, let us discuss an example: a system of $N=4$ bosons with $n=3$ components.  Say that we want to construct the basis space of wave functions for an interaction at $L=4$. The general form of the basis wave functions is $|L=4, {\mathfrak{q}} \rangle$. In this case, the possible symmetry types that the wave function can have are $\lambda = [4], [3,1], [2,2]$, and $[2,1,1]$. Looking at \tref{tableDimensionsOfPolynomialSpaces}, the wave functions with these symmetry types have  the respective polynomial space dimensions of $p(4,4,[4])=2$, $p(4,4,[3,1])=2$, $p(4,4,[2,2])=2$, and $p(4,4,[2,1,1])=1$. Further, the corresponding dimensions of the space of \generalizedspin{s} are given by counting all of the primitive \generalizedspin{s} in \tref{tableGeneralizedSpinWaveFunctions}, which are respectively 15, 15, 6, and 3. The total dimension of the space of $L=4$ pseudopotentials, that is, the total number, $p(N,L)$,  of possible values of $\mathfrak{q}$, is, thus, $15 \times 2 + 15 \times 2 + 6 \times 2 + 3 \times 1 = 75$.  Each of these basis wave functions is obtained by multiplying the spatial part of the wave function by the  \generalizedspin{} and symmetrizing (for bosons in this case).   The pseudopotential matrix at $L=4$ would then be a 75-dimensional Hermitian matrix!

In a system with $n$ components, if the Hamiltonian has full SU($n$) symmetry, then the pseudopotential Hamiltonian, \eref{eqGeneralizedProjectionHamiltonian}, will be diagonal in both $\lambda$ and $\lambda_z$ variables.

\subsection{Systems with Spin Rotational Symmetry}

Often when we consider particles with $n$ internal states (components), these multiple states may have arisen from particles that had spin $J = (n-1)/2$.   If the Hamiltonian is spin-rotationally invariant, then it is very useful to consider a set of basis wave functions that are eigenfunctions of $S^2$ and $S_z$ such that the Hamiltonian \eref{eqGenericManyBodyHamiltonian} is diagonal in these variables.  In fact, even if the Hamiltonian is not quite spin-rotationally invariant, it may be very convenient to work in such a spin basis.   We, thus, would like to construct basis states which correspond to both a given symmetry type and to a given $S^2$ eigenvalue via linear combinations of the primitive  \generalizedspin{s} given in \tref{tableGeneralizedSpinWaveFunctions}. Mathematically speaking, this method corresponds to the decomposition of representations of the group SU($n$) into representations of SU(2); the procedure for doing the decomposition is described in Ref.~\onlinecite{hamermeshbook}. For a system constructed from spins of magnitude $J$, it is to be understood that our notation should be interpreted as $\alpha$ corresponding to the highest spin state $J$, $\beta$ corresponding to spin $J-1$, and so on. The total $z$-component of spin angular-momentum eigenvalue is given by
\[
S_z = J N_\alpha + (J-1) N_\beta +(J-2) N_\gamma \ldots -J N_\zeta.
\]
For spin-1 systems, appropriate linear combinations of the primitive \generalizedspin{s} forming spin-1 eigenfunctions are listed in \tref{tableSpinOneEigenfunctions}. Note that these spin eigenfunctions are listed such that $S=S_z$, and so additional spin eigenfunctions can be constructed by explicitly lowering the $S_z$ eigenvalues using the $\hat S^-$ operator. Due to the increased spin degree of freedom, we find that there can now be several values of $S$ associated with each symmetry type $\lambda$ (by contrast, in the spin-1/2 case where the spin eigenvalue $S$ corresponds to a unique symmetry type). In general, it is also possible for multiple occurrences of the same $S$ value to occur for a given symmetry type; however, this feature only occurs for $N>5$. For systems constructed from spin-3/2 particles the procedure is along similar lines and we shall outline the general method for constructing such spin eigenfunctions for arbitrary spin in \aref{appendixGeneralizedSpinWaveFunctions}.

The method to construct a basis wave function is once again the same as in the two-component case. This time we start by choosing a one of the linear combinations of primitive \generalizedspin{s} from \tref{tableSpinOneEigenfunctions} corresponding to a given $\lambda$ and a given $S^2$ eigenvalue. The linear combination of \generalizedspin{s} will be associated with a polynomial corresponding to that symmetry $\lambda$ for bosons (or conjugate symmetry $ \tilde \lambda$ in the case of fermions) and at a given order $L$. The complete basis wave function is given by symmetrizing (or antisymmetrizing) the product of spin and spatial parts. We shall explain how these results were derived in Appendices~\ref{appendixConstructionOfMulticomponentWaveFunctions} and \ref{appendixGeneralizedSpinWaveFunctions}.

\subsection{Spin and Valley: the Case of Graphene}

More generally, a particle may have several different internal degrees of freedom.  A particularly important example of this is graphene which has two spin states and two valley states resulting in four possible internal states.    The Hamiltonian may have full SU(4) symmetry or, for example, it could be symmetric only under the SU(2) rotations of the spin. \cite{toke2007,goerbig2007,lim2011} 

It is useful always to think of this in terms of a full SU(4) symmetry (which may be broken). We have already enumerated all possible spatial wave functions with symmetry types corresponding to SU(4) (for $N \le 5$): the dimensions of the corresponding polynomial spaces are all contained within \tref{tableDimensionsOfPolynomialSpaces}, and corresponding spatial wave functions for these symmetries are in Tables~\ref{tablePrimitivePolynomialsFermions} and \ref{tablePrimitivePolynomialsBosons}, respectively, for fermions and bosons.  However, in the case where the system has an SU(2) spin rotational symmetry in each valley [but not the full SU(4) symmetry] it is useful to decompose these states into their $S^2$ eigenstates to exploit the symmetries of the problem as much as possible. Mathematically speaking we are decomposing SU(4) representations into representations of SU(2) $\times$ SU(2). 

In our \generalizedspin{} notation, it is to be understood that in this situation we interpret $\alpha$ as corresponding to spin-up in the first valley, $\beta$ as corresponding to spin-down in the first valley, and, similarly, $\gamma$ and $\delta$ as representing spin-up and spin-down states in the second valley. The total $z$-component of spin angular momentum eigenvalue is, thus, given by
\[
S_z = {1\over{2}} (N_\alpha +N_\gamma) - {1\over{2}} (N_\beta + N_\delta).
\]
We can then construct  spin eigenfunctions by choosing appropriate linear combinations of primitive \generalizedspin{s}.   In this case there are necessarily multiple linearly independent $S^2$ eigenstates corresponding to each SU(4) symmetry type.   The appropriate linear combinations of primitive \generalizedspin{s} are given in \tref{tableGrapheneBasis}.   

To clarify, each primitive \generalizedspin{} is labeled by $\lambda$, $S$, and $S_z$, and, as shown in the table, there may be several basis wave functions with the same value of all of these variables and we can distinguish amoung these cases by introducing a further index $k$. Note that in the table only states with $S=S_z$ are shown; other values of $S_z$ are obtained by applying a lowering operator. 

Given any primitive \generalizedspin{} in \tref{tableGrapheneBasis} we follow the same procedure as in the above cases; for bosons (fermions) we multiply it this primitive  \generalizedspin{} by a polynomial with the same (conjugate) symmetry type $\lambda$ ($\tilde \lambda$) and then (anti-) symmetrize over all variables.

In \aref{appendixGeneralizedSpinWaveFunctions} we shall explain how we derived the spin eigenfunctions in  \tref{tableGrapheneBasis}. In that appendix we shall also explain further possible extensions to the cases given here. For example, we shall explain how to go about constructing an appropriate pseudopotential basis to describe systems where there are more than two interacting degrees of freedom per particle or systems where the degrees of freedom themselves have more than two possible values (i.e., are higher spin than 1/2).

\subsection{Tables}

All of the results presented in the tables here have been generated using a {MATHEMATICA}-based computer program. \cite{multicomponentWaveFunctionProgram} The program is capable of producing spatial and \generalizedspin{s} for systems of up to $N=6$ and for an arbitrary number of components, although we only have space to give a selection of results for the most relevant cases here.

%%%%%%%%%%%%%%%%%%%%%%%%%%%%%%%%%%

\clearpage

\onecolumngrid
\begin{center}
\begin{table}[t]
\centering
\begin{tabular*}{\textwidth}{@{\extracolsep{\fill}} c @{\extracolsep{\fill}} c @{\extracolsep{\fill}} c @{\extracolsep{\fill} c  } 
@{\extracolsep{\fill}} c  @{\extracolsep{\fill}} c  @{\extracolsep{\fill}} c  @{\extracolsep{\fill}} c  @{\extracolsep{\fill}} c  @{\extracolsep{\fill}} c  @{\extracolsep{\fill}} c  @{\extracolsep{\fill}} c  @{\extracolsep{\fill}} c  @{\extracolsep{\fill}} c  @{\extracolsep{\fill}} c  @{\extracolsep{\fill}} c  @{\extracolsep{\fill}} c  @{\extracolsep{\fill}} c  @{\extracolsep{\fill}} c  @{\extracolsep{\fill}} c  @{\extracolsep{\fill}} c  @{\extracolsep{\fill}}}
\hline \hline
\multicolumn{1}{c}{\multirow{2}{*}{$N$}} & \multicolumn{1}{c}{\multirow{2}{*}{ $\lambda$}} & \multicolumn{1}{c}{\multirow{2}{*}{$\sfermion$}} & \multicolumn{1}{c|}{\multirow{2}{*}{$\sboson$}} &  \multicolumn{16}{c}{L} \\
 \multicolumn{4}{c|}{} &  \multicolumn{1}{c}{0} & 1 & 2 & 3 & 4 & 5 & 6 & 7 & 8 & 9 & 10 & 11 & 12 & 13 & 14 & 15 \\
 \hline
\multicolumn{1}{c}{\multirow{2}{*}{2}} & \multicolumn{1}{c}{[2]}  & \multicolumn{1}{c}{0} & \multicolumn{1}{c|}{1} & 1 & 0 & 1 & 0 & 1 & 0 & 1 & 0 & 1 & 0 & 1 & 0 & 1 & 0 & 1 & \multicolumn{1}{c}{0}\\
\multicolumn{1}{c}{}  & \multicolumn{1}{c}{[1,1]} & \multicolumn{1}{c}{1} & \multicolumn{1}{c|}{0} & 0 & 1 & 0 & 1 & 0 & 1 & 0 & 1 & 0 & 1 & 0 & 1 & 0 & 1 & 0 & \multicolumn{1}{c}{1}\\
 \hline
\multicolumn{1}
{c}{\multirow{3}{*}{3}} & \multicolumn{1}{c}{[3]} & \multicolumn{1}{c}{--} & \multicolumn{1}{c|}{3/2} & 1 & 0 & 1 & 1 & 1 & 1 & 2 & 1 & 2 & 2 & 2 & 2 & 3 & 2 & 3  &\multicolumn{1}{c}{3}\\
\multicolumn{1}{c}{}   & \multicolumn{1}{c}{[2,1]} & \multicolumn{1}{c}{1/2} & \multicolumn{1}{c|}{1/2} & 0 & 1 & 1 & 1 & 2 & 2 & 2 & 3 & 3 & 3 & 4 & 4 & 4 & 5 & 5 & \multicolumn{1}{c}{5}\\
\multicolumn{1}{c}{} & \multicolumn{1}{c}{[1,1,1]} & \multicolumn{1}{c}{3/2} & \multicolumn{1}{c|}{--} & 0 & 0 & 0 & 1 & 0 & 1 & 1 & 1 & 1 & 2 & 1 & 2 & 2 & 2 & 2 &  \multicolumn{1}{c}{3}\\
 \hline
\multicolumn{1}{c}{\multirow{5}{*}{4}} & \multicolumn{1}{c}{[4]} & \multicolumn{1}{c}{--} & \multicolumn{1}{c|}{2}& 1 & 0 & 1 & 1 & 2 & 1 & 3 & 2 & 4 & 3 & 5 & 4 & 7 & 5 & 8 &\multicolumn{1}{c}{7}\\
\multicolumn{1}{c}{} & \multicolumn{1}{c}{[3,1]} & \multicolumn{1}{c}{--} & \multicolumn{1}{c|}{1}& 0 & 1 & 1 & 2 & 2 & 4 & 4 & 6 & 6 & 9 & 9 & 12 & 12 & 16 & 16 & \multicolumn{1}{c}{20}\\
\multicolumn{1}{c}{}   & \multicolumn{1}{c}{[2,2]} & \multicolumn{1}{c}{0} & \multicolumn{1}{c|}{0}& 0 & 0 & 1 & 0 & 2 & 1 & 3 & 2 & 5 & 3 & 7 & 5 & 9 & 7 & 12 & \multicolumn{1}{c}{9}\\
\multicolumn{1}{c}{}   & \multicolumn{1}{c}{[2,1,1]} & \multicolumn{1}{c}{1} & \multicolumn{1}{c|}{--} & 0 & 0 & 0 & 1 & 1 & 2 & 2 & 4 & 4 & 6 & 6 & 9 & 9 & 12 & 12 & \multicolumn{1}{c}{16}\\
\multicolumn{1}{c}{} & \multicolumn{1}{c}{[1,1,1,1]} & \multicolumn{1}{c}{2} & \multicolumn{1}{c|}{--} & 0 & 0 & 0 & 0 & 0 & 0 & 1 & 0 & 1 & 1 & 2 & 1 & 3 & 2 & 4 & \multicolumn{1}{c}{3}\\
 \hline
\multicolumn{1}{c}{\multirow{6}{*}{5}} & \multicolumn{1}{c}{[5]} & \multicolumn{1}{c}{--} & \multicolumn{1}{c|}{5/2} & 1 & 0 & 1 & 1 & 2 & 2 & 3 & 3 & 5 & 5 & 7 & 7 & 10 & 10 & 13 & \multicolumn{1}{c}{14}\\
\multicolumn{1}{c}{}  & \multicolumn{1}{c}{[4,1]} & \multicolumn{1}{c}{--} & \multicolumn{1}{c|}{3/2} & 0 & 1 & 1 & 2 & 3 & 4 & 6 & 8 & 10 & 13 & 16 & 20 & 24 & 29 & 34  & \multicolumn{1}{c}{40}\\
\multicolumn{1}{c}{}  & \multicolumn{1}{c}{[3,2]} & \multicolumn{1}{c}{--} & \multicolumn{1}{c|}{1/2} & 0 & 0 & 1 & 1 & 2 & 3 & 5 & 6 & 9 & 11 & 15 & 18 & 23 & 27 & 34  & \multicolumn{1}{c}{39}\\
\multicolumn{1}{c}{} & \multicolumn{1}{c}{[3,1,1]} & \multicolumn{1}{c}{1/2} & \multicolumn{1}{c|}{--} & 0 & 0 & 0 & 1 & 1 & 3 & 3 & 6 & 7 & 11 & 13 & 18 & 21 & 28 & 32 & \multicolumn{1}{c}{41}\\
\multicolumn{1}{c}{} & \multicolumn{1}{c}{[2,1,1,1]} & \multicolumn{1}{c}{3/2} & \multicolumn{1}{c|}{--} & 0 & 0 & 0 & 0 & 0 & 0 & 1 & 1 & 2 & 3 & 4 & 6 & 8 & 10 & 13 & \multicolumn{1}{c}{16}\\
\multicolumn{1}{c}{}  & \multicolumn{1}{c}{[1,1,1,1,1]} & \multicolumn{1}{c}{5/2} & \multicolumn{1}{c|}{--} & 0 & 0 & 0 & 0 & 0 & 0 & 0 & 0 & 0 & 0 & 1 & 0 & 1 & 1 & 2 & \multicolumn{1}{c}{2}\\
 \hline  \hline
\end{tabular*}
\caption{Dimensions of the polynomial spaces: the number of independent parameters occurring in the spatial part of the wave function at degree $L$ for quantum Hall states containing $N$ particles and with symmetry type $\lambda$. Where applicable to the two-component case, which is described in terms of spin eigenfunctions, we list the spin eigenvalues $S$ corresponding to either a fermionic ($\sfermion$) or a bosonic ($\sboson$) wave function. For each $S$ listed there will be $2S+1$ possible values of $S_z$. The symmetry types labeled by $\lambda$ are introduced in \sref{subPseudoSymmetryTypes}. We explain in the text that not all symmetry types $\lambda$ correspond to a spin eigenvalue $\sfermion$ or $\sboson$, and we indicate these cases  by a dash  in the table.}
\label{tableDimensionsOfPolynomialSpaces}
\end{table}
\end{center}

%%%%%%%%%%%%%%%%%%%%%%%%%%%%%%%%%%

\begin{table}[H]
\centering
\vspace{-1em}
\subfloat[][$N=2$]{
\begin{tabular}{ccl}
\hline \hline
\multicolumn{1}{c}{$\lambda$ }  &  $\sfermion$ &  \multicolumn{1}{|c}{Primitive Polynomials}\\[2pt]
\hline
\multicolumn{1}{c}{$[2]$} &  0 &    $\begin{array}{l}  1 \end{array}$   \\
\multicolumn{1}{c}{$[1,1]$}  &  1 &   $\begin{array}{l} J_{12} \end{array}$  \\
\hline \hline
\end{tabular}
}
\\[1em]
\subfloat[$N=3$]{
\begin{tabular}{cc|l}
\hline \hline
\multicolumn{1}{c}{$\lambda$}  &  $\sfermion$ &  \multicolumn{1}{|c}{Primitive Polynomials}   \\[2pt]
\hline
\multicolumn{1}{c}{$[3]$} & -- &   $\begin{array}{l} 1  \end{array}$    \\
\hline
\multicolumn{1}{c}{$[2,1]$} & 1/2 &    $\begin{array}{l} \beta_1 = J_{12} \\ \beta_2 = J_{12}(e_{1,12}) \end{array}$   \\
\hline
\multicolumn{1}{c}{$[1,1,1]$}  & 3/2 & $\begin{array}{l} J_{123} \end{array}$  \\
\hline \hline
\end{tabular}
}
\qquad
\subfloat[$N=4$]{
\begin{tabular}{cc|l}
\hline \hline
\multicolumn{1}{c}{$\lambda$}  &  $\sfermion$ &  \multicolumn{1}{|c}{Primitive Polynomials}   \\[2pt]
\hline
\multicolumn{1}{c}{$[4]$} & -- &   $\begin{array}{l}  1 \end{array}$    \\
\hline
\multicolumn{1}{c}{$[3,1]$} & -- &   $\begin{array}{l} \beta_1 = J_{12} \\ \beta_2 = J_{12}(e_{1,12}) \\ \beta_3 = J_{12} \left(3 \tilde z_3 e_{1,12}+e_{1,12}^2-e_{2,12}+3 \tilde z_3^2\right) \end{array}$    \\[1.6em]
\hline
\multicolumn{1}{c}{$[2,2]$}  & 0 &  $\begin{array}{l}  \beta_2 = J_{12} J_{34} \\ \beta_4 = J_{12} J_{34} (e_{1,12}^2) \end{array}$ \\
\hline
\multicolumn{1}{c}{$[2,1,1]$}  & 1 &  $\begin{array}{l} \beta_3 = J_{123} \\ \beta_4 = J_{123}(e_{1,123})\\ \beta_5 =J_{123}(e_{2,123}) \end{array}$ \\
\hline
\multicolumn{1}{c}{$[1,1,1,1]$}  & 2 &  $\begin{array}{l}  J_{1234} \end{array}$ \\
\hline \hline
\end{tabular}
}
\\
\subfloat[$N=5$]{
\begin{tabular}{cc|l}
\hline \hline
\multicolumn{1}{c}{$\lambda$}  &  $S_{\mbox{\small fermion}}$ &  \multicolumn{1}{|c}{Primitive Polynomials}   \\[2pt]
\hline
\multicolumn{1}{c}{$[5]$} & -- &   $\begin{array}{l} 1 \end{array}$    \\
\hline
\multicolumn{1}{c}{$[4,1]$} & -- &  $\begin{array}{l} \beta_1 = J_{12} \\ \beta_2 = J_{12}( e_{1,12})\\ \beta_3 = J_{12} \left(3 e_{1,12}^2-3 e_{2,12}+2 e_{2,345}\right)  \\ \beta_4 = J_{12} \left(5 e_{1,12}^3-4 e_{2,12} e_{1,12}-6 e_{2,345} e_{1,12}-6 e_{3,345}\right)\end{array}$    \\[2.3em]
\hline
\multicolumn{1}{c}{$[3,2]$}  & -- & $\begin{array}{l} \beta_2 = J_{12} J_{34} \\ \beta_3 =J_{12} J_{34} \left(e_{1,12}+e_{1,34}\right) \\ \beta_4 = J_{12} J_{34} \left(e_{1,12}^2+2 e_{1,34} e_{1,12}+e_{1,34}^2-e_{2,12}-e_{2,34}\right) \\ \beta_5 =J_{12} J_{34} \left(e_{1,12}^3+5 e_{1,34} e_{1,12}^2+5 e_{1,34}^2 e_{1,12}+4 e_{2,12} e_{1,12}+e_{2,34} e_{1,12}+e_{1,34}^3+e_{1,34} e_{2,12}+4 e_{1,34} e_{2,34}\right) \\ \beta_6=J_{12} J_{34} \left(\begin{array}{l} 2 e_{1,12}^4+6 e_{1,34} e_{1,12}^3+8 e_{1,34}^2 e_{1,12}^2-e_{2,12} e_{1,12}^2-2 e_{2,34} e_{1,12}^2+6 e_{1,34}^3 e_{1,12}+2 e_{2,12} e_{2,34}\\-e_{1,34} e_{2,12} e_{1,12}-e_{1,34} e_{2,34} e_{1,12}+2 e_{1,34}^4+2 e_{2,12}^2+2 e_{2,34}^2-2 e_{1,34}^2 e_{2,12}-e_{1,34}^2 e_{2,34}\end{array} \right) \\ \end{array}$ \\[3.8em]
\hline
\multicolumn{1}{c}{$[3,1,1]$}  & -- & $\begin{array}{l} \beta_3 = J_{123} \\ \beta_4 = J_{123}\left( e_{1,123}\right)\\ \beta_5 = J_{123}\left(A_1 e_{1,123}^2-2 A_1 e_{2,45}+2 A_2 e_{2,123}\right)\\ \beta_6 = J_{123} \left(3 e_{1,123}^3+5 e_{2,45} e_{1,123}-e_{2,123} e_{1,123}-2 e_{3,123}\right) \\ \beta_7 =J_{123}\left(5 e_{1,123}^4-6 e_{2,45} e_{1,123}^2-3 e_{3,123} e_{1,123}+6 e_{2,45}^2-6 e_{2,45} e_{2,123}\right) \end{array}$ \\[3em]
\hline
\multicolumn{1}{c}{$[2,2,1]$}  & 1/2 &  $\begin{array}{l}  \beta_4 = J_{123} J_{45} \\ \beta_5 = J_{123} J_{45}\left(e_{1,123}\right)\\ \beta_6 = J_{123} J_{45} \left(e_{1,123}^2+e_{2,45}-e_{2,123}\right)\\ \beta_7 = J_{123} J_{45} \left(3 e_{1,123}^3-2 e_{2,45} e_{1,123}+e_{3,123}\right)\\ \beta_8 =J_{123} J_{45} \left(e_{1,123}^4+3 e_{2,45} e_{1,123}^2+e_{2,123} e_{1,123}^2-e_{3,123} e_{1,123}+e_{2,123}^2-e_{2,45} e_{2,123}\right)\end{array}$ \\[3em]
\hline
\multicolumn{1}{c}{$[2,1,1,1]$}  & 3/2 &  $\begin{array}{l}  \beta_6 = J_{1234} \\ \beta_7 = J_{1234}\left( e_{1,1234}\right) \\ \beta_8 = J_{1234}\left(e_{1,1234}^2\right) \\ \beta_9 = J_{1234}\left( e_{1,1234}^3\right)  \end{array}$ \\[2.3em]
\hline
\multicolumn{1}{c}{$[1,1,1,1,1]$}  & 5/2 & $\begin{array}{l} J_{12345} \end{array}$   \\
\hline \hline
\end{tabular}
}
\caption{Primitive polynomials $\beta_L$ of degree $L$ for construction of \emph{fermion} wave functions, listed by their symmetry type $\lambda$. Note that for compactness we use the notation $e_{m,i_1 i_2  \ldots i_p}$ defined in \eref{eqShortHandSymmetricPolynomial} to be the elementary symmetric polynomial of degree $m$ in the $p$ variables $\tilde z_{i_1}, \ldots, \tilde z_{i_p}$, and we similarly write Jastrow factors on $p$ variables as $J_{i_1 \ldots i_p}$ as defined in \eref{eqShortHandJastrow}. The column labeled $\sfermion$ is in reference to the case of spin-1/2 (two component) particles only; entries labeled with a dash do not occur for the spin-1/2 case.  The symmetry type $\lambda$ pertains more generally.}
\label{tablePrimitivePolynomialsFermions}
\end{table}

%%%%%%%%%%%%%%%%%%%%%%%%%%%%%%%%%%

\begin{table}[H]
\centering
\subfloat[$N=2$]{
\begin{tabular}{cc|l}
\hline \hline
\multicolumn{1}{c}{$\lambda$}  &  $\sboson$ &  \multicolumn{1}{|c}{Primitive Polynomials} \\[2pt]
\hline
\multicolumn{1}{c}{$[2]$} &  1 &  $\begin{array}{l} 1 \end{array}$     \\
\multicolumn{1}{c}{$[1,1]$}  &  0 &   $\begin{array}{l}  J_{12} \end{array}$   \\
\hline \hline
\end{tabular}
}
\qquad
\subfloat[$N=3$]{
\begin{tabular}{cc|l}
\hline \hline
\multicolumn{1}{c}{$\lambda$}  &  $\sboson$ &  \multicolumn{1}{|c}{Primitive Polynomials}   \\[2pt]
\hline
\multicolumn{1}{c}{$[3]$} & 3/2 &   $\begin{array}{l} 1 \end{array}$   \\
\hline
\multicolumn{1}{c}{$[2,1]$} & 1/2 &  $\begin{array}{l}  \beta_1 = e_{1,12} \\ \beta_2 =  e_{1,12}^2+2 e_{2,12}\end{array}$    \\[0.8em]
\hline
\multicolumn{1}{c}{$[1,1,1]$}  & -- & $\begin{array}{l}  J_{123} \end{array}$ \\
\hline \hline
\end{tabular}
}
\\
\subfloat[$N=4$]{
\begin{tabular}{cc|l}
\hline \hline
\multicolumn{1}{c}{$\lambda$}  &  $\sboson$ &  \multicolumn{1}{|c}{Primitive Polynomials}   \\[2pt]
\hline
\multicolumn{1}{c}{$[4]$} & 2 &   $\begin{array}{l} 1 \end{array}$ \\
\hline
\multicolumn{1}{c}{$[3,1]$} & 1 &    $\begin{array}{l} \beta_1 = e_{1,123} \\ \beta_2 = \left(e_{1,123}^2+e_{2,123}\right) \\ \beta_3 =  \left(e_{1,123} e_{2,123}+3 e_{3,123}\right)\end{array}$   \\
\hline
\multicolumn{1}{c}{$[2,2]$}  & 0 & $\begin{array}{l}  \beta_2 = \left(e_{1,12}^2+2 e_{2,12}+2 e_{2,34}\right) \\ \beta_4 = \left(e_{1,12}^4+4 e_{2,12} e_{1,12}^2+4 e_{2,34} e_{1,12}^2-2 e_{2,12}^2-2 e_{2,34}^2-12 e_{2,12} e_{2,34}\right) \end{array}$   \\[1em]
\hline
\multicolumn{1}{c}{$[2,1,1]$}  & -- & $\begin{array}{l}  \beta_3 = 3 \tilde z_3^2 e_{1,12}- \tilde z_3 e_{1,12}^2+2 \tilde z_3 e_{2,12}-e_{1,12}^3+e_{1,12} e_{2,12}+2 \tilde z_3^3 \\ \beta_4 =   \tilde z_3^3 e_{1,12}-3  \tilde z_3^2 e_{1,12}^2- \tilde z_3 e_{1,12}^3+6  \tilde z_3 e_{1,12} e_{2,12}+3 e_{1,12}^2 e_{2,12} \\ \beta_5 =  \tilde z_3 e_{1,12}^4+3  \tilde z_3^2 e_{1,12}^3+2  \tilde z_3^3 e_{1,12}^2-4  \tilde z_3 e_{2,12} e_{1,12}^2 \\ -6 \tilde z_3^2 e_{2,12} e_{1,12}-4  \tilde z_3 e_{2,12}^2-4  \tilde z_3^3 e_{2,12}-e_{2,12} e_{1,12}^3-2 e_{2,12}^2 e_{1,12} \end{array}$  \\[2.5em]
\hline
\multicolumn{1}{c}{$[1,1,1,1]$}  & -- & $\begin{array}{l}  J_{1234} \end{array}$  \\
\hline \hline
\end{tabular}
}
\\
\subfloat[$N=5$]{
\noindent\makebox[\textwidth]{%	centered figures wider than text width
\begin{tabular}{cc|l}
\hline \hline
\multicolumn{1}{c}{$\lambda$}  &  $S_{\mbox{\small boson}}$ &  \multicolumn{1}{|c}{Primitive Polynomials}   \\[2pt]
 \hline
\multicolumn{1}{c}{$[5]$} & 5/2 &   $\begin{array}{l}  1\end{array}$   \\
 \hline
\multicolumn{1}{c}{$[4,1]$} & 3/2 &  $\begin{array}{l}  \beta_1 = e_{1,1234} \\ \beta_2 = \left(3 e_{1,1234}^2+2 e_{2,1234}\right) \\ \beta_3 =  \left(5 e_{1,1234}^3-3 e_{2,1234} e_{1,1234}+3 e_{3,1234}\right) \\ \beta_4 = \left(3 e_{1,1234}^4+4 e_{2,1234} e_{1,1234}^2-4 e_{3,1234} e_{1,1234}-2 e_{2,1234}^2+4 e_{4,1234}\right) \end{array}$  \\[2.3em]
 \hline
\multicolumn{1}{c}{$[3,2]$}  & 1/2 &  $\begin{array}{l}   \beta_2 = \left(e_{1,123}^2+3 e_{2,45}+e_{2,123}\right) \\ \beta_3 = \left(e_{1,123} e_{2,45}+e_{1,123} e_{2,123}+3 e_{3,123}\right)\\ \beta_4 = \left(e_{1,123}^4+6 e_{2,45} e_{1,123}^2+2 e_{2,123} e_{1,123}^2-2 e_{3,123} e_{1,123}-3 e_{2,45}^2-e_{2,123}^2-12 e_{2,45} e_{2,123}\right) \\ \beta_5 = \left(e_{2,45} e_{1,123}^3+e_{2,123} e_{1,123}^3+7 e_{3,123} e_{1,123}^2-e_{2,45}^2 e_{1,123}-e_{2,123}^2 e_{1,123}-12 e_{2,45} e_{3,123}-3 e_{2,123} e_{3,123}\right) \\ \beta_6= \left( \begin{array}{l} 2 e_{1,123}^6+18 e_{2,45} e_{1,123}^4+6 e_{2,123} e_{1,123}^4-6 \
e_{3,123} e_{1,123}^3-27 e_{2,45}^2 e_{1,123}^2-9 e_{2,123}^2 \
e_{1,123}^2\\-60 e_{2,45} e_{2,123} e_{1,123}^2-60 e_{2,45} e_{3,123} \
e_{1,123}+12 e_{2,123} e_{3,123} e_{1,123}+6 e_{2,45}^3+2 \
e_{2,123}^3\\+60 e_{2,45} e_{2,123}^2-3 e_{3,123}^2+60 e_{2,45}^2 \
e_{2,123}\end{array} \right) \end{array}$  \\[4.5em]
 \hline
\multicolumn{1}{c}{$[3,1,1]$}  & -- &  $\begin{array}{l}  \beta_3 = J_{45} \left(2 e_{1,123}^2+3 e_{2,45}-2 e_{2,123}\right)\\ \beta_4 = J_{45}\left(2 e_{1,123} e_{2,45}+3 e_{1,123} e_{2,123}-3 e_{3,123}\right)\\ \beta_5 = J_{45} \left(\begin{array}{l} 20 A_1 e_{2,45} e_{1,123}^3-2 A_2 e_{2,45} e_{1,123}^3+20 A_1 e_{2,123} e_{1,123}^3+2 A_2 e_{2,123} e_{1,123}^3-4 A_5 e_{3,123} e_{1,123}^2\\-10 A_1 e_{2,45}^2 e_{1,123}-3 A_2 e_{2,45}^2 e_{1,123}-20 A_1 e_{2,123}^2 e_{1,123}-2 A_2 e_{2,123}^2 e_{1,123}-40 A_1 e_{2,45} e_{2,123} e_{1,123}\\+4 A_2 e_{2,45} e_{2,123} e_{1,123}+40 A_1 e_{2,123} e_{3,123}\end{array}\right) \\ \beta_6 = J_{45}\left(2 e_{2,45} e_{1,123}^3+2 e_{2,123} e_{1,123}^3-e_{2,45}^2 e_{1,123}-2 e_{2,123}^2 e_{1,123}-4 e_{2,45} e_{2,123} e_{1,123}+4 e_{2,123} e_{3,123}\right) \\ \beta_7 = J_{45} \left( \begin{array}{l} 2 e_{1,123}^6+7 e_{2,45} e_{1,123}^4-2 e_{2,123} e_{1,123}^4+20 e_{3,123} e_{1,123}^3-12 e_{2,45}^2 e_{1,123}^2\\+10 e_{2,123}^2 e_{1,123}^2-4 e_{2,45} e_{2,123} e_{1,123}^2-50 e_{2,45} e_{3,123} e_{1,123}-15 e_{2,123} e_{3,123} e_{1,123}\\+3 e_{2,45}^3-10 e_{2,123}^3-10 e_{2,45} e_{2,123}^2+15 e_{3,123}^2+18 e_{2,45}^2 e_{2,123} \end{array} \right) \end{array}$ \\[5.8em]
 \hline  \hline
\end{tabular}
}
}
\caption[]{Primitive polynomials $\beta_L$ of degree $L$  for construction of \emph{boson} wave functions, listed by their symmetry type $\lambda$. As in \tref{tablePrimitivePolynomialsFermions},  the symbol $e_{m,i_1 i_2 \ldots i_p}$ is defined in \eref{eqShortHandSymmetricPolynomial} and $J_{i_1 \ldots i_p}$ is defined in \eref{eqShortHandJastrow}.   The column labeled $\sboson$ is in reference to the case of spin-1/2 (two component) particles only; entries labeled with a dash do not occur for the spin-1/2 case.  The symmetry type $\lambda$ pertains more generally.}

\label{tablePrimitivePolynomialsBosons}
\end{table}

\begin{table}[H]
\ContinuedFloat
\centering
\noindent\makebox[\textwidth]{%	centered figures wider than text width
\subfloat[$N=5 $ (cont.)]{
\begin{tabular*}{\textwidth}{ @{\extracolsep{\fill}} c @{\extracolsep{\fill}} c| @{\extracolsep{\fill}}  l }
\hline  \hline
\multicolumn{1}{c}{$\lambda$}  &  $S_{\mbox{\small boson}}$ &  \multicolumn{1}{|c}{Primitive Polynomials}   \\[2pt]
\hline
\multicolumn{1}{c}{$[2,2,1]$}  & -- &  $\begin{array}{l} \beta_4 = \left( \begin{array}{l} e_{1,34} e_{1,12}^3-6 e_{1,34}^2 e_{1,12}^2+6 e_{2,12} e_{1,12}^2+4 e_{2,34} e_{1,12}^2-3 e_{1,34}^3 e_{1,12}\\+13 e_{1,34} e_{2,12} e_{1,12}+13 e_{1,34} e_{2,34} e_{1,12}+4 e_{1,34}^2 e_{2,12}+6 e_{1,34}^2 e_{2,34}+8 e_{2,12} e_{2,34} \end{array} \right) \\[1.2em] \beta_5 = \left( \begin{array}{l}e_ {1,34} e_ {1,12}^4+5 e_ {1,34}^2 e_ {1,12}^3-2 e_ {2,12} e_ {1,12}^3-6 e_ {2,34} e_ {1,12}^3+5 e_ {1,34}^3 e_ {1,12}^2-6 e_ {1,34} e_ {2,12} e_ {1,12}^2\\-9 e_ {1,34} e_ {2,34} e_ {1,12}^2+e_ {1,34}^4 e_ {1,12}-4 e_ {2,12}^2 e_ {1,12}-6 e_ {2,34}^2 e_ {1,12}-9 e_ {1,34}^2 e_ {2,12} e_ {1,12}-6 e_ {1,34}^2 e_ {2,34} e_ {1,12}\\+6 e_ {2,12} e_ {2,34} e_ {1,12}-6 e_ {1,34} e_ {2,12}^2-4 e_ {1,34} e_ {2,34}^2-6 e_ {1,34}^3 e_ {2,12}-2 e_ {1,34}^3 e_ {2,34}+6 e_ {1,34} e_ {2,12} e_ {2,34}\end{array} \right) \\[1.8em] \beta_6 =  \left( \begin{array}{l}2 e_ {2,34} e_ {1,12}^4-e_ {1,34} e_ {2,12} e_ {1,12}^3+4 e_ {1,34} e_ {2,34} e_ {1,12}^3-2 e_ {2,34}^2 e_ {1,12}^2+e_ {1,34}^2 e_ {2,12} e_ {1,12}^2+e_ {1,34}^2 e_ {2,34} e_ {1,12}^2\\-10 e_ {2,12} e_ {2,34} e_ {1,12}^2-2 e_ {1,34} e_ {2,12}^2 e_ {1,12}-2 e_ {1,34} e_ {2,34}^2 e_ {1,12}+4 e_ {1,34}^3 e_ {2,12} e_ {1,12}-e_ {1,34}^3 e_ {2,34} e_ {1,12}\\-18 e_ {1,34} e_ {2,12} e_ {2,34} e_ {1,12}-2 e_ {1,34}^2 e_ {2,12}^2-4 e_ {2,12} e_ {2,34}^2+2 e_ {1,34}^4 e_ {2,12}-4 e_ {2,12}^2 e_ {2,34}-10 e_ {1,34}^2 e_ {2,12} e_ {2,34} \end{array} \right)\\[1.8em] \beta_7 = \left( \begin{array}{l}  -3 e_{1,34}^2 e_{1,12}^5+6 e_{2,34} e_{1,12}^5-9 e_{1,34}^3 e_{1,12}^4+19 e_{1,34} e_{2,34} e_{1,12}^4-9 e_{1,34}^4 e_{1,12}^3+6 e_{2,12}^2 e_{1,12}^3+6 e_{2,34}^2 e_{1,12}^3\\+7 e_{1,34}^2 e_{2,12} e_{1,12}^3+20 e_{1,34}^2 e_{2,34} e_{1,12}^3-14 e_{2,12} e_{2,34} e_{1,12}^3-3 e_{1,34}^5 e_{1,12}^2+16 e_{1,34} e_{2,12}^2 e_{1,12}^2\\+13 e_{1,34} e_{2,34}^2 e_{1,12}^2+20 e_{1,34}^3 e_{2,12} e_{1,12}^2+7 e_{1,34}^3 e_{2,34} e_{1,12}^2-42 e_{1,34} e_{2,12} e_{2,34} e_{1,12}^2+13 e_{1,34}^2 e_{2,12}^2 e_{1,12}\\+16 e_{1,34}^2 e_{2,34}^2 e_{1,12}-6 e_{2,12} e_{2,34}^2 e_{1,12}+19 e_{1,34}^4 e_{2,12} e_{1,12}+2 e_{2,12}^2 e_{2,34} e_{1,12}-42 e_{1,34}^2 e_{2,12} e_{2,34} e_{1,12}\\+6 e_{1,34}^3 e_{2,12}^2+6 e_{1,34}^3 e_{2,34}^2+2 e_{1,34} e_{2,12} e_{2,34}^2+6 e_{1,34}^5 e_{2,12}-6 e_{1,34} e_{2,12}^2 e_{2,34}-14 e_{1,34}^3 e_{2,12} e_{2,34}\end{array} \right)\\[3em] \beta_8 = \left( \begin{array}{l} 3 e_{1,34} e_{1,12}^7+13 e_{1,34}^2 e_{1,12}^6-6 e_{2,12} e_{1,12}^6-22 e_{2,34} e_{1,12}^6+30 e_{1,34}^3 e_{1,12}^5\\-23 e_{1,34} e_{2,12} e_{1,12}^5-84 e_{1,34} e_{2,34} e_{1,12}^5+40 e_{1,34}^4 e_{1,12}^4+10 e_{2,12}^2 e_{1,12}^4\\+80 e_{2,34}^2 e_{1,12}^4-81 e_{1,34}^2 e_{2,12} e_{1,12}^4-145 e_{1,34}^2 e_{2,34} e_{1,12}^4+138 e_{2,12} e_{2,34} e_{1,12}^4\\+30 e_{1,34}^5 e_{1,12}^3+49 e_{1,34} e_{2,12}^2 e_{1,12}^3+160 e_{1,34} e_{2,34}^2 e_{1,12}^3\\-150 e_{1,34}^3 e_{2,12} e_{1,12}^3-150 e_{1,34}^3 e_{2,34} e_{1,12}^3+360 e_{1,34} e_{2,12} e_{2,34} e_{1,12}^3+13 e_{1,34}^6 e_{1,12}^2-10 e_{2,12}^3 e_{1,12}^2\\-22 e_{2,34}^3 e_{1,12}^2+114 e_{1,34}^2 e_{2,12}^2 e_{1,12}^2+114 e_{1,34}^2 e_{2,34}^2 e_{1,12}^2-130 e_{2,12} e_{2,34}^2 e_{1,12}^2-145 e_{1,34}^4 e_{2,12} e_{1,12}^2\\-81 e_{1,34}^4 e_{2,34} e_{1,12}^2-72 e_{2,12}^2 e_{2,34} e_{1,12}^2+440 e_{1,34}^2 e_{2,12} e_{2,34} e_{1,12}^2+3 e_{1,34}^7 e_{1,12}-37 e_{1,34} e_{2,12}^3 e_{1,12}\\-37 e_{1,34} e_{2,34}^3 e_{1,12}+160 e_{1,34}^3 e_{2,12}^2 e_{1,12}+49 e_{1,34}^3 e_{2,34}^2 e_{1,12}-270 e_{1,34} e_{2,12} e_{2,34}^2 e_{1,12}\\-84 e_{1,34}^5 e_{2,12} e_{1,12}-23 e_{1,34}^5 e_{2,34} e_{1,12}-270 e_{1,34} e_{2,12}^2 e_{2,34} e_{1,12}+360 e_{1,34}^3 e_{2,12} e_{2,34} e_{1,12}\\-22 e_{1,34}^2 e_{2,12}^3-10 e_{1,34}^2 e_{2,34}^3-44 e_{2,12} e_{2,34}^3+80 e_{1,34}^4 e_{2,12}^2+10 e_{1,34}^4 e_{2,34}^2-160 e_{2,12}^2 e_{2,34}^2\\-72 e_{1,34}^2 e_{2,12} e_{2,34}^2-22 e_{1,34}^6 e_{2,12}-6 e_{1,34}^6 e_{2,34}\\-44 e_{2,12}^3 e_{2,34}-130 e_{1,34}^2 e_{2,12}^2 e_{2,34}+138 e_{1,34}^4 e_{2,12} e_{2,34}\end{array} \right) \end{array}$  \\[17.5em]
\hline
\multicolumn{1}{c}{$[2,1,1,1]$}  & -- &  $\begin{array}{l}  \beta_6 = J_{345} \left(2 e_{1,12}^3-5 e_{2,12} e_{1,12}+e_{2,345} e_{1,12}-2 e_{3,345}\right)\\ \beta_7 = J_{345} \left(2 e_{2,12} e_{1,12}^2-e_{3,345} e_{1,12}-2 e_{2,12}^2+2 e_{2,12} e_{2,345}\right)\\[0.2em] \beta_8 = J_{345} \left( \begin{array}{l}2 e_{1,12}^5-7 e_{2,12} e_{1,12}^3-e_{2,345} e_{1,12}^3-7 e_{3,345} e_{1,12}^2+20 e_{2,12}^2 e_{1,12}-e_{2,345}^2 e_{1,12}\\+9 e_{2,12} e_{2,345} e_{1,12}+12 e_{2,12} e_{3,345}+2 e_{2,345} e_{3,345}\end{array} \right)\\[1.2em] \beta_9 = J_{345} \left( \begin{array}{l}8 e_{2,12} e_{1,12}^4+2 e_{2,345} e_{1,12}^4-8 e_{3,345} e_{1,12}^3-10 e_{2,12}^2 e_{1,12}^2+e_{2,345}^2 e_{1,12}^2-e_{2,12} e_{2,345} e_{1,12}^2\\+17 e_{2,12} e_{3,345} e_{1,12}+4 e_{2,345} e_{3,345} e_{1,12}+20 e_{2,12}^3-10 e_{2,12} e_{2,345}^2-2 e_{3,345}^2-10 e_{2,12}^2 e_{2,345}\end{array} \right) \end{array}$  \\[4.3em]
\hline
\multicolumn{1}{c}{$[1,1,1,1,1]$}  & -- &  $\begin{array}{l}  J_{12345} \end{array}$  \\
\hline \hline
\end{tabular*}
}
}
\caption{Continued}
\end{table}

%%%%%%%%%%%%%%%%%%%%%%%%%%%%%%%%%%

\begin{table}[H]
\setcounter{subtable}{0}
\centering
\vspace{-1em}
\subfloat[Three-component $N=2$]{
\begin{tabular}{c|c|c}
\hline \hline
\multicolumn{1}{c|}{$\lambda$} & Set of Primitive Generalized Spin Wave Functions & Counting \\
\hline
\multicolumn{1}{c|}{$[2]$} & $\left| {\alpha \alpha} \right\rangle , \left| {\alpha \beta} \right\rangle , \left| {\alpha \gamma} \right\rangle , \left| {\beta \beta} \right\rangle , \left| {\beta \gamma} \right\rangle , \left| {\gamma \gamma} \right\rangle$ & 6\\[2pt]
\hline
\multicolumn{1}{c|}{$[1,1]$} & $\left| {\alpha \beta} \right\rangle , \left| {\alpha \gamma} \right\rangle , \left| {\beta \gamma} \right\rangle$ & 3\\[2pt]
\hline \hline
\end{tabular}
}
\\
\subfloat[Three-component $N=3$]{
\begin{tabular}{c|c|c}
\hline \hline
\multicolumn{1}{c|}{$\lambda$} & Set of Primitive Generalized Spin Wave Functions & Counting \\
\hline
\multicolumn{1}{c|}{$[3]$} & $\left| {\alpha \alpha \alpha} \right\rangle , \left| {\alpha \alpha \beta} \right\rangle , \left| {\alpha \alpha \gamma} \right\rangle, \left| {\alpha \beta \beta} \right\rangle , \left| {\alpha \beta \gamma} \right\rangle , \left| {\alpha \gamma \gamma} \right\rangle , \left| {\beta \beta \beta } \right\rangle , \left| {\beta \beta \gamma } \right\rangle , \left| {\beta \gamma \gamma} \right\rangle, \left| {\gamma \gamma \gamma} \right\rangle$ & 10\\[2pt]
\hline
\multicolumn{1}{c|}{$[2,1]$} & $\left| {\alpha \alpha \beta} \right\rangle , \left| {\alpha \alpha \gamma} \right\rangle , \left| {\alpha \beta \beta} \right\rangle  , \left| {\alpha \beta \gamma} \right\rangle , \left| {\alpha \gamma \beta} \right\rangle , \left| {\alpha \gamma \gamma} \right\rangle , \left| {\beta \beta \gamma } \right\rangle , \left| {\beta \gamma \gamma} \right\rangle $ & $16 = 8 \times 2$ \\[2pt]
\hline
\multicolumn{1}{c|}{$[1,1,1]$} & $\left| {\alpha \beta \gamma} \right\rangle$ & 1\\[2pt]
\hline \hline
\end{tabular}
}
\\
\subfloat[Three-component $N=4$]{
\begin{tabular}{c|c|c}
\hline \hline
\multicolumn{1}{c|}{$\lambda$} & Set of  Primitive Generalized Spin Wave Functions & Counting\\
\hline
\multicolumn{1}{c|}{$[4]$} & $\begin{array}{l} \left| {\alpha \alpha \alpha \alpha }\right\rangle , \left|
{\alpha \alpha \alpha \beta }\right\rangle , \left| {\alpha
\alpha \alpha \gamma }\right\rangle , \left| {\alpha \alpha
\beta \beta }\right\rangle , \left| {\alpha \alpha \beta \gamma
}\right\rangle , \left| {\alpha \alpha \gamma \gamma }\right\rangle , \left| {\alpha \beta \beta \beta }\right\rangle ,\\ \left| {
\alpha \beta \beta \gamma }\right\rangle , \left| {\alpha \beta
\gamma \gamma }\right\rangle , \left| {\alpha \gamma \gamma
\gamma }\right\rangle , \left| {\beta \beta \beta \beta }\right\rangle , \left| {\beta \beta \beta \gamma }\right\rangle , \left| {
\beta \beta \gamma \gamma }\right\rangle , \left| {\beta \gamma
\gamma \gamma }\right\rangle , \left| {\gamma \gamma \gamma
\gamma }\right\rangle \end{array}$ & 15\\[1em]
\hline
\multicolumn{1}{c|}{$[3,1]$} & $\begin{array}{l} \left| {\alpha  \alpha  \alpha  \beta }\right\rangle , \left| {\alpha  \alpha  \alpha  \gamma }\right\rangle , \left| {\alpha  \alpha  \beta  \beta }\right\rangle , \left| {\alpha  \alpha  \beta  \gamma }\right\rangle , \left| {\alpha  \alpha  \gamma  \beta }\right\rangle , \left| {\alpha  \alpha  \gamma  \gamma }\right\rangle , \left| {\alpha  \beta  \beta  \beta }\right\rangle , \left| {\alpha  \beta  \beta  \gamma }\right\rangle ,\\ \left| {\alpha  \beta  \gamma  \beta }\right\rangle , \left| {\alpha  \beta  \gamma  \gamma }\right\rangle , \left| {\alpha  \gamma  \gamma  \beta }\right\rangle , \left| {\alpha  \gamma  \gamma  \gamma }\right\rangle , \left| {\beta  \beta  \beta  \gamma }\right\rangle , \left| {\beta  \beta  \gamma  \gamma }\right\rangle , \left| {\beta  \gamma  \gamma  \gamma }\right\rangle \end{array}$ & $45 = 15 \times 3$\\[1em]
\hline
\multicolumn{1}{c|}{$[2,2]$} & $\left| {\alpha  \alpha  \beta  \beta }\right\rangle , \left| {\alpha  \alpha  \beta  \gamma }\right\rangle , \left| {\alpha  \alpha  \gamma  \gamma }\right\rangle , \left| {\alpha  \beta  \beta  \gamma }\right\rangle , \left| {\alpha  \beta  \gamma  \gamma }\right\rangle , \left| {\beta  \beta  \gamma  \gamma }\right\rangle  $ & $12 = 6 \times 2$\\[2pt]
\hline
\multicolumn{1}{c|}{$[2,1,1]$} & $\begin{array}{l}   \left| {\alpha \alpha \beta \gamma
}\right\rangle , \left| {
\alpha \beta \beta \gamma }\right\rangle , \left| {\alpha \beta
\gamma \gamma }\right\rangle  \end{array}$ &$9 = 3 \times 3$\\[2pt]
\hline \hline
\end{tabular}
}
\\
\subfloat[Three-Component $N=5$]{
\begin{tabular}{c|c|c}
\hline \hline
\multicolumn{1}{c|}{$\lambda$} & Set of Primitive Generalized Spin Wave Functions & Counting\\
 \hline
\multicolumn{1}{c|}{$[5]$} & $\begin{array}{l} \left| {\alpha \alpha \alpha \alpha \alpha }\right\rangle ,
\left| {\alpha \alpha \alpha \alpha \beta }\right\rangle , \left|
{\alpha \alpha \alpha \alpha \gamma }\right\rangle , \left|
{\alpha \alpha \alpha \beta \beta }\right\rangle , \left| {\alpha
\alpha \alpha \beta \gamma }\right\rangle , \left| {\alpha
\alpha \alpha \gamma \gamma }\right\rangle , \left| {\alpha
\alpha \beta \beta \beta }\right\rangle , \\ \left| {\alpha \alpha
\beta \beta \gamma }\right\rangle , \left| {\alpha \alpha \beta
\gamma \gamma }\right\rangle , \left| {\alpha \alpha \gamma
\gamma \gamma }\right\rangle , \left| {\alpha \beta \beta \beta
\beta }\right\rangle ,  \left| {\alpha \beta \beta \beta \gamma }
\right\rangle , \left| {\alpha \beta \beta \gamma \gamma }\right\rangle , \left| {\alpha \beta \gamma \gamma \gamma }\right\rangle , \\
\left| {\alpha \gamma \gamma \gamma \gamma }\right\rangle ,
\left| {\beta \beta \beta \beta \beta }\right\rangle , \left|
{\beta \beta \beta \beta \gamma }\right\rangle , \left| {\beta
\beta \beta \gamma \gamma }\right\rangle , \left| {\beta \beta
\gamma \gamma \gamma }\right\rangle , \left| {\beta \gamma \gamma
\gamma \gamma }\right\rangle , \left| {\gamma \gamma \gamma
\gamma \gamma }\right\rangle \end{array}$ & 21\\[1.7em]
 \hline
\multicolumn{1}{c|}{$[4,1]$} & $\begin{array}{l}
\left| {\alpha  \alpha  \alpha  \alpha  \beta }\right\rangle , \left| {\alpha  \alpha  \alpha  \alpha  \gamma }\right\rangle , \left| {\alpha  \alpha  \alpha  \beta  \beta }\right\rangle , \left| {\alpha  \alpha  \alpha  \beta  \gamma }\right\rangle , \left| {\alpha  \alpha  \alpha  \gamma  \beta }\right\rangle , \left| {\alpha  \alpha  \alpha  \gamma  \gamma }\right\rangle , \left| {\alpha  \alpha  \beta  \beta  \beta }\right\rangle , \left| {\alpha  \alpha  \beta  \beta  \gamma }\right\rangle ,\\ \left| {\alpha  \alpha  \beta  \gamma  \beta }\right\rangle , \left| {\alpha  \alpha  \beta  \gamma  \gamma }\right\rangle , \left| {\alpha  \alpha  \gamma  \gamma  \beta }\right\rangle , \left| {\alpha  \alpha  \gamma  \gamma  \gamma }\right\rangle , \left| {\alpha  \beta  \beta  \beta  \beta }\right\rangle , \left| {\alpha  \beta  \beta  \beta  \gamma }\right\rangle , \left| {\alpha  \beta  \beta  \gamma  \beta }\right\rangle , \left| {\alpha  \beta  \beta  \gamma  \gamma }\right\rangle ,\\ \left| {\alpha  \beta  \gamma  \gamma  \beta }\right\rangle , \left| {\alpha  \beta  \gamma  \gamma  \gamma }\right\rangle , \left| {\alpha  \gamma  \gamma  \gamma  \beta }\right\rangle , \left| {\alpha  \gamma  \gamma  \gamma  \gamma }\right\rangle , \left| {\beta  \beta  \beta  \beta  \gamma }\right\rangle , \left| {\beta  \beta  \beta  \gamma  \gamma }\right\rangle , \left| {\beta  \beta  \gamma  \gamma  \gamma }\right\rangle , \left| {\beta  \gamma  \gamma  \gamma  \gamma }\right\rangle  \end{array}$ & $96 = 24 \times 4$\\[1.7em]
 \hline
\multicolumn{1}{c|}{$[3,2]$} & $\begin{array}{l} \left| {\alpha  \alpha  \alpha  \beta  \beta }\right\rangle , \left| {\alpha  \alpha  \alpha  \beta  \gamma }\right\rangle , \left| {\alpha  \alpha  \alpha  \gamma  \gamma }\right\rangle , \left| {\alpha  \alpha  \beta  \beta  \beta }\right\rangle , \left| {\alpha  \alpha  \beta  \beta  \gamma }\right\rangle , \left| {\alpha  \alpha  \beta  \gamma  \gamma }\right\rangle , \left| {\alpha  \alpha  \gamma  \beta  \beta }\right\rangle , \left| {\alpha  \alpha  \gamma  \beta  \gamma }\right\rangle ,\\ \left| {\alpha  \alpha  \gamma  \gamma  \gamma }\right\rangle , \left| {\alpha  \beta  \beta  \beta  \gamma }\right\rangle , \left| {\alpha  \beta  \beta  \gamma  \gamma }\right\rangle , \left| {\alpha  \beta  \gamma  \beta  \gamma }\right\rangle , \left| {\alpha  \beta  \gamma  \gamma  \gamma }\right\rangle , \left| {\beta  \beta  \beta  \gamma  \gamma }\right\rangle , \left| {\beta  \beta  \gamma  \gamma  \gamma }\right\rangle \end{array}$ & $75 = 15 \times 5$\\[1em]
 \hline
\multicolumn{1}{c|}{$[3,1,1]$} & $\left| {\alpha  \alpha  \alpha  \beta  \gamma }\right\rangle , \left| {\alpha  \alpha  \beta  \beta  \gamma }\right\rangle , \left| {\alpha  \alpha  \gamma  \beta  \gamma }\right\rangle , \left| {\alpha  \beta  \beta  \beta  \gamma }\right\rangle , \left| {\alpha  \beta  \gamma  \beta  \gamma }\right\rangle , \left| {\alpha  \gamma  \gamma  \beta  \gamma }\right\rangle$ & $36 = 6 \times 6$\\[2pt]
 \hline
\multicolumn{1}{c|}{$[2,2,1]$} & $\left| {\alpha  \alpha  \beta  \beta  \gamma }\right\rangle , \left| {\alpha  \alpha  \beta  \gamma  \gamma }\right\rangle , \left| {\alpha  \beta  \beta  \gamma  \gamma }\right\rangle$ &$15 = 3 \times 5$\\[2pt]
 \hline  \hline
\end{tabular}
}
\\
\subfloat[Four-component $N=2$]{
\begin{tabular}{c|c|c}
\hline \hline
\multicolumn{1}{c|}{$\lambda$} & Set of Primitive Generalized Spin Wave Functions & Counting\\
\hline
\multicolumn{1}{c|}{$[2]$} & $\left| {\alpha \alpha }\right\rangle , \left| {\alpha \beta }\right\rangle , \left| {\alpha \gamma }\right\rangle , \left| {\alpha \delta }\right\rangle , \left| {\beta \beta }\right\rangle , \left| {\beta \gamma }\right\rangle , \left| {\beta \delta }\right\rangle
 \left| {\gamma \gamma }\right\rangle , \left| {\gamma \delta }\right\rangle , \left| {\delta \delta }\right\rangle$ & 10\\[2pt]
\hline
\multicolumn{1}{c|}{$[1,1]$} & $\left| {\alpha \beta }\right\rangle , \left| {\alpha \gamma }\right\rangle , \left| {\alpha \delta }\right\rangle ,\left| {\beta \gamma }\right\rangle , \left| {\beta \delta }\right\rangle
 \left| {\gamma \delta }\right\rangle $ & 6\\[2pt]
\hline \hline
\end{tabular}
}
\\
\subfloat[Four-component $N=3$]{
\begin{tabular}{c|c|c}
\hline \hline
\multicolumn{1}{c|}{$\lambda$} & Set of Primitive Generalized Spin Wave Functions & Counting\\
\hline
\multicolumn{1}{c|}{$[3]$} & $\begin{array}{l} \left| {\alpha  \alpha  \alpha }\right\rangle , \left| {\alpha  \alpha  \beta }\right\rangle , \left| {\alpha  \alpha  \gamma }\right\rangle , \left| {\alpha  \alpha  \delta }\right\rangle , \left| {\alpha  \beta  \beta }\right\rangle , \left| {\alpha  \beta  \gamma }\right\rangle , \left| {\alpha  \beta  \delta }\right\rangle , \left| {\alpha  \gamma  \gamma }\right\rangle , \left| {\alpha  \gamma  \delta }\right\rangle , \left| {\alpha  \delta  \delta }\right\rangle ,\\ \left| {\beta  \beta  \beta }\right\rangle , \left| {\beta  \beta  \gamma }\right\rangle , \left| {\beta  \beta  \delta }\right\rangle , \left| {\beta  \gamma  \gamma }\right\rangle , \left| {\beta  \gamma  \delta }\right\rangle , \left| {\beta  \delta  \delta }\right\rangle , \left| {\gamma  \gamma  \gamma }\right\rangle , \left| {\gamma  \gamma  \delta }\right\rangle , \left| {\gamma  \delta  \delta }\right\rangle , \left| {\delta  \delta  \delta }\right\rangle \end{array}$ & 20\\[1em]
\hline
\multicolumn{1}{c|}{$[2,1]$} & $\begin{array}{l} \left| {\alpha  \alpha  \beta }\right\rangle , \left| {\alpha  \alpha  \gamma }\right\rangle , \left| {\alpha  \alpha  \delta }\right\rangle , \left| {\alpha  \beta  \beta }\right\rangle , \left| {\alpha  \beta  \gamma }\right\rangle , \left| {\alpha  \beta  \delta }\right\rangle , \left| {\alpha  \gamma  \beta }\right\rangle , \left| {\alpha  \gamma  \gamma }\right\rangle , \left| {\alpha  \gamma  \delta }\right\rangle , \left| {\alpha  \delta  \beta }\right\rangle ,\\ \left| {\alpha  \delta  \gamma }\right\rangle , \left| {\alpha  \delta  \delta }\right\rangle , \left| {\beta  \beta  \gamma }\right\rangle , \left| {\beta  \beta  \delta }\right\rangle , \left| {\beta  \gamma  \gamma }\right\rangle , \left| {\beta  \gamma  \delta }\right\rangle , \left| {\beta  \delta  \gamma }\right\rangle , \left| {\beta  \delta  \delta }\right\rangle , \left| {\gamma  \gamma  \delta }\right\rangle , \left| {\gamma  \delta  \delta }\right\rangle \end{array}$ & $40 = 20 \times 2$\\[1em]
\hline
\multicolumn{1}{c|}{$[1,1,1]$} & $\begin{array}{l}   \left| {\alpha \beta \gamma }\right\rangle , \left| {\alpha \beta \delta }\right\rangle, \left| {\alpha \gamma \delta }\right\rangle , \left| {\beta \gamma \delta }\right\rangle \end{array}$ & 4\\[2pt]
\hline \hline
\end{tabular}
}
\caption{The minimal set of primitive \generalizedspin{s} specifying a complete basis, classified by symmetry type $\lambda$. In order to count the number of states, we count the number of primitive \generalizedspin{s} in each symmetry type, multiplied by the dimension of the corresponding symmetric group representation (see  \aref{appendixGeneralizedSpinWaveFunctions} for details). The total number should be equal to $n^N$ for an $n$ component system of $N$ particles.  }
\label{tableGeneralizedSpinWaveFunctions}
\end{table}

%%%%%%%%%%%%%%%%%%%%%%%%%%%%%%%%%%

\begin{table}[H]
\centering
\ContinuedFloat
\subfloat[Four-Component $N=4$]{
\begin{tabular}{c|c|c}
\hline \hline
\multicolumn{1}{c|}{$\lambda$} & Set of Primitive Generalized Spin Wave Functions & Counting\\
\hline
\multicolumn{1}{c|}{$[4]$} & $\begin{array}{l} \left| {\alpha \alpha \alpha \alpha }\right\rangle , \left| {\alpha \alpha \alpha \beta }\right\rangle , \left| {\alpha \alpha \alpha \gamma }\right\rangle , \left| {\alpha \alpha \alpha \delta }\right\rangle , \left| {\alpha \alpha \beta \beta }\right\rangle , \left| {\alpha \alpha \beta \gamma }\right\rangle , \left| {\alpha \alpha \beta \delta }\right\rangle , \left| {\alpha \alpha \gamma \gamma }\right\rangle , \left| {\alpha \alpha \gamma \delta }\right\rangle , \left| {\alpha \alpha \delta \delta }\right\rangle , \left| {\alpha \beta \beta \beta }\right\rangle ,\\ \left| {\alpha \beta \beta \gamma }\right\rangle , \left| {\alpha \beta \beta \delta }\right\rangle , \left| {\alpha \beta \gamma \gamma }\right\rangle , \left| {\alpha \beta \gamma \delta }\right\rangle , \left| {\alpha \beta \delta \delta }\right\rangle , \left| {\alpha \gamma \gamma \gamma }\right\rangle , \left| {\alpha \gamma \gamma \delta }\right\rangle , \left| {\alpha \gamma \delta \delta }\right\rangle , \left| {\alpha \delta \delta \delta }\right\rangle , \left| {\beta \beta \beta \beta }\right\rangle , \left| {\beta \beta \beta \gamma }\right\rangle , \left| {\beta \beta \beta \delta }\right\rangle ,\\ \left| {\beta \beta \gamma \gamma }\right\rangle ,\left| {\beta \beta \gamma \delta }\right\rangle , \left| {\beta \beta \delta \delta }\right\rangle , \left| {\beta \gamma \gamma \gamma }\right\rangle , \left| {\beta \gamma \gamma \delta }\right\rangle , \left| {\beta \gamma \delta \delta }\right\rangle , \left| {\beta \delta \delta \delta }\right\rangle , \left| {\gamma \gamma \gamma \gamma }\right\rangle , \left| {\gamma \gamma \gamma \delta }\right\rangle , \left| {\gamma \gamma \delta \delta }\right\rangle , \left| {\gamma \delta \delta \delta }\right\rangle , \left| {\delta \delta \delta \delta }\right\rangle \end{array}$ &35\\[1.7em]
\hline
\multicolumn{1}{c|}{$[3,1]$} & $\begin{array}{l}  \left| {\alpha  \alpha  \alpha  \beta }\right\rangle , \left| {\alpha  \alpha  \alpha  \gamma }\right\rangle , \left| {\alpha  \alpha  \alpha  \delta }\right\rangle , \left| {\alpha  \alpha  \beta  \beta }\right\rangle , \left| {\alpha  \alpha  \beta  \gamma }\right\rangle , \left| {\alpha  \alpha  \beta  \delta }\right\rangle , \left| {\alpha  \alpha  \gamma  \beta }\right\rangle , \left| {\alpha  \alpha  \gamma  \gamma }\right\rangle , \left| {\alpha  \alpha  \gamma  \delta }\right\rangle , \left| {\alpha  \alpha  \delta  \beta }\right\rangle ,\\ \left| {\alpha  \alpha  \delta  \gamma }\right\rangle , \left| {\alpha  \alpha  \delta  \delta }\right\rangle , \left| {\alpha  \beta  \beta  \beta }\right\rangle , \left| {\alpha  \beta  \beta  \gamma }\right\rangle , \left| {\alpha  \beta  \beta  \delta }\right\rangle , \left| {\alpha  \beta  \gamma  \beta }\right\rangle , \left| {\alpha  \beta  \gamma  \gamma }\right\rangle , \left| {\alpha  \beta  \gamma  \delta }\right\rangle , \left| {\alpha  \beta  \delta  \beta }\right\rangle , \left| {\alpha  \beta  \delta  \gamma }\right\rangle ,\\ \left| {\alpha  \beta  \delta  \delta }\right\rangle , \left| {\alpha  \gamma  \gamma  \beta }\right\rangle , \left| {\alpha  \gamma  \gamma  \gamma }\right\rangle , \left| {\alpha  \gamma  \gamma  \delta }\right\rangle , \left| {\alpha  \gamma  \delta  \beta }\right\rangle , \left| {\alpha  \gamma  \delta  \gamma }\right\rangle , \left| {\alpha  \gamma  \delta  \delta }\right\rangle , \left| {\alpha  \delta  \delta  \beta }\right\rangle , \left| {\alpha  \delta  \delta  \gamma }\right\rangle , \left| {\alpha  \delta  \delta  \delta }\right\rangle ,\\ \left| {\beta  \beta  \beta  \gamma }\right\rangle , \left| {\beta  \beta  \beta  \delta }\right\rangle , \left| {\beta  \beta  \gamma  \gamma }\right\rangle , \left| {\beta  \beta  \gamma  \delta }\right\rangle , \left| {\beta  \beta  \delta  \gamma }\right\rangle , \left| {\beta  \beta  \delta  \delta }\right\rangle , \left| {\beta  \gamma  \gamma  \gamma }\right\rangle , \left| {\beta  \gamma  \gamma  \delta }\right\rangle , \left| {\beta  \gamma  \delta  \gamma }\right\rangle , \left| {\beta  \gamma  \delta  \delta }\right\rangle ,\\ \left| {\beta  \delta  \delta  \gamma }\right\rangle , \left| {\beta  \delta  \delta  \delta }\right\rangle , \left| {\gamma  \gamma  \gamma  \delta }\right\rangle , \left| {\gamma  \gamma  \delta  \delta }\right\rangle , \left| {\gamma  \delta  \delta  \delta }\right\rangle\end{array}$ & $135 = 45 \times 3$\\[3.1em]
\hline
\multicolumn{1}{c|}{$[2,2]$} & $\begin{array}{l} \left| {\alpha  \alpha  \beta  \beta }\right\rangle , \left| {\alpha  \alpha  \beta  \gamma }\right\rangle , \left| {\alpha  \alpha  \beta  \delta }\right\rangle , \left| {\alpha  \alpha  \gamma  \gamma }\right\rangle , \left| {\alpha  \alpha  \gamma  \delta }\right\rangle , \left| {\alpha  \alpha  \delta  \delta }\right\rangle , \left| {\alpha  \beta  \beta  \gamma }\right\rangle , \left| {\alpha  \beta  \beta  \delta }\right\rangle , \left| {\alpha  \beta  \gamma  \gamma }\right\rangle , \left| {\alpha  \beta  \gamma  \delta }\right\rangle ,\\ \left| {\alpha  \beta  \delta  \delta }\right\rangle , \left| {\alpha  \gamma  \beta  \delta }\right\rangle , \left| {\alpha  \gamma  \gamma  \delta }\right\rangle , \left| {\alpha  \gamma  \delta  \delta }\right\rangle , \left| {\beta  \beta  \gamma  \gamma }\right\rangle , \left| {\beta  \beta  \gamma  \delta }\right\rangle , \left| {\beta  \beta  \delta  \delta }\right\rangle , \left| {\beta  \gamma  \gamma  \delta }\right\rangle , \left| {\beta  \gamma  \delta  \delta }\right\rangle , \left| {\gamma  \gamma  \delta  \delta }\right\rangle  \end{array}$ & $40 = 20 \times 2$\\[1em]
\hline
\multicolumn{1}{c|}{$[2,1,1]$} & $\begin{array}{l}\left| {\alpha  \alpha  \beta  \gamma }\right\rangle , \left| {\alpha  \alpha  \beta  \delta }\right\rangle , \left| {\alpha  \alpha  \gamma  \delta }\right\rangle , \left| {\alpha  \beta  \beta  \gamma }\right\rangle , \left| {\alpha  \beta  \beta  \delta }\right\rangle , \left| {\alpha  \beta  \gamma  \delta }\right\rangle , \left| {\alpha  \gamma  \beta  \gamma }\right\rangle , \left| {\alpha  \gamma  \beta  \delta }\right\rangle , \left| {\alpha  \gamma  \gamma  \delta }\right\rangle , \left| {\alpha  \delta  \beta  \gamma }\right\rangle ,\\ \left| {\alpha  \delta  \beta  \delta }\right\rangle , \left| {\alpha  \delta  \gamma  \delta }\right\rangle , \left| {\beta  \beta  \gamma  \delta }\right\rangle , \left| {\beta  \gamma  \gamma  \delta }\right\rangle , \left| {\beta  \delta  \gamma  \delta }\right\rangle   \end{array}$ & $45 = 15 \times 3$\\[1em]
\hline
\multicolumn{1}{c|}{$[1,1,1,1]$} & $\left| {\alpha \beta \gamma \delta}\right\rangle $ &1\\[2pt]
\hline \hline
\end{tabular}
}
\\
\subfloat[Four-Component $N=5$]{
\begin{tabular}{c|c|c}
\hline \hline
\multicolumn{1}{c|}{$\lambda$} & Set of  Primitive Generalized Spin Wave Functions & Counting\\
\hline
\multicolumn{1}{c|}{$[5]$} & $\begin{array}{l} \left| {\alpha \alpha \alpha \alpha \alpha }\right\rangle , \left| {\alpha \alpha \alpha \alpha \beta }\right\rangle , \left| {\alpha \alpha \alpha \alpha \gamma }\right\rangle , \left| {\alpha \alpha \alpha \alpha \delta }\right\rangle , \left| {\alpha \alpha \alpha \beta \beta }\right\rangle , \left| {\alpha \alpha \alpha \beta \gamma }\right\rangle , \left| {\alpha \alpha \alpha \beta \delta }\right\rangle , \left| {\alpha \alpha \alpha \gamma \gamma }\right\rangle , \left| {\alpha \alpha \alpha \gamma \delta }\right\rangle ,\\ \left| {\alpha \alpha \alpha \delta \delta }\right\rangle , \left| {\alpha \alpha \beta \beta \beta }\right\rangle , \left| {\alpha \alpha \beta \beta \gamma }\right\rangle  \left| {\alpha \alpha \beta \beta \delta }\right\rangle , \left| {\alpha \alpha \beta \gamma \gamma }\right\rangle , \left| {\alpha \alpha \beta \gamma \delta }\right\rangle , \left| {\alpha \alpha \beta \delta \delta }\right\rangle , \left| {\alpha \alpha \gamma \gamma \gamma }\right\rangle , \left| {\alpha \alpha \gamma \gamma \delta }\right\rangle ,\\ \left| {\alpha \alpha \gamma \delta \delta }\right\rangle , \left| {\alpha \alpha \delta \delta \delta }\right\rangle , \left| {\alpha \beta \beta \beta \beta }\right\rangle , \left| {\alpha \beta \beta ,\beta \gamma }\right\rangle , \left| {\alpha \beta \beta \beta \delta }\right\rangle , \left| {\alpha \beta \beta \gamma \gamma }\right\rangle , \left| {\alpha \beta \beta \gamma \delta }\right\rangle , \left| {\alpha \beta \beta \delta \delta }\right\rangle , \left| {\alpha \beta \gamma \gamma \gamma }\right\rangle ,\\ \left| {\alpha ,\beta ,\gamma \gamma \delta }\right\rangle , \left| {\alpha \beta \gamma \delta \delta }\right\rangle , \left| {\alpha \beta \delta \delta \delta }\right\rangle , \left| {\alpha \gamma \gamma \gamma \gamma }\right\rangle , \left| {\alpha \gamma \gamma \gamma \delta }\right\rangle , \left| {\alpha \gamma \gamma \delta \delta }\right\rangle , \left| {\alpha \gamma \delta \delta \delta }\right\rangle , \left| {\alpha \delta \delta \delta \delta }\right\rangle , \left| {\beta \beta \beta \beta \beta }\right\rangle ,\\ \left| {\beta \beta \beta \beta \gamma }\right\rangle , \left| {\beta \beta \beta \beta \delta }\right\rangle , \left| {\beta \beta \beta \gamma \gamma }\right\rangle , \left| {\beta \beta \beta \gamma \delta }\right\rangle , \left| {\beta \beta \beta \delta \delta }\right\rangle , \left| {\beta \beta \gamma \gamma \gamma }\right\rangle ,\left| {\beta \beta \gamma \gamma \delta }\right\rangle , \left| {\beta \beta \gamma \delta \delta }\right\rangle , \left| {\beta \beta \delta \delta \delta }\right\rangle ,\left| {\beta \gamma \gamma \gamma \gamma }\right\rangle ,\\ \left| {\beta \gamma \gamma \gamma \delta }\right\rangle , \left| {\beta \gamma \gamma \delta \delta }\right\rangle , \left| {\beta \gamma \delta \delta \delta }\right\rangle , \left| {\beta \delta \delta \delta \delta }\right\rangle , \left| {\gamma \gamma \gamma \gamma \gamma }\right\rangle , \left| {\gamma \gamma \gamma \gamma \delta }\right\rangle , \left| {\gamma \gamma \gamma \delta \delta }\right\rangle , \left| {\gamma \gamma \delta \delta \delta }\right\rangle , \left| {\gamma \delta \delta \delta \delta }\right\rangle , \left| {\delta \delta \delta \delta \delta }\right\rangle \end{array}$& 56\\[3.7em]
\hline
\multicolumn{1}{c|}{$[4,1]$} & $\begin{array}{l} \left| {\alpha  \alpha  \alpha  \alpha  \beta }\right\rangle , \left| {\alpha  \alpha  \alpha  \alpha  \gamma }\right\rangle , \left| {\alpha  \alpha  \alpha  \alpha  \delta }\right\rangle , \left| {\alpha  \alpha  \alpha  \beta  \beta }\right\rangle , \left| {\alpha  \alpha  \alpha  \beta  \gamma }\right\rangle , \left| {\alpha  \alpha  \alpha  \beta  \delta }\right\rangle , \left| {\alpha  \alpha  \alpha  \gamma  \beta }\right\rangle , \left| {\alpha  \alpha  \alpha  \gamma  \gamma }\right\rangle , \left| {\alpha  \alpha  \alpha  \gamma  \delta }\right\rangle , \left| {\alpha  \alpha  \alpha  \delta  \beta }\right\rangle ,\\ \left| {\alpha  \alpha  \alpha  \delta  \gamma }\right\rangle , \left| {\alpha  \alpha  \alpha  \delta  \delta }\right\rangle , \left| {\alpha  \alpha  \beta  \beta  \beta }\right\rangle , \left| {\alpha  \alpha  \beta  \beta  \gamma }\right\rangle , \left| {\alpha  \alpha  \beta  \beta  \delta }\right\rangle , \left| {\alpha  \alpha  \beta  \gamma  \beta }\right\rangle , \left| {\alpha  \alpha  \beta  \gamma  \gamma }\right\rangle , \left| {\alpha  \alpha  \beta  \gamma  \delta }\right\rangle , \left| {\alpha  \alpha  \beta  \delta  \beta }\right\rangle , \left| {\alpha  \alpha  \beta  \delta  \gamma }\right\rangle ,\\ \left| {\alpha  \alpha  \beta  \delta  \delta }\right\rangle , \left| {\alpha  \alpha  \gamma  \gamma  \beta }\right\rangle , \left| {\alpha  \alpha  \gamma  \gamma  \gamma }\right\rangle , \left| {\alpha  \alpha  \gamma  \gamma  \delta }\right\rangle , \left| {\alpha  \alpha  \gamma  \delta  \beta }\right\rangle , \left| {\alpha  \alpha  \gamma  \delta  \gamma }\right\rangle , \left| {\alpha  \alpha  \gamma  \delta  \delta }\right\rangle , \left| {\alpha  \alpha  \delta  \delta  \beta }\right\rangle , \left| {\alpha  \alpha  \delta  \delta  \gamma }\right\rangle , \left| {\alpha  \alpha  \delta  \delta  \delta }\right\rangle ,\\ \left| {\alpha  \beta  \beta  \beta  \beta }\right\rangle , \left| {\alpha  \beta  \beta  \beta  \gamma }\right\rangle , \left| {\alpha  \beta  \beta  \beta  \delta }\right\rangle , \left| {\alpha  \beta  \beta  \gamma  \beta }\right\rangle , \left| {\alpha  \beta  \beta  \gamma  \gamma }\right\rangle , \left| {\alpha  \beta  \beta  \gamma  \delta }\right\rangle , \left| {\alpha  \beta  \beta  \delta  \beta }\right\rangle , \left| {\alpha  \beta  \beta  \delta  \gamma }\right\rangle , \left| {\alpha  \beta  \beta  \delta  \delta }\right\rangle , \left| {\alpha  \beta  \gamma  \gamma  \beta }\right\rangle ,\\ \left| {\alpha  \beta  \gamma  \gamma  \gamma }\right\rangle , \left| {\alpha  \beta  \gamma  \gamma  \delta }\right\rangle , \left| {\alpha  \beta  \gamma  \delta  \beta }\right\rangle , \left| {\alpha  \beta  \gamma  \delta  \gamma }\right\rangle , \left| {\alpha  \beta  \gamma  \delta  \delta }\right\rangle , \left| {\alpha  \beta  \delta  \delta  \beta }\right\rangle , \left| {\alpha  \beta  \delta  \delta  \gamma }\right\rangle , \left| {\alpha  \beta  \delta  \delta  \delta }\right\rangle , \left| {\alpha  \gamma  \gamma  \gamma  \beta }\right\rangle , \left| {\alpha  \gamma  \gamma  \gamma  \gamma }\right\rangle ,\\ \left| {\alpha  \gamma  \gamma  \gamma  \delta }\right\rangle , \left| {\alpha  \gamma  \gamma  \delta  \beta }\right\rangle , \left| {\alpha  \gamma  \gamma  \delta  \gamma }\right\rangle , \left| {\alpha  \gamma  \gamma  \delta  \delta }\right\rangle , \left| {\alpha  \gamma  \delta  \delta  \beta }\right\rangle , \left| {\alpha  \gamma  \delta  \delta  \gamma }\right\rangle , \left| {\alpha  \gamma  \delta  \delta  \delta }\right\rangle , \left| {\alpha  \delta  \delta  \delta  \beta }\right\rangle , \left| {\alpha  \delta  \delta  \delta  \gamma }\right\rangle , \left| {\alpha  \delta  \delta  \delta  \delta }\right\rangle ,\\ \left| {\beta  \beta  \beta  \beta  \gamma }\right\rangle , \left| {\beta  \beta  \beta  \beta  \delta }\right\rangle , \left| {\beta  \beta  \beta  \gamma  \gamma }\right\rangle , \left| {\beta  \beta  \beta  \gamma  \delta }\right\rangle , \left| {\beta  \beta  \beta  \delta  \gamma }\right\rangle , \left| {\beta  \beta  \beta  \delta  \delta }\right\rangle , \left| {\beta  \beta  \gamma  \gamma  \gamma }\right\rangle , \left| {\beta  \beta  \gamma  \gamma  \delta }\right\rangle , \left| {\beta  \beta  \gamma  \delta  \gamma }\right\rangle , \left| {\beta  \beta  \gamma  \delta  \delta }\right\rangle ,\\ \left| {\beta  \beta  \delta  \delta  \gamma }\right\rangle , \left| {\beta  \beta  \delta  \delta  \delta }\right\rangle , \left| {\beta  \gamma  \gamma  \gamma  \gamma }\right\rangle , \left| {\beta  \gamma  \gamma  \gamma  \delta }\right\rangle , \left| {\beta  \gamma  \gamma  \delta  \gamma }\right\rangle , \left| {\beta  \gamma  \gamma  \delta  \delta }\right\rangle , \left| {\beta  \gamma  \delta  \delta  \gamma }\right\rangle , \left| {\beta  \gamma  \delta  \delta  \delta }\right\rangle , \left| {\beta  \delta  \delta  \delta  \gamma }\right\rangle , \left| {\beta  \delta  \delta  \delta  \delta }\right\rangle ,\\ \left| {\gamma  \gamma  \gamma  \gamma  \delta }\right\rangle , \left| {\gamma  \gamma  \gamma  \delta  \delta }\right\rangle , \left| {\gamma  \gamma  \delta  \delta  \delta }\right\rangle , \left| {\gamma  \delta  \delta  \delta  \delta }\right\rangle \end{array}$ &$336 = 84 \times 4$\\[5.8em]
\hline
\multicolumn{1}{c|}{$[3,2]$} & $\begin{array}{l} \left| {\alpha  \alpha  \alpha  \beta  \beta }\right\rangle , \left| {\alpha  \alpha  \alpha  \beta  \gamma }\right\rangle , \left| {\alpha  \alpha  \alpha  \beta  \delta }\right\rangle , \left| {\alpha  \alpha  \alpha  \gamma  \gamma }\right\rangle , \left| {\alpha  \alpha  \alpha  \gamma  \delta }\right\rangle , \left| {\alpha  \alpha  \alpha  \delta  \delta }\right\rangle , \left| {\alpha  \alpha  \beta  \beta  \beta }\right\rangle , \left| {\alpha  \alpha  \beta  \beta  \gamma }\right\rangle , \left| {\alpha  \alpha  \beta  \beta  \delta }\right\rangle , \left| {\alpha  \alpha  \beta  \gamma  \gamma }\right\rangle ,\\ \left| {\alpha  \alpha  \beta  \gamma  \delta }\right\rangle , \left| {\alpha  \alpha  \beta  \delta  \delta }\right\rangle , \left| {\alpha  \alpha  \gamma  \beta  \beta }\right\rangle , \left| {\alpha  \alpha  \gamma  \beta  \gamma }\right\rangle , \left| {\alpha  \alpha  \gamma  \beta  \delta }\right\rangle , \left| {\alpha  \alpha  \gamma  \gamma  \gamma }\right\rangle , \left| {\alpha  \alpha  \gamma  \gamma  \delta }\right\rangle , \left| {\alpha  \alpha  \gamma  \delta  \delta }\right\rangle , \left| {\alpha  \alpha  \delta  \beta  \beta }\right\rangle , \left| {\alpha  \alpha  \delta  \beta  \gamma }\right\rangle ,\\ \left| {\alpha  \alpha  \delta  \beta  \delta }\right\rangle , \left| {\alpha  \alpha  \delta  \gamma  \gamma }\right\rangle , \left| {\alpha  \alpha  \delta  \gamma  \delta }\right\rangle , \left| {\alpha  \alpha  \delta  \delta  \delta }\right\rangle , \left| {\alpha  \beta  \beta  \beta  \gamma }\right\rangle , \left| {\alpha  \beta  \beta  \beta  \delta }\right\rangle , \left| {\alpha  \beta  \beta  \gamma  \gamma }\right\rangle , \left| {\alpha  \beta  \beta  \gamma  \delta }\right\rangle , \left| {\alpha  \beta  \beta  \delta  \delta }\right\rangle , \left| {\alpha  \beta  \gamma  \beta  \gamma }\right\rangle ,\\ \left| {\alpha  \beta  \gamma  \beta  \delta }\right\rangle , \left| {\alpha  \beta  \gamma  \gamma  \gamma }\right\rangle , \left| {\alpha  \beta  \gamma  \gamma  \delta }\right\rangle , \left| {\alpha  \beta  \gamma  \delta  \delta }\right\rangle , \left| {\alpha  \beta  \delta  \beta  \gamma }\right\rangle , \left| {\alpha  \beta  \delta  \beta  \delta }\right\rangle , \left| {\alpha  \beta  \delta  \gamma  \gamma }\right\rangle , \left| {\alpha  \beta  \delta  \gamma  \delta }\right\rangle , \left| {\alpha  \beta  \delta  \delta  \delta }\right\rangle , \left| {\alpha  \gamma  \gamma  \beta  \delta }\right\rangle ,\\ \left| {\alpha  \gamma  \gamma  \gamma  \delta }\right\rangle , \left| {\alpha  \gamma  \gamma  \delta  \delta }\right\rangle , \left| {\alpha  \gamma  \delta  \beta  \delta }\right\rangle , \left| {\alpha  \gamma  \delta  \gamma  \delta }\right\rangle , \left| {\alpha  \gamma  \delta  \delta  \delta }\right\rangle , \left| {\beta  \beta  \beta  \gamma  \gamma }\right\rangle , \left| {\beta  \beta  \beta  \gamma  \delta }\right\rangle , \left| {\beta  \beta  \beta  \delta  \delta }\right\rangle , \left| {\beta  \beta  \gamma  \gamma  \gamma }\right\rangle , \left| {\beta  \beta  \gamma  \gamma  \delta }\right\rangle ,\\ \left| {\beta  \beta  \gamma  \delta  \delta }\right\rangle , \left| {\beta  \beta  \delta  \gamma  \gamma }\right\rangle , \left| {\beta  \beta  \delta  \gamma  \delta }\right\rangle , \left| {\beta  \beta  \delta  \delta  \delta }\right\rangle , \left| {\beta  \gamma  \gamma  \gamma  \delta }\right\rangle , \left| {\beta  \gamma  \gamma  \delta  \delta }\right\rangle , \left| {\beta  \gamma  \delta  \gamma  \delta }\right\rangle , \left| {\beta  \gamma  \delta  \delta  \delta }\right\rangle , \left| {\gamma  \gamma  \gamma  \delta  \delta }\right\rangle , \left| {\gamma  \gamma  \delta  \delta  \delta }\right\rangle
\end{array}$ &$300 = 60 \times 5$\\[3.8em]
\hline
\multicolumn{1}{c|}{$[3,1,1]$} & $\begin{array}{l} \left| {\alpha  \alpha  \alpha  \beta  \gamma }\right\rangle , \left| {\alpha  \alpha  \alpha  \beta  \delta }\right\rangle , \left| {\alpha  \alpha  \alpha  \gamma  \delta }\right\rangle , \left| {\alpha  \alpha  \beta  \beta  \gamma }\right\rangle , \left| {\alpha  \alpha  \beta  \beta  \delta }\right\rangle , \left| {\alpha  \alpha  \beta  \gamma  \delta }\right\rangle , \left| {\alpha  \alpha  \gamma  \beta  \gamma }\right\rangle , \left| {\alpha  \alpha  \gamma  \beta  \delta }\right\rangle , \left| {\alpha  \alpha  \gamma  \gamma  \delta }\right\rangle , \left| {\alpha  \alpha  \delta  \beta  \gamma }\right\rangle ,\\ \left| {\alpha  \alpha  \delta  \beta  \delta }\right\rangle , \left| {\alpha  \alpha  \delta  \gamma  \delta }\right\rangle , \left| {\alpha  \beta  \beta  \beta  \gamma }\right\rangle , \left| {\alpha  \beta  \beta  \beta  \delta }\right\rangle , \left| {\alpha  \beta  \beta  \gamma  \delta }\right\rangle , \left| {\alpha  \beta  \gamma  \beta  \gamma }\right\rangle , \left| {\alpha  \beta  \gamma  \beta  \delta }\right\rangle , \left| {\alpha  \beta  \gamma  \gamma  \delta }\right\rangle , \left| {\alpha  \beta  \delta  \beta  \gamma }\right\rangle , \left| {\alpha  \beta  \delta  \beta  \delta }\right\rangle ,\\ \left| {\alpha  \beta  \delta  \gamma  \delta }\right\rangle , \left| {\alpha  \gamma  \gamma  \beta  \gamma }\right\rangle , \left| {\alpha  \gamma  \gamma  \beta  \delta }\right\rangle , \left| {\alpha  \gamma  \gamma  \gamma  \delta }\right\rangle , \left| {\alpha  \gamma  \delta  \beta  \gamma }\right\rangle , \left| {\alpha  \gamma  \delta  \beta  \delta }\right\rangle , \left| {\alpha  \gamma  \delta  \gamma  \delta }\right\rangle , \left| {\alpha  \delta  \delta  \beta  \gamma }\right\rangle , \left| {\alpha  \delta  \delta  \beta  \delta }\right\rangle , \left| {\alpha  \delta  \delta  \gamma  \delta }\right\rangle ,\\ \left| {\beta  \beta  \beta  \gamma  \delta }\right\rangle , \left| {\beta  \beta  \gamma  \gamma  \delta }\right\rangle , \left| {\beta  \beta  \delta  \gamma  \delta }\right\rangle , \left| {\beta  \gamma  \gamma  \gamma  \delta }\right\rangle , \left| {\beta  \gamma  \delta  \gamma  \delta }\right\rangle , \left| {\beta  \delta  \delta  \gamma  \delta }\right\rangle \end{array}$ &$216 = 36 \times 6$\\[2.4em]
\hline
\multicolumn{1}{c|}{$[2,2,1]$} & $\begin{array}{l} \left| {\alpha  \alpha  \beta  \beta  \gamma }\right\rangle , \left| {\alpha  \alpha  \beta  \beta  \delta }\right\rangle , \left| {\alpha  \alpha  \beta  \gamma  \gamma }\right\rangle , \left| {\alpha  \alpha  \beta  \gamma  \delta }\right\rangle , \left| {\alpha  \alpha  \beta  \delta  \gamma }\right\rangle , \left| {\alpha  \alpha  \beta  \delta  \delta }\right\rangle , \left| {\alpha  \alpha  \gamma  \gamma  \delta }\right\rangle , \left| {\alpha  \alpha  \gamma  \delta  \delta }\right\rangle , \left| {\alpha  \beta  \beta  \gamma  \gamma }\right\rangle , \left| {\alpha  \beta  \beta  \gamma  \delta }\right\rangle ,\\ \left| {\alpha  \beta  \beta  \delta  \gamma }\right\rangle , \left| {\alpha  \beta  \beta  \delta  \delta }\right\rangle , \left| {\alpha  \beta  \gamma  \gamma  \delta }\right\rangle , \left| {\alpha  \beta  \gamma  \delta  \delta }\right\rangle , \left| {\alpha  \gamma  \beta  \delta  \gamma }\right\rangle , \left| {\alpha  \gamma  \beta  \delta  \delta }\right\rangle , \left| {\alpha  \gamma  \gamma  \delta  \delta }\right\rangle , \left| {\beta  \beta  \gamma  \gamma  \delta }\right\rangle , \left| {\beta  \beta  \gamma  \delta  \delta }\right\rangle , \left| {\beta  \gamma  \gamma  \delta  \delta }\right\rangle\end{array}$ &$100 = 20 \times 5$\\[1em]
\hline
\multicolumn{1}{c|}{$[2,1,1,1]$} & $\left| {\alpha  \alpha  \beta  \gamma  \delta }\right\rangle , \left| {\alpha  \beta  \beta  \gamma  \delta }\right\rangle , \left| {\alpha  \gamma  \beta  \gamma  \delta }\right\rangle , \left| {\alpha  \delta  \beta  \gamma  \delta }\right\rangle$ &$16 = 4 \times 4$\\[2pt]
\hline \hline
\end{tabular}
}
\caption{Continued}
\end{table}

%%%%%%%%%%%%%%%%%%%%%%%%%%%%%%%%%%

\begin{table}[H]
\setcounter{subtable}{0}
\centering
\subfloat[$N=3$]{
\begin{tabular}{cccc}
\hline \hline
\multirow{2}{*}{$S_z$}  & \multicolumn{3}{|c}{$\lambda$} \\
 \multicolumn{1}{c|}{}  & \multicolumn{1}{c}{$[3]$} & \multicolumn{1}{c}{$[2,1]$} & \multicolumn{1}{c}{$[1,1,1]$}   \\[2pt]
\hline
\multicolumn{1}{c|}{3} & $\left| {\alpha  \alpha  \alpha }\right \rangle$ & -- &  --   \\

\multicolumn{1}{c|}{2}  & -- & $\left| {\alpha  \alpha  \beta }\right\rangle$& --   \\

\multicolumn{1}{c|}{1}  & $ \left| {\alpha  \beta  \beta }\right \rangle-2 \left| {\alpha  \alpha  \gamma }\right \rangle
 $ & $ 2\left| {\alpha  \beta  \beta }\right\rangle -\left| {\alpha  \alpha  \gamma }\right\rangle$ & --  \\

\multicolumn{1}{c|}{0}  & -- & -- & $\left| {\alpha  \beta  \gamma }\right\rangle$  \\
 \hline
\multicolumn{1}{c|}{Allowed $S$} & 1,3 & 1,2 & 0  \\
\multicolumn{1}{c|}{Counting} & $10 = 7+3$ & $8 = 5+3$ & 1  \\
\hline \hline
\end{tabular}
}
\\
\subfloat[$N=4$]{
\begin{tabular}{ccccc}
\hline \hline
\multirow{2}{*}{$S_z$}  & \multicolumn{4}{|c}{$\lambda$} \\
\multicolumn{1}{c|}{}   & \multicolumn{1}{c}{$[4]$} &  \multicolumn{1}{c}{$[3,1]$} &  \multicolumn{1}{c}{$[2,2]$} &  \multicolumn{1}{c}{$[2,1,1]$}  \\[2pt]
\hline
\multicolumn{1}{c|}{4}  & $\left| {\alpha  \alpha  \alpha  \alpha }\right\rangle$ & -- & --  & -- \\

\multicolumn{1}{c|}{3} & -- & $\left| {\alpha  \alpha  \alpha  \beta }\right\rangle$& --  & -- \\

\multicolumn{1}{c|}{2}  & $ \left| {\alpha  \alpha  \beta  \beta }\right\rangle - 2\left| {\alpha  \alpha  \alpha  \gamma }\right\rangle$ & $ \left| \
{\alpha  \alpha  \beta  \beta }\right\rangle - {2\over{3}} \left| {\alpha  \alpha  \alpha  \gamma }\right\rangle $ & $\left| {\alpha  \alpha  \beta  \beta }\right\rangle$  &-- \\

\multicolumn{1}{c|}{1}  & -- & $4\left| {\alpha  \beta  \beta  \beta }\right\rangle -\left| {\alpha  \alpha  \beta  \gamma }\right\rangle  -3\left| {\alpha  \alpha  \gamma  \beta }\right\rangle  $ & -- & $ \left| {\alpha  \alpha  \beta  \gamma }\right\rangle$  \\

\multicolumn{1}{c|}{0}  & $4\left| {\alpha  \alpha  \gamma  \gamma }\right\rangle -4\left| {\alpha  \beta  \beta  \gamma }\right\rangle + \left| {\beta  \beta  \beta  \beta }\right\rangle$ & -- & $4\left| {\alpha  \beta  \beta  \gamma }\right\rangle - \left| {\alpha  \alpha  \gamma  \gamma }\right\rangle$ &--  \\
 \hline
 \multicolumn{1}{c|}{Allowed $S$} & 0,2,4 & 1,2,3 & 0,2 & 1  \\
\multicolumn{1}{c|}{Counting} & $15 = 9+5+1$ & $15 = 7+5+3$ & $6 = 5+1$ & 3 \\
\hline \hline
\end{tabular}
}
\\
\subfloat[$N=5$]{
\begin{tabular}{cccccc}
\hline \hline
\multirow{2}{*}{$S_z$} & \multicolumn{5}{|c}{$\lambda$} \\
\multicolumn{1}{c|}{}  & \multicolumn{1}{c}{$[5]$} &\multicolumn{1}{c}{$[4,1]$} & \multicolumn{1}{c}{$[3,2]$} & \multicolumn{1}{c}{$[3,1,1]$} & \multicolumn{1}{c}{$[2,2,1]$}  \\[2pt]
\hline
\multicolumn{1}{c|}{5}  & $\left| {\alpha  \alpha  \alpha  \alpha  \alpha }\right\rangle$ & -- & --  & -- & --\\
\multicolumn{1}{c|}{4}  & -- & $\left| {\alpha  \alpha  \alpha  \alpha  \beta }\right\rangle$& --  & -- & --\\

\multicolumn{1}{c|}{3}  & $\left| {\alpha  \alpha  \alpha  \beta  \beta }\right\rangle -2\left| {\alpha  \alpha  \alpha  \alpha  \gamma }\right\rangle $  & $\left| {\alpha  \alpha  \alpha  \beta  \beta }\right\rangle - {3 \over{4} }\left| {\alpha  \alpha  \alpha  \alpha  \gamma }\right\rangle
$& $\left| {\alpha  \alpha  \alpha  \beta  \beta }\right\rangle$  & -- & --  \\

\multicolumn{1}{c|}{2}  &-- & $\begin{array}{l} \left| {\alpha  \alpha  \beta  \beta  \beta }\right\rangle - {4 \over{15}} \left| {\alpha  \alpha  \alpha  \beta  \gamma }\right\rangle \\- {16 \over{15}}\left| {\alpha  \alpha  \alpha  \gamma  \beta }\right\rangle \end{array}$& $ \left| {\alpha  \alpha  \beta  \beta  \beta }\right\rangle - {1 \over{3}}\left| {\alpha  \alpha  \alpha  \beta  \gamma }\right\rangle$  &$\left| {\alpha  \alpha  \alpha  \beta  \gamma }\right\rangle$&-- \\

\multicolumn{1}{c|}{1}   & $\begin{array}{l} 4\left| {\alpha  \alpha  \alpha  \gamma  \gamma }\right\rangle -4\left| {\alpha  \alpha  \beta  \beta  \gamma }\right\rangle \\+ \left| {\alpha  \beta  \beta  \beta  \beta }\right\rangle \end{array}$& $ \begin{array}{l} \left| {\alpha  \beta  \beta  \beta  \beta }\right\rangle - \left| {\alpha  \alpha  \beta  \gamma  \beta }\right\rangle \\ - {1 \over{2}} \left| {\alpha  \alpha  \beta  \beta  \gamma }\right\rangle +{2 \over{3}} \left| {\alpha  \alpha  \alpha  \gamma  \gamma }\right\rangle \end{array}$ & $\begin{array}{l} \left| {\alpha  \alpha  \gamma  \beta  \beta }\right\rangle + {2 \over{3}}\left| {\alpha  \alpha  \alpha  \gamma  \gamma }\right\rangle  \\-\left| {\alpha  \alpha  \beta  \beta  \gamma }\right\rangle \end{array}$ &  -- &$\left| {\alpha  \alpha  \beta  \beta  \gamma }\right\rangle$ \\

\multicolumn{1}{c|}{0}  & -- & -- & -- & $\left| {\alpha  \beta  \beta  \beta  \gamma }\right\rangle - \left| {\alpha  \alpha  \gamma  \beta  \gamma }\right\rangle $  &-- \\
 \hline
\multicolumn{1}{c|}{Allowed $S$} & 1,3,5 & 1,2,3,4 & 1,2,3 & 0,2 & 1  \\
\multicolumn{1}{c|}{Counting} & $21 = 11+7+3$ & $24 = 9+7+5+3$ & $15 = 7+5+3$ & $6 = 5+1$ & 3 \\
\hline \hline
\end{tabular}
}
\caption{Linear combinations of primitive three-component \generalizedspin{s} that lead to spin-1 eigenfunctions. In the spin language the notation should be interpreted as: $\alpha, \beta$ and $\gamma$ being $s_z = +1,0$ and $-1$ respectively. Note that only highest possible $S_z$ states of a given $S$ series are listed; additional states with the same $S$ eigenvalue can be found by applying a spin lowering operator to any of the states listed here. Additionally we have listed the total number of states for each symmetry type (including the $S_z$ degeneracy): the total should equal the number of three-component \generalizedspin{s} for that symmetry type given in \tref{tableGeneralizedSpinWaveFunctions}.}
\label{tableSpinOneEigenfunctions}
\end{table}

\begin{table}[H]
\centering
\subfloat[$N=2$]{
\begin{tabular}{c|c|c}
\hline \hline
\multirow{2}{*}{S} & \multicolumn{2}{|c}{$\lambda$} \\
\multicolumn{1}{c|}{}  & \multicolumn{1}{c}{$[2]$} &  \multicolumn{1}{c}{$[1,1]$}   \\[2pt]
\hline
\multicolumn{1}{c|}{1} & $\left| {\gamma  \gamma }\right\rangle , \left| {\alpha  \gamma }\right \rangle,\left| {\alpha  \alpha }\right \rangle$ & $\left| {\alpha  \gamma }\right\rangle$\\
\multicolumn{1}{c|}{0}  & $\left| {\beta  \gamma }\right \rangle - \left| {\alpha  \delta }\right \rangle$ & $\left| {\gamma  \delta }\right\rangle , \left| {\beta  \gamma }\right \rangle - \left| {\alpha  \delta }\right \rangle , \left| {\alpha  \beta }\right \rangle$ \\
\hline
\multicolumn{1}{c|}{Total No.} & $10 = 3 \times 3 +1$ & $6 = 1 \times 3 +3 \times 1$ \\
\hline \hline
\end{tabular}
}
\\
\subfloat[$N=3$]{
\begin{tabular}{c|c|c|c}
\hline \hline
\multirow{2}{*}{S}& \multicolumn{3}{|c}{$\lambda$} \\
\multicolumn{1}{c|}{}  & \multicolumn{1}{c}{$[3]$} & \multicolumn{1}{c}{$[2,1]$} & \multicolumn{1}{c}{$[1,1,1]$}   \\[2pt]
\hline
\multicolumn{1}{c|}{3/2} & $\left| {\gamma  \gamma  \gamma }\right\rangle , \left| {\alpha  \gamma  \gamma }\right \rangle , \left| {\alpha  \alpha  \gamma }\right \rangle , \left| {\alpha  \alpha  \alpha }\right \rangle$ & $\left| {\alpha  \gamma  \gamma }\right\rangle , \left| {\alpha  \alpha  \gamma }\right \rangle$ & -- \\[4pt]
\multicolumn{1}{c|}{1/2}  & $\left| {\alpha  \alpha  \delta }\right \rangle + 2\left| {\alpha  \beta  \gamma }\right \rangle , \left| {\alpha  \alpha  \beta }\right \rangle $ & $\begin{array}{l}  \left| {\gamma  \gamma  \delta }\right\rangle , \left| {\beta  \gamma  \gamma }\right \rangle - \left| {\alpha  \gamma  \delta }\right \rangle , \left| {\alpha  \delta  \gamma }\right \rangle - \left| {\alpha  \gamma  \delta }\right \rangle \\ {1 \over{2}} \left| {\alpha  \alpha  \delta }\right \rangle + \left| {\alpha  \gamma  \beta }\right \rangle , \left| {\alpha  \beta  \gamma }\right \rangle - \left| {\alpha  \alpha  \delta }\right \rangle, \left| {\alpha  \alpha  \beta }\right\rangle  \end{array}$ & $\left| {\alpha  \gamma  \delta }\right\rangle , \left| {\alpha  \beta  \gamma }\right \rangle$ \\[1.1em]
\multicolumn{1}{c|}{Counting} & $ 20 = 4 \times 4 + 2 \times 2$ & $20 = 2 \times 4 + 6 \times 2$ & $4 = 2 \times 2$ \\
\hline \hline
\end{tabular}
}
\\
\subfloat[$N=4$]{
\begin{tabular}{c|c|c|c|c|c}
\hline \hline
\multirow{2}{*}{S} & \multicolumn{5}{|c}{$\lambda$} \\
\multicolumn{1}{c|}{}  &  \multicolumn{1}{c}{$[4]$} &  \multicolumn{1}{c}{$[3,1]$} &  \multicolumn{1}{c}{$[2,2]$} &  \multicolumn{1}{c}{$[2,1,1]$} &  \multicolumn{1}{c}{$[1,1,1,1]$}   \\[2pt]
\hline
\multicolumn{1}{c|}{2} & $\begin{array}{l}\left| {\gamma  \gamma  \gamma  \gamma }\right\rangle , \left| {\alpha  \gamma  \gamma  \gamma }\right \rangle ,\\ \left| {\alpha  \alpha  \gamma  \gamma }\right \rangle , \left| {\alpha  \alpha  \alpha  \gamma }\right \rangle ,\\ \left| {\alpha  \alpha  \alpha  \alpha }\right \rangle \end{array}$ & $\left| {\alpha  \gamma  \gamma  \gamma }\right\rangle , \left| {\alpha  \alpha  \gamma  \gamma }\right \rangle , \left| {\alpha  \alpha  \alpha  \gamma }\right \rangle $ & $\left| {\alpha  \alpha  \gamma  \gamma }\right\rangle$ & -- & --  \\
\hline
\multicolumn{1}{c|}{1} & $\begin{array}{l} \left| {\beta  \gamma  \gamma  \gamma }\right \rangle - \left| {\alpha  \gamma  \gamma  \delta }\right \rangle ,\\ \left| {\alpha  \beta  \gamma  \gamma }\right \rangle - \left| {\alpha  \alpha  \gamma  \delta }\right \rangle ,\\ \left| {\alpha  \alpha  \beta  \gamma }\right \rangle - \left| {\alpha  \alpha  \alpha  \delta }\right \rangle \end{array}$ & $\begin{array}{l} \left| {\gamma  \gamma  \gamma  \delta }\right\rangle , \left| {\alpha  \alpha  \alpha  \beta }\right\rangle,\\ \left| {\beta  \gamma  \gamma  \gamma }\right\rangle -\left| {\alpha  \gamma  \gamma  \delta }\right\rangle , \\ \left|{\alpha  \gamma \delta  \gamma }\right\rangle - \left| {\alpha  \gamma  \gamma  \delta }\right\rangle ,\\ \left| {\alpha  \alpha  \gamma  \delta }\right\rangle + \left|
{\alpha  \gamma  \gamma  \beta }\right\rangle ,\\  \left| {\alpha  \beta  \gamma  \gamma }\right\rangle - \left| {\alpha  \alpha  \gamma  \delta }\right\rangle ,\\ \left| \
{\alpha  \alpha  \delta  \gamma }\right\rangle - \left| {\alpha  \alpha  \gamma  \delta }\right\rangle , \\ \frac{1}{3}\left| {\alpha  \alpha  \alpha  \delta }\right\rangle + \left| {\alpha  \alpha  \gamma  \beta }\right\rangle ,\\ \left| {\alpha  \alpha  \beta  \gamma }\right\rangle -\left| {\alpha  \alpha  \alpha  \delta }\right\rangle \\ \end{array}$ & $\begin{array}{l} \left| {\alpha  \gamma  \gamma  \delta }\right\rangle ,\\ \left| {\alpha  \beta  \gamma  \gamma }\right\rangle -\left| {\alpha  \alpha  \gamma  \delta }\right\rangle,\\ \left| {\alpha  \alpha  \beta  \gamma }\right\rangle\end{array}$ & $\begin{array}{l} \left| {\alpha  \gamma  \gamma  \delta }\right\rangle , \\ \left| {\alpha  \gamma  \beta  \gamma }\right\rangle , \\ \left| {\alpha  \alpha  \gamma  \delta }\right\rangle , \\ \left| {\alpha  \alpha  \beta  \gamma }\right\rangle \end{array}$ & -- \\[5.3em]
\hline
\multicolumn{1}{c|}{0}  & $\left| {\beta \beta  \gamma  \gamma }\right \rangle - 2\left| {\alpha  \beta  \gamma  \delta }\right \rangle$ & $\begin{array}{l} \left| {\alpha  \gamma  \delta  \delta }\right\rangle -\left| {\alpha  \delta  \delta  \gamma }\right\rangle  \\-\left| {\beta  \gamma  \gamma  \delta }\right\rangle + \left| {\beta  \gamma  \delta  \gamma }\right\rangle , \\ \left| {\alpha  \alpha  \delta  \delta }\right\rangle  -\left| \ {\alpha  \beta  \gamma  \delta }\right\rangle \\- \left| {\alpha  \beta  \delta  \gamma }\right\rangle + \left| {\beta  \beta  \gamma \gamma }\right\rangle , \\ -\frac{1}{6}\left| {\alpha  \alpha  \beta  \delta }\right\rangle  -\frac{1}{2}\left| {\alpha  \alpha  \delta  \beta }\right\rangle \\+ \frac{1}{3}\left| {\alpha  \beta  \beta  \gamma }\right\rangle + \left| {\alpha  \beta  \gamma  \beta }\right\rangle\end{array}$ & $\begin{array}{l} \left| {\gamma  \gamma  \delta  \delta }\right\rangle ,\left| {\alpha  \alpha  \beta  \beta }\right\rangle , \\
\left| {\beta  \gamma  \gamma  \delta }\right\rangle -\frac{1}{2}\left| {\alpha  \gamma  \delta  \delta }\right\rangle ,\\ \left| {\alpha  \alpha  \delta  \delta }\right\rangle  -2\left| {\alpha  \beta  \gamma  \delta }\right\rangle + \left| {\beta  \beta  \gamma  \gamma }\right\rangle, \\ \frac{1}{2}\left| {\alpha  \beta  \gamma  \delta }\right\rangle + \left| {\alpha  \gamma  \beta  \delta }\right\rangle , \\ \left| {\alpha  \beta  \beta  \gamma }\right\rangle -\frac{1}{2}\left| {\alpha  \alpha  \beta  \delta }\right\rangle \end{array}$ & $\begin{array}{l} \left| {\beta  \gamma  \gamma  \delta }\right\rangle -\left| {\alpha  \delta  \gamma  \delta }\right\rangle , \\ \left| {\alpha \delta  \beta  \gamma }\right\rangle -\left| {\alpha  \beta  \gamma  \delta }\right\rangle -\left| {\alpha  \gamma  \beta  \delta }\right\rangle , \\ \left| {\alpha  \beta  \beta  \gamma }\right\rangle -\frac{1}{2}\left| {\alpha  \alpha  \beta  \delta }\right\rangle  \end{array}$ & $\left| {\alpha  \beta  \gamma  \delta }\right\rangle$ \\[4em]
\hline
\multicolumn{1}{c|}{Count.} & $35 = 5 \times 5 + 3 \times 3 +1 $ & $45 = 3 \times 5 + 9 \times 3 +3 $ & $20 = 1 \times 5 + 3 \times 3 +6 $ & $15 = 4 \times 3 +3$ & 1 \\
\hline \hline
\end{tabular}
}
\\
\subfloat[$N=5$. In this table we simply state the number of cases.]{
\begin{tabular}{*{1}{c}*{6}{c}}
\hline \hline
\multirow{2}{*}{S} & \multicolumn{6}{|c}{$\lambda$} \\
\multicolumn{1}{c|}{}  & $[5]$ & $[4,1]$ & $[3,2]$ & $[3,1,1]$ & $[2,2,1]$ & $[2,1,1,1]$  \\[2pt]
\hline
\multicolumn{1}{c|}{5/2} & 6 & 4 & 2 & 0 & 0 & 0 \\

\multicolumn{1}{c|}{3/2} & 4 & 12 & 6 & 6 & 2 & 0\\

\multicolumn{1}{c|}{1/2}  & 2 & 6 & 12 & 6 & 6 & 2 \\
\hline
\multicolumn{1}{c|}{Counting} & 56 & 84 & 60  & 36 & 20 & 4\\
\hline \hline
\end{tabular}
}
\caption{Decomposition of SU(4) appropriate for the cases such as graphene which has SU(2) $\times$ SU(2) internal states (spin $\times$ valley).   In the spin language the notation should be interpreted as  $\alpha$ is $s_z=+1/2$ in first valley and $\beta$ is  $s_z =-1/2$ in first valley; $\gamma$ is $s_z=+1/2$ in the second valley and $\delta$ is  $s_z =-1/2$ in the second valley. Note that only the highest possible $S_z$ states of a given $S$ are listed: additional states with the same $S$ eigenvalue can be found by applying a spin lowering operator to any of the states listed here. We have also listed the total number of states for each symmetry type (including the $S_z$ degeneracy): the total should equal the number of four-component \generalizedspin{s} for that symmetry type given in \tref{tableGeneralizedSpinWaveFunctions}.}
\label{tableGrapheneBasis}
\end{table}

\twocolumngrid

%%%%%%%%%%%%%%%%%%%%%%%%%%%%%%%%%%
%%%%%%%%%%%%%%%%%%%%%%%%%%%%%%%%%%

\section{Discussion}
\label{secDiscussion}

The main result of this work is the enumeration of the space of $N$-particle wave functions for bosons or fermions with internal degrees of freedom such as spin (i.e., multiple components).  These wave functions can be used as a basis for writing pseudopotential coefficients to parametrize physical problems, as, for example, in Refs.~\onlinecite{yang2008,bishara2009}.   More importantly, however, determining such a complete basis strongly suggests much of the physics of wave functions that can be generated with model interactions.

To elaborate on this last point, let us return to the spinless (single component) case for a moment, and for simplicity, let us consider bosons. In the case of $N=3$ body interaction, there are three-body wave functions with $L=0,2,3,4, \ldots$ (this can be read off from the table in Ref.~\onlinecite{simon2007a}, which is identical to the $N=3$ fully symmetric $\lambda=[3]$ case of \tref{tableDimensionsOfPolynomialSpaces} here). One can consider a family of model Hamiltonians that successively gives each of these terms some positive energy. \cite{simon2007c} For example, if $L=0$ is given positive energy (i.e., there is a nonzero pseudopotential coefficient $V_{L=0,N=3}$) and no other potential energy term is in the Hamiltonian, then it can be demonstrated that the highest density zero-energy ground state is the Moore--Read wave function, which has the property that when three particles approach each other, the wave function vanishes as $L=2$ powers (since $L=1$ is forbidden). \cite{moore1991} One can then consider adding a positive energy for $L=2$ as well (a positive pseudopotential $V_{L=2, N=3}$), which results in the Gaffnian \cite{simon2007b} as its ground-state wave function, where the wave function vanishes as $L=3$ powers when three particles approach each other. One can continue by adding positive energy for $L=3$, resulting in the Haffnian \cite{green2002} which then vanishes as $L=4$ powers when three particles approach each other.  

For higher $L$ the situation becomes somewhat more complicated for several reasons. \cite{simon2007c,jackson2013}  First, there can be several wave functions with the same value of $L$  (e.g., two such $N=3$ wave functions at $L=6$), so specifying $L$ alone (equivalent to specifying a thin-torus limit \cite{bergholtz2007,bergholtz2008,ardonne2008, seidel2011} or a pattern of zeros \cite{wen2008b}) does not fully specify the wave function. Second, it is possible that a Hamiltonian that forbids all $L < L_0$ for clusters of $N$ particles can have a ground-state wave function where $L > L_0$ for clusters of $N$ particles rather than $L=L_0$ (see, for example, the discussion in Ref.~\onlinecite{simon2007c}). Nonetheless, the general idea that one can dictate the vanishing properties of a wave function by appropriately choosing nonzero pseudopotentials remains a powerful approach both to understanding the properties of quantum Hall wave functions and to generating new and interesting trial states.

To generalize this approach to the multicomponent case, if one is interested in two component (spin-1/2) quantum Hall states, for, say, bosons, we can look at \tref{tableDimensionsOfPolynomialSpaces} and quickly see what kind of pseudopotentials are possible, which then also suggests what kind of wave functions might occur.  In the simplest case we might consider two-body interactions. Trivially, in this case (looking in the $N=2$ rows of the table), we see that even $L$ occurs in the triplet channel ($\sboson=1$), whereas odd $L$ occurs in the singlet channel ($\sboson=0$). Choosing to give positive energy (positive pseudopotential coefficient) to all $L < m$  (triplet) even and all $L < n$ (singlet) odd generates the Halperin $mmn$ ground state. \cite{halperin1983}

Let us now move on to three-body interactions and, for simplicity, let us still consider spin-1/2 bosons.  In the spin polarized ($\sboson=3/2$) channel we see exactly the same structure we did for spinless bosons; wave functions occurring at $L=0,2,3,4, \ldots$  [$N=3, S=3/2$ line of \tref{tableDimensionsOfPolynomialSpaces}]. In the $\sboson=1/2$ channel, on the other hand, we see wave functions at $L=1,2,3, \ldots$. A simple example of a three-body Hamiltonian is one that forbids $L=0$ in the polarized ($\sboson=3/2$) channel (a positive $V_{L=0,N=3}^{S=3/2}$ only);  such a Hamiltonian generates the $k=2$ NASS state as its ground state. \cite{ardonne1999,ardonne2001} As one might suspect, this wave function vanishes as $L=2$ powers when three particles come together in the $\sboson=3/2$ channel and vanishes as $L=1$ power in the $\sboson=1/2$ channel.  These are the lowest powers on the table that are not explicitly forbidden by the Hamiltonian.

By similarly examining the tables presented in the current work we can easily propose new wave functions that generalize those already discussed, and we may even be able to guess at the model Hamiltonians that generate these wave functions as their ground states.   For example, generalizing the $k=2$ NASS state we might propose to forbid the lowest powers that are not forbidden by the NASS Hamiltonian, i.e., we choose to forbid $N=3, S=3/2, L=2$ and the $N=3 ,S=1/2, L=1$ in addition to the $N=3, S=3/2, L=0$  of the NASS wave function.  We might guess that the resulting wave function should vanish  $L=3$ powers when three particles come together in the $S=3/2$ channel and as $L=2$ powers when three come together in the $S=1/2$ channel.  Such a wave function, a spin-singlet generalization of the Gaffnian, has been discussed recently in Refs.~\onlinecite{estienne2012,davenport2013a}. 

The examples discussed here are only a fraction of the model Hamiltonians that have been discussed in the literature. \cite{haldane1988,moore1991,read1999,ardonne1999,ardonne2001,ardonne2002,green2002,reijnders2002,reijnders2004,simon2007b,hormozi2012} Since the language of multicomponent pseudopotentials is arbitrarily general, all of these model Hamiltonians can be rephrased into this language.  In many cases it is quite obvious from looking at our tables what vanishing behaviors for $N$-particle cluster are being forbidden and what the resulting properties of the corresponding ground-state wave functions should therefore be (albeit rigorous proofs of these correspondences may be more tricky).

Once one starts considering multicomponent wave functions, there are clearly quite a few model Hamiltonians and corresponding model wave functions that can be considered (see Refs.~\onlinecite{ardonne2011,estienne2012} for recent progress in these directions).   For example, in Ref.~\onlinecite{estienne2012} spin-singlet versions of Jack wave functions have been constructed.   However, a much larger variety of states may be constructed by relaxing some of the restrictions of that work.  One possible direction is the generation of wave functions that mix $N$ and $N'$ body multicomponent interactions; for example, the double Pfaffian wave function discussed in Ref.~\onlinecite{hormozi2012} is the ground state of the sum of a three-body interaction and a two-body interaction.   Because of the added richness of the multicomponent case over the single component case there are certainly a far wider variety of possible model Hamiltonians and model wave functions to be explored.  Potentially we may even find some new physically realizable multicomponent wave functions with interesting braiding properties that can be exploited for quantum information processing. \cite{nayak2008}

Even in the simple spinless (single component) case there remain quite a few open questions about model wave functions, such as Jacks, and model Hamiltonians; and these questions generalize to similar questions about the multicomponent cases.  For example,  it remains uncertain in many cases whether simple (physical) Hamiltonians exist that generate these special wave functions as their ground states.  Further, outside of the simplest Jack wave functions (the Read--Rezayi series and the Laughlin wave functions), all of the spin-polarized Jacks correspond to nonunitary conformal field theories \cite{estienne2009,bernevig2009}  which excludes them from describing gapped states of matter. \cite{Read2009}   In at least some cases, \cite{simon2007b,green2002} it is suspected that these special wave functions may correspond to critical points between gapped phases; however, a general understanding of the criticality of these special wave functions is lacking.   There have been, however, some attempts to understand some of these properties for the spinless case by understanding the relationship to the multicomponent cases first. \cite{hermanns2011a} This provides yet another motivation for exploring multicomponent Hamiltonians and wave functions.

\begin{acknowledgments}
First, we acknowledge the hospitality of the Institut Henri Poincar\'e, Paris, during the final stages of the preparation of this work. We also thank Jacob Katriel, who has been very helpful in pointing us in the direction of some key references concerning the construction of spin eigenfunctions and \generalizedspin{s}. Finally, we thank Eddy Ardonne for some useful discussions. This research was supported by the EPSRC Grant No. EP/I032487/1. 
\end{acknowledgments}

%**************************************************************************************************************************************************%

%%%%%%%%%%%%%%%%%%%%%%%%%%%%%%%%%%
%%%%%%%%%%%%%%%%%%%%%%%%%%%%%%%%%%

\appendix

\vspace*{20pt}
\begin{center}
{ {\small {\bf APPENDICES}}}
\end{center}

In this set of appendices we shall set out to provide a detailed exposition of the linear vector space inhabited by (multicomponent) Haldane pseudopotentials, that is, the vector space defined in \eref{eqDefinitionGeneralizaedPseudopotential},
\[
|L, {\mathfrak{q}}\rangle.
\]
In doing so, the key realization is that vector space of multicomponent pseudopotentials is \emph{identical} to the vector space of coordinate wave functions describing a multicomponent quantum Hall system. We can, thus, map the problem of discovering a suitable basis in which to describe pseudopotentials in the LLL to a problem of constructing multiparticle coordinate wave functions in the LLL with an internal degree of freedom. The properties required of these wave functions are that they should be homogeneous of degree $L$, translationally invariant, and overall (anti-) symmetric for (fermions) bosons.  These conditions are equivalent to the statement that each wave function would be a rotationally invariant state on an appropriately sized sphere. 

In these appendices we shall provide a description of how to construct such wave functions $\psi \equiv |L, {\mathfrak{q}}\rangle$. 
Very generally, systems with $n$ internal degrees of freedom are characterized by irreducible representations of the special unitary group SU($n$). The fundamental objective of this work is, thus, to decompose the space of wave functions $\psi$ in terms of a basis of wave functions corresponding to irreducible representations of SU($n$).

The layout of the appendices will be as follows: in \aref{appendixMathPreliminaries} we shall allude to some important mathematical preliminaries, which are to be employed in the remaining appendices; in \aref{appendixConstructionOfMulticomponentWaveFunctions} we shall describe the general procedure for the construction of the basis of wave functions from a combination of spatial functions and \generalizedspin{s}; in \aref{appendixGeneralizedSpinWaveFunctions} we shall concentrate on the details of the \generalizedspin{s} and we shall derive the results presented in \tref{tableGeneralizedSpinWaveFunctions}; in \aref{appendixSpatialWaveFunctions} we shall concentrate on the details of the spatial parts of the wave functions; in particular, we shall apply the procedure of \aref{appendixConstructionOfMulticomponentWaveFunctions} in the context of quantum Hall wave functions, and we shall derive the coordinate wave functions listed in  \tref{tablePrimitivePolynomialsFermions} and \tref{tablePrimitivePolynomialsBosons} and the dimensions of the vector space of polynomials  listed in \tref{tableDimensionsOfPolynomialSpaces}; finally, \aref{appendixFurtherMaths} deals with some more advanced mathematical underpinnings of these ideas. 

%%%%%%%%%%%%%%%%%%%%%%%%%%%%%%%%%%
%%%%%%%%%%%%%%%%%%%%%%%%%%%%%%%%%%

\section{Mathematical Prelimaries}
\label{appendixMathPreliminaries}

Our goal in this appendix will be to deal with the underlying mathematical concepts and theorems that are to be employed in later appendices. The underlying mathematics is that of integer partitions, of Young tableaux and of the representation theory of the symmetric group and special unitary group SU($n$).

\subsection{Integer Partitions}

The concept of an \emph{integer partition} is frequently used in this work. An integer partition $\lambda$ is defined as follows \cite{andrewsbook}: for a positive integer $N$, an integer partition is a way of writing $N$ as a list of $k$ positive integer summands $\lambda = [N_1,N_2,\ldots,N_k]$ for which $\sum_{i=1}^k N_i = N$. The list of summands is typically presented in weakly descending order (that is, $N_1 \ge N_2 \ge N_3 \ldots\ge N_k$). For example, the set of integer partitions for the integer 5 are $[5]$, $[4,1]$, $[3,2]$, $[3,1,1]$, $[2,2,1]$, $[2,1,1,1]$, $[1,1,1,1,1]$.

When writing down a set of integer partitions such as that given in the above example, it is clearly important to specify an ordering system. In this paper we shall make use of \emph{lexicographic ordering} \cite{andrewsbook}: when comparing two integer partitions $\lambda$ and $\mu$ we say that $\lambda$ occurs before $\mu$, denoted $\lambda \ge \mu$,   if $\lambda_i = \mu_i$ for $i=1,\ldots,j$ for some integer $j$ between 1 and $k-1$, and then $\lambda_{j+1} > \mu_{j+1}$. In the above example we have used lexicographic ordering, for instance, $[3,2]$ occurs before $[3,1,1]$. In Appendix~\ref{appendixFurtherMaths} we shall discuss a more general ordering method known as \emph{dominance ordering}. 
 
\subsection{Young Tableaux}

The concept of \emph{Young tableaux} is an important tool for categorizing group representations, which we shall need to do shortly. Before discussing specific group representations, let us, first, explain how to construct Young tableaux in general. 

To draw a Young tableau, one first draws a \emph{Young frame} or diagram. A Young frame is constructed as follows: given an integer partition  $\lambda$, we build a frame by placing a series of empty boxes in rows and columns. There is one row for each of the $k$ summands in the integer partition: the $i$th row contains $\lambda_i$ boxes. We shall use the convention that a new row is started beneath the previous row and that the rows are in the order specified by the integer partition; hence, the length of a row will always be less than or equal to the length of the row above. In addition, the rows are to be left justified (we conform to the English notation).  For example, the Young frame corresponding to the integer partition $\lambda = [3, 2,1,1]$ would be
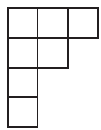
\[
\yng(3,2,1,1)
\]

A Young tableau can be constructed from a Young frame by placing a series of $N$ integers in the boxes. The set of integers can be chosen in several ways; of particular significance to our work are the arrangements called \emph{standard tableaux} and the arrangements called \emph{semistandard tableaux}. 

In a standard tableau, the integers 1 to $N$ are placed in the boxes of a Young frame of size $N$, with each number occurring precisely once. Further, we place the numbers such that the integer placed in any box is strictly less than both the integer placed in the box immediately to the right and the integer placed in the box immediately below. For example, here is a possible standard tableau constructed from the Young frame given above:
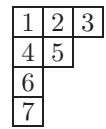
\[
\young(123,45,6,7)
\]

There can, of course, be multiple ways to satisfy these conditions, in other words, if the shape of a Young tableau is specified by an integer partition $\lambda$ then there is a corresponding set of size $f^{\lambda}$ of admissible standard tableaux. We introduce an index $r$, which runs from 1 to $f^{\lambda}$, to distinguish between the possible standard tableaux $T_r$ in the set. 

In order to specify one particular standard tableau in the set of $T_r$ we must first order that set: to do so we shall again employ a kind of lexicographic ordering system. For a given standard tableau we generate a list of numbers starting with the number in the top left box (for the above tableau that would be 1), then working along each row from left to right taking the numbers from these boxes and then adjoining the numbers from each subsequent row in the same fashion. Doing so will generate a list of $N$ numbers $N_1$ to $N_N$ (in the above tableau that would be $[1,2,3,4,5,6,7]$) and we then simply apply the procedure of lexicographic ordering to these lists.

Let us demonstrate the definitions given in the last two paragraphs via the following example: the integer partition $\lambda = [3,1]$ has $f^{[3,1]} = 3$, and the lexicographically ordered set of standard tableaux are
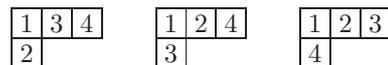
\[
\young(134,2) \qquad \young(124,3) \qquad \young(123,4)
\]

A semistandard tableau is constructed along similar lines: in this case, however, one is allowed to place any of the numbers 1 to $n$ in the boxes of a Young frame of size $N$, where $N$ does not have to equal to $n$. It is not necessary to include every number from 1 to $n$ and the numbers can also be repeated. Further, we place the numbers such that they are nondecreasing from left to right and strictly increasing from top to bottom. For example, here is a possible semistandard tableau of type $n=4$ and $N=8$:
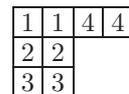
\[
\young(1144,22,33)
\]
The complete set of semistandard tableaux for a given $\lambda$ is enumerated by listing every possible arrangement of subsets of the numbers 1 to $n$ in the $N$ boxes and satisfying the conditions described above. As with the standard tableaux, there are multiple ways to satisfy these conditions and we, therefore, have a set of  $h^\lambda$ admissible semistandard tableaux. Semistandard tableaux can also be ordered lexicographically, using the same procedure as for the standard tableaux. For example, with $\lambda=[2,1]$ and with $n=3$ we have $h^{[2,1]}=8$ and the set of lexicographically ordered semistandard tableaux are
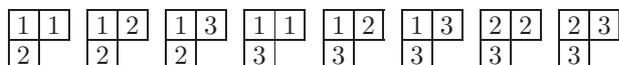
\[
\young(11,2) \,\, \,\, \young(12,2) \,\, \,\, \young(13,2) \,\, \,\, \young(11,3) \,\, \,\, \young(12,3) \,\, \,\, \young(13,3) \,\, \,\, \young(22,3) \,\, \,\, \young(23,3)
\]

Finally, we shall introduce the term \emph{conjugate} to refer to any tableaux related by a reflection along the diagonal, for example, the following tableau of shape $\lambda=[3,1]$:
\[
\young(134,2)
\]
is conjugate to the following tableau of shape $\tilde{\lambda} = [2,1,1]$:
\[
\young(12,3,4)
\]

\subsection{Irreducible Representations of the Symmetric Group}

The \emph{symmetric group} ${\rm{S}}_N$ is the group of permutations on $N$ objects, $\left\{\pi_1,\ldots,\pi_{N!}\right\}$ . A permutation $\pi_i $ of labels $(123\ldots N)$ to $({\rm{i}_1\rm{i}_2\rm{i}_3\ldots\rm{i}}_N)$ is denoted by $({\rm{i}_1\rm{i}_2\rm{i}_3\ldots\rm{i}}_N)$. A transposition (pair interchange) of label $\rm{i}$ and label $\rm{j}$ with all other labels unchanged is denoted by $(\rm{i};\rm{j})$. 

The classification of multicomponent wave functions depends essentially on the properties of ${\rm{S}}_N$ and, in particular, on the understanding of irreducible representations of ${\rm{S}}_N$. Throughout this appendix we shall assume that the reader is familiar with the fundamental concepts of representation theory; a more detailed discussion can be found in Ref.~\onlinecite{hamermeshbook}. The correspondence between wave functions and representations of the symmetric group manifests itself in the concept of a symmetry type, as described in \sref{subPseudoSymmetryTypes}.  The interpretation in terms of symmetry types is said to be the physicists' interpretation of the mathematics of the irreducible representations. \cite{hamermeshbook}

In order to construct multicomponent wave functions, we will find that we need to construct an appropriate \emph{symmetric group algebra}. A well-known example of a symmetric group algebra is the algebra of Young operators (for their definition, see \aref{appendixFurtherMaths}). There is, very generally, a direct correspondence between a representation of a group and a representation of the group algebra \cite{hamermeshbook}. For our purposes it will be convenient to construct the symmetric group algebra corresponding to the \emph{orthogonal} representation of the symmetric group (the reasoning behind this choice is explained in \aref{appendixFurtherMaths}; essentially it will give rise to multicomponent wave functions with convenient orthogonality properties). We shall define the orthogonal representation of the symmetric group shortly.  In order to construct such an algebra we will then make use of the \emph{matric} basis of the symmetric group algebra, which is a very general method to construct a basis directly from a group representation. \cite{paunczbook,rutherfordbook}

\subsubsection{Young's Orthogonal Representation of the Symmetric Group}

A fundamental principle of the representation theory of the symmetric group is that the irreducible representations of the symmetric group are in one-to-one correspondence with the standard tableaux. \cite{hamermeshbook} Given the set of Young tableaux of shape $\lambda$ we can always construct some kind of irreducible representation matrices $U\left( {{\rm{\pi }}_{i} } \right)^\lambda$ of dimension $f^{\lambda}$ for ${\rm{\pi }}_i \in {\rm{S}}_N $. In order for the representation to be an orthogonal representation, all that is required is that the representation matrices are themselves orthogonal matrices, meaning that they satisfy $U\left( {{\rm{\pi }}_{i}^{ - 1} } \right)_{rs}  = U\left( {{\rm{\pi }}_{i} } \right)_{sr}$.

In fact, orthogonal representation matrices can be constructed directly from standard tableaux using a measure called the \emph{axial distance}. \cite{rutherfordbook} Suppose that the number p appears in the $i_{\rm{p}}$th row and the $j_{\rm{p}}$th column of a given standard tableau and that the number q appears in the $i_{\rm{q}}$th row and the $j_{\rm{q}}$th column. The axial distance from p to q in tableau $T_r$ is defined as
\[
d_{{\rm{p,q}}}^{\,r}  = \left( {i_{\rm{p}}  - j_{\rm{p}} } \right) - \left( {i_{\rm{q}}  - j_{\rm{q}} } \right).
\]
Using this definition one can construct orthogonal representation matrices $U\left( {\rm{\pi }}_i  \right)$  for ${\rm{\pi }}_i \in {\rm{S}}_N $ as follows: first, we define the elements of the representation matrix $U\left[ \left( {\rm{k} - 1} ; {\rm{k}}\right) \right]$ of transpositions $\left( {{\rm{k}} - 1;{\rm{k}}} \right) $ for ${\rm{k}} = 1 \ldots N$:
\begin{align*}
&U\left[ \left( {{\rm{k}} - 1;{\rm{k}}} \right) \right]_{rs}  = -1 / {d_{{\rm{k}} - 1,{\rm{k}}}^r } = \rho_r  \,, \\
&U\left[ \left( {{\rm{k}} - 1; {\rm{k}}} \right) \right]_{rs}  =  \left\{ \begin{array}{l} 0 \qquad \qquad \qquad \,\,\, {\rm{if}}\,\,T_s  \ne \left( {{\rm{k}} - 1; {\rm{k}}} \right)T_r  \\ \left( {1 - \rho_r^2 } \right)^{1/2} \qquad {\rm{if}}\,\,T_r  = \left( {{\rm{k}} - 1; {\rm{k}}} \right)T_r\end{array} \right. 
\end{align*}
We then note that every permutation can be expressed as a product of transpositions of the form $\left( {\rm{k}} - 1 ; {\rm{k}}\right)$;  the proof is given in Ref.~\onlinecite{rutherfordbook}, for example. Using this result we can generate the representation matrices corresponding to every other element of the symmetric group by simple matrix multiplication. This is Young's orthogonal representation. Some of these orthogonal matrices are tabulated in Ref.~\onlinecite{hamermeshbook}.

Representations of the symmetric group corresponding to conjugate Young tableau shapes are known as \emph{contragradient} (or dual) representations (see Ref.~\onlinecite{paunczbook}). We can explicitly construct contragradient orthogonal representation matrices $V \left( {{\rm{\pi }}_{i} } \right)^{\tilde \lambda}$, which are related to the original orthogonal representation matrices $U\left( {{\rm{\pi }}_{i} } \right)^\lambda$ by
\[
V\left( {{\rm{\pi }}_i } \right)_{rs}^{\tilde \lambda}   = {{\mathop{\rm sgn}} \left( {{\rm{\pi }}_i}  \right)} U\left( {{\rm{\pi }}_i } \right)_{rs}^\lambda  .
\]
Here we have used the sign of a permutation, ${{\mathop{\rm sgn}} \left( {{\rm{\pi }}_i}  \right)}$, which is defined as follows: writing a permutation as a product of transpositions, if there are $m$ transpositions in the product then the sign of the permutation is ${{\mathop{\rm sgn}} \left( {{\rm{\pi }}_i}  \right)} = (-1)^m$.

\subsubsection{The Matric Basis of the Symmetric Group Algebra}
\label{subPseudoMatricBasis}

For any given set of symmetric group representation matrices $U\left( {{\rm{\pi }}_i } \right)^\lambda$ of dimension $f^{\lambda}$ we can, in general, form the following elements of the corresponding symmetric group algebra, called \emph{matric units} \cite{paunczbook} or \emph{seminormal units} \cite{rutherfordbook}:
\begin{equation}
{\hat{e}}_{rs}^\lambda   = \frac{{f ^{\lambda}  }}{{N!}}\sum\limits_{i = 1}^{N!} {U\left( {{\rm{\pi }}_i^{ - 1} } \right)_{sr}^\lambda  {\rm{\pi }}_i } \,\,\,\,\,\,\,\,\,\,\,\,\,r,s = 1,\ldots,f^{\lambda}.
\label{eqMatricBasis}
\end{equation}
These elements are linearly independent (the proof of which is given in Ref.~\onlinecite{rutherfordbook}) and can be chosen as a basis for the symmetric group algebra (see Ref.~\onlinecite{paunczbook}). They satisfy the orthogonality condition (with no summation implied):
\begin{equation}
{\hat{e}}_{rs}^\lambda  {\hat{e}}_{uv}^\mu   = \delta _{\lambda \mu } \delta _{su} {\hat{e}}_{rv}^\lambda .
\label{eqMatrixUnitOrthogonality}
\end{equation}
The diagonal elements, $\hat e_{rr}$, are idempotent and mutually orthogonal. Some explicit forms of these operators are tabulated in Ref.~\onlinecite{davenport2012a}.

Using the matric units, we can also define a basis for the group algebra of a cotragradient representation,
\begin{equation}
{\hat{e}}_{rs}^{\tilde \lambda}   = \frac{{f ^{\lambda}  }}{{N!}}\sum\limits_{i = 1}^{N!} {V\left( {{\rm{\pi }}_i } \right)_{rs}^{\tilde \lambda}  {\rm{\pi }}_i } \,.
\label{eqContragradientMatricBasis}
\end{equation}
and these elements satisfy the same condition as above, namely
\[
{\hat{e}}_{rs}^{\tilde \lambda}  {\hat{e}}_{uv}^{\tilde \mu}   = \delta _{\tilde \lambda \tilde \mu } \delta _{su} {\hat{e}}_{rv}^{\tilde \lambda}.
\]

\subsection{Irreducible Representations of SU($n$)}

The special unitary group, denoted SU($n$), is the group of $n \times n$ unitary matrices with determinant 1. SU($n$) also describes the structure of multicomponent systems. Irreducible representations of SU($n$) are in one-to-one correspondence with certain subsets of the irreducible representations of the symmetric group; specifically, those representations associated with the set of standard Young tableaux restricted to having at most $n$ rows (see Ref.~\onlinecite{hamermeshbook} for more details). 

\subsection{The Lie algebra of SU($n$)}

In this work we describe wave functions with internal degrees of freedom such as spin. The reader will likely be familiar with the fact that quantum states labeled by spins are classified by the irreducible representations of the Lie algebra of SU(2), that is, the algebra of the $\hat S^2$ and $\hat S_z$ operators. More generally, a \generalizedspin{} will be classified by representations of the Lie algebra of SU($n$). Irreducible representations of the Lie algebra of SU($n$) are sometimes called \emph{multiplets} and each multiplet contains a vector space of associated functions or states. \cite{pfeiferbook} For example, for SU(2) the multiplets correspond to the values of the quantum numbers $S$ and the states within each multiplet are labeled by the possible values of $S_z$; for SU(3) the multiplets correspond to the classification of baryonic particles in the quark model. \cite{halzenbook} The general form of the SU($n$) multiplets is described in detail in Ref.~\onlinecite{pfeiferbook}. Since the multiplets correspond to the irreducible representations of SU($n$), they are labeled in accordance with the set of Young tableaux of shape $\lambda$. In other words, a multiplet labeled by $\lambda$ corresponds to a particular symmetry type $\lambda$.

In order to label a particular state in a given multiplet, one must specify the component content of that state: for example, in the two-component case, the number of spin-up and spin-down particles (or, alternatively, the $S_z$ eigenvalue); in the three-component case, the quark content of the state, and so on. We shall introduce the notation $\lambda_z$, which will specify the component content of a given state and, thus, distinguish between different states in the same multiplet (the nomenclature is inspired by $S_z$). In addition, there can be multiple states that occur for the same value of $\lambda_z$; the number of such states is called the multiplicity $M(\lambda,\lambda_z)$. 

There is a one-to-one correspondence between the states of a SU($n$) multiplet and the set of semistandard tableaux. \cite{hamermeshbook} For a given $\lambda$ we construct the set of semistandard tableaux containing up to $n$ different numbers; these can be interpreted as the set of components $\lambda_z$ for that state. Recall our earlier example of the set of SU(3) semistandard tableaux: 
\[
\young(11,2)
\]
corresponds to $\lambda_z = \left\{2,1,0\right\}$, and since there is only one occurrence of this type of tableau we have $M([2,1],\left\{2,1,0\right\})=1$. Similarly, we have
\[
\young(12,3) \qquad \young(13,2) 
\]
which both correspond to $\lambda_z=\left\{1,1,1\right\}$. In this case we clearly have $M([2,1],\left\{1,1,1\right\})=2$ for SU(3). The total number of states in each multiplet, including the multiplicity, is the SU($n$) dimension of the representation; it is calculated by enumerating the complete set of semistandard Young tableaux for a given $\lambda$, for example, with $\lambda=[2,1]$ and $n=3$, the dimension is 8. 

A more complete description of the Lie algebras of SU($n$) and their representations can be found in Ref.~\onlinecite{pfeiferbook}.

\subsection{Tensor Products of SU($n$) Multiplets}

The problem of constructing a vector space of all possible $N$-particle \generalizedspin{s} is equivalent to the mathematical problem of forming tensor products of $N$ \emph{fundamental} multiplets. A fundamental multiplet is the irreducible representation corresponding to a Young tableau containing a single box---so for SU($n$) the dimension of the fundamental multiplet is equal to $n$. Such tensor products can be decomposed into a direct sum of multiplets. \cite{pfeiferbook} An important feature is that in this direct sum it is possible to have multiple occurrences of the same multiplet (we shall give an example shortly). In our general construction we shall therefore introduce an index $r$ which will distinguish between such repeated multiplets. This decomposition enables us to determine the possible multiplets and multiplet states which fully span our vector space. 

For illustrative purposes we shall go through the three-component case in detail here. A three-component system can be decomposed into SU(3) multiplets. We can determine the allowed set of \generalizedspin{s} by forming tensor products of SU(3) representations of a single-particle state.

We start with a single-particle state with three components, which corresponds to the fundamental SU(3) multiplet. To deduce the vector space of two-particle \generalizedspin{s} we form a tensor product of two fundamental multiplets (here we label the multiplets by the shapes of the corresponding young frames):
\[
\yng(1) \otimes \yng(1) =\yng(2) \oplus \yng(1,1)
\]
This product is evaluated according to the Littlewood-Richardson rule (see, for example, Ref.~\onlinecite{hamermeshbook}) to give a direct product of SU(3) multiplets. The [2] representation is six-dimensional, while the [1,1] representation is three-dimensional (cf. the $N=2$ case in \tref{tableGeneralizedSpinWaveFunctions}).

If we want to include a third particle then we construct
\[
\yng(1) \otimes \yng(1) \otimes \yng(1) =\yng(3) \oplus \yng(2,1) \oplus \yng(2,1) \oplus \yng(1,1,1)
\]
We see that there are two copies of the [2,1] multiplet, which has dimension 8. The two copies are distinguished by the index $r$ and the states within the two multiplets are mutually orthogonal to each other (so in this example we can have $r=1$ or $r=2$). The [3] multiplet has dimension 10, and the [1,1,1] has dimension 1: so we see that the total dimension is $10+2 \times 8+1=27  = 3^3$ (cf. the $N=3$ case in \tref{tableGeneralizedSpinWaveFunctions}). The total dimensions calculated in this way are consistent with the counting of states listed in \tref{tableGeneralizedSpinWaveFunctions}.

We can also apply the Littlewood--Richardson rule to the construction of standard tableaux, thus, we observe
\[
\young(1) \otimes \young(2) \otimes \young(3) = \young(123) \oplus \young(13,2) \oplus \young(12,3) \oplus \young(1,2,3)
\]
We see that the two copies of the [2,1] multiplet can be distinguished by their association to standard Young tableaux. This result comes about because of the underlying connection between tensor representations of groups and irreducible representations of the symmetric group: specifically the symmetry property of the indices of the tensor itself forms a representation of the symmetric group.  

More generally, therefore, the number of repeated occurrences of a particular Young tableau shape $\lambda$ simply corresponds to the dimension $f^\lambda$ of the symmetric group representation that would also be denoted by a Young tableau of that shape (a proof of this result is given, for example, in Ref.~\onlinecite{hamermeshbook}). For example, the symmetric group representation dimension of the [2,1] tableau is 2 (recall that the symmetric group representation dimension is given by the number of standard tableaux for a given tableau shape).  Thus, one can associate a specific standard Young tableau shape with each occurrence of an SU($n$) multiplet in our decomposition. 

To summarize, the tensor product of $N$ fundamental multiplets will contain every admissible irreducible representation of the Lie algebra of SU($n$), that is, the irreducible representations corresponding to every possible standard tableaux of shape $\lambda$ with size $N$ and with no more than $n$ rows, \cite{hamermeshbook} and in such a decomposition the multiplet described by $\lambda$ will occur $f^{\lambda}$ times, with the occurrences being distinguished by an association to standard Young tableaux. 

%%%%%%%%%%%%%%%%%%%%%%%%%%%%%%%%%%
%%%%%%%%%%%%%%%%%%%%%%%%%%%%%%%%%%

\section{General Procedure for Construction of multicomponent Wave functions}
\label{appendixConstructionOfMulticomponentWaveFunctions}

In this appendix we shall describe the solution to the very general problem of constructing wave functions for multicomponent systems by decomposing into a basis of spatial wave functions and \generalizedspin{s}. In the following appendix, \aref{appendixGeneralizedSpinWaveFunctions}, we shall describe the forms of the  \generalizedspin{s} in more detail. Finally, in \aref{appendixSpatialWaveFunctions}, we shall describe how to apply the general arguments for construction of spatial wave functions presented in the current appendix to the special case of construction of translationally invariant LLL quantum Hall wave functions with fixed relative angular momentum.

The vector space of wave functions describing a system of identical particles with internal degrees of freedom can always be written as a tensor product of the spaces of spatial and \generalizedspin{s}:
\[ 
 \left| {\psi } \right\rangle \,\, \in \,\, \left| {\Phi _{{\rm{spatial}}}} \right\rangle  \otimes \left| {X_{{\rm{\gssubscript{}}}}} \right\rangle .
\]
Thus, the dimension of the Hilbert space (with no symmetry restriction) is equal to the product of the dimensions of the space of spatial and \generalizedspin{s}.   Additionally, such a wave function must conform to the symmetry condition required for a system of identical fermions (bosons)---namely the wave function must be antisymmetric (symmetric) under any exchanges of the particle labels. 

If we are trying to determine the dimension of the space of wave functions with $N$ particles of some degree $L$ that obeys the proper symmetry condition, we can, in principle, start with the most general spatial function $\Phi _{{\rm{spatial}}}$ of this degree and the most general \generalizedspin{} $X_{{\rm{\gssubscript{}}}}$  (where a ``most general wave function'' contains one arbitrary coefficient multiplying each of the basis functions in the space). The most general $\Phi _{{\rm{spatial}}}$ is given by a linear combination of all possible arrangements of $N$ particles into $N_{\mbox{\tiny orbitals}}$ single-particle orbitals $\phi_1$ \ldots $\phi_N$ (including multiple occupancy) with a fixed combined degree:
\begin{equation}
\Phi_{{\rm{spatial}}}    = \sum\limits_{i_1,\ldots, i_N = 1}^{N_{\mbox{\tiny orbitals}}} {a_{i_1,\ldots,i_N} \left[ \phi_{i_1} (r_1) \ldots \phi_{i_N}(r_N)\right] } ,
\label{eqPrimitiveSpatial}
\end{equation}
where $a_{i_1,\ldots,i_N}$ are arbitrary coefficients. Similarly the most general $X_{{\rm{\gssubscript{}}}}$ is given by
\[
X_{{\rm{\gssubscript{}}}} = \sum\limits_{j=1}^{n^N} x_j \vartheta_j ,
\]
where $x_j$ are arbitrary coefficients and where $\vartheta_j$ are the complete set of primitive \generalizedspin{s} of the form given in \eref{eqPrimitiveGeneralizedSpinWaveFunction}, of which there are $n^N$ for an $N$-particle, $n$-component system. The most general combined wave function is given by the product of the \generalizedspin{} and spatial wave function: $X_{{\rm{\gssubscript{}}}} \Phi_{{\rm{spatial}}}$. 

Next, we impose that the wave function $\psi$ is overall antisymmetric (for fermions) by applying an antisymmetrizer $\mathcal{\hat A}$ to the most general construction,
\begin{equation}  
\psi = \mathcal{\hat A}{\Phi _{{\rm{spatial}}}}{X_{{\rm{\gssubscript{}}}}},
\label{eqMostGeneralConstruction} 
\end{equation}
where
\[
\mathcal{\hat A}  = \frac{1}{\sqrt{N!}} \sum\limits_{i=1}^{N!} {{\mathop{\rm sgn}} \left( {{\rm{\pi }}_i}  \right)}  {\rm{\pi}}_i.
\] 
For bosons we simply replace the antisymmetrizer by a symmetrizer $\mathcal{\hat S}$:
\[
\mathcal{\hat S}  = \frac{1}{\sqrt{N!}} \sum\limits_{i=1}^{N!}   {\rm{\pi}}_i.
\] 
After applying one of these symmetry operators to the most general construction we will be left with a wave function containing a certain number of linearly independent coefficients: the total number of those coefficients is the size of the Hilbert space. One can always deduce by brute force the number of linearly independent coefficients in $\psi$ by determining the linearly independent combinations of the $a_{i_1,\ldots,i_N}$ and $x_j$ coefficients remaining after the symmetry operator is applied. In general, however, this method is extremely inefficient and becomes essentially algebraically impossible for more than a very few particles and very low degrees. A much cleaner approach is to fully exploit the symmetries of the problem, which is what we shall do. 

We explained in \aref{appendixMathPreliminaries} that \generalizedspin{s} are classified by irreducible representations of the Lie algebra of SU($n$) given by forming tensor products of $N$ fundamental multiplets. Specifically, we can decompose the vector space of  \generalizedspin{s} into subspaces labeled by Young tableaux of shape $\lambda$ and in such a decomposition the multiplet labeled by $\lambda$ occurs $f^{\lambda}$ times.  We can, thus, write a basis of \generalizedspin{s} $X_{{\rm{\gssubscript{}}}} \left( N, \lambda, \lambda_z, r\right) $ labeled in the same way as states within SU($n$) multiplets, where $N$ is the number of particles, $\lambda$ is the symmetry type, and where $r$ runs from 1 to $f^{\lambda}$ and distinguishes between SU($n$) multiplets occurring in the tensor product with the same $\lambda$. We shall use the convention that wave functions with different $r$ values are mutually orthogonal.   Recall from \sref{subMultiComponentResults} (and see below in \sref{subGeneralizedSpinWaveFunctions}) that $\lambda_z$ describes how many particles in the \generalizedspin{} are of each type $\alpha$, $\beta$, $\gamma, \ldots$.  Note that it may or may not be possible to construct a  wave function in representation $\lambda$ given a particular $N$ and $\lambda_z$ as we will describe further in \sref{subGeneralizedSpinWaveFunctions} below.

Given the above decomposition of the  \generalizedspin{} into representations, we can analogously decompose the combined wave function $\psi$ based on the same irreducible representation of $SU(n)$ . This follows because the spatial function is independent of the component values. Thus, without loss of generality, we may divide the space of possible $N$-particle wave functions $\psi$ into subspaces indexed by $\lambda$ and $\lambda_z$. In order to decompose our wave functions into a basis of irreducible representations while remaining of the general form given in \eref{eqMostGeneralConstruction}, we propose the following alternative general form for a basis for $\psi$, which loses no generality:
\begin{equation}
\psi   \left(N,\lambda,\lambda_z\right) = \sum\limits_{r=1}^{f^\lambda} c_r \mathcal{\hat A} \,  \Phi _{{\rm{spatial}}} X_{{\rm{\gssubscript{}}}} \left( {N,\lambda,\lambda_z ,r} \right),
\label{eqAlternativeGeneralConstruction}
\end{equation}
where $c_r$ are arbitrary coefficients. For bosons we simply use a symmetrization operator $\mathcal {\hat S}$ instead of an antisymmetrization operator $\mathcal {\hat A}$.

In this expression we are free to choose the most general possible $\Phi _{{\rm{spatial}}}$.   However, here not all different choices of $\Phi_{\rm{spatial}}$ will correspond to different wave functions $\psi$. In \sref{subGeneralizedSpinWaveFunctions} we will first construct the \generalizedspin{s} $X_{{\rm{\gssubscript{}}}} \left( N, \lambda, \lambda_z, r \right)$, then, in \sref{subCombinationSpinSpatial} we will (use group theory to) determine how to construct a complete (but not over-complete) basis for the spatial wave functions $\Phi_{\rm{spatial}}$. 

\subsection{Introduction to Generalized Spin Wave functions}
\label{subGeneralizedSpinWaveFunctions}

The \generalizedspin{s} $X_{{\rm{\gssubscript{}}}} \left( N, \lambda, \lambda_z, r \right)$ describing a system with $n$ components correspond to particular irreducible representations of SU($n$): those given by forming tensor products of $N$ fundamental multiplets.  Such representations are in correspondence with the standard tableaux of shape $\lambda$ and, hence, with a subset of the symmetric group. We must specify the tableau shape $\lambda$, where $\lambda$ is a partition of the integer $N$ into at most $n$ parts, the standard tableau index $r$, and the component composition of the function, which is denoted by $\lambda_z$. In order to enforce that $X_{{\rm{\gssubscript{}}}} \left( N, \lambda, \lambda_z, r \right)$ are mutually orthogonal in $\lambda$, $\lambda_z$ and $r$, we shall construct our SU($n$) representations based on the orthogonal representation of the symmetric group.  

The  \generalizedspin{s} can be written as linear combinations of primitive \generalizedspin{s}: these are defined  by specifying a list of the values that each component can take; $\alpha$, $\beta$, $\gamma$, and so on. For example, $\vartheta_j = \left| { \alpha \gamma  \beta \beta \alpha} \right\rangle $ is a primitive \generalizedspin{}.  One can order these wave functions according to a lexicographic ordering scheme, so the first primitive \generalizedspin{} would be 
\[
\vartheta_1 \left(N,\lambda_z \right)= \left| { \alpha  \alpha  \alpha \ldots \beta \beta \ldots  \zeta  \zeta} \right\rangle .
\]
The index $\lambda_z$ represents the list $\left\{ N_{\alpha,} N_{\beta},\ldots,N_{\zeta}\right\} $ of $N$ that corresponds to the number of times each type of component value occurs in the primitive \generalizedspin{}. In general, we have wave functions $\vartheta_j$, where $j$ runs from 1 to the number of primitive \generalizedspin{s} $d \left(\lambda_z\right)$ of all possible orderings of the components and indexes the position of $\vartheta_j$ within an ordered list of those wave functions. 

The explicit forms of the \generalizedspin{s} can be deduced as follows: first, we write down the orthogonal representation matrices $U({\rm{\pi}}_i)^{\lambda}$ corresponding to the subset of standard tableaux of shape $\lambda$; from this orthogonal representation we then construct a corresponding group algebra using the matric basis, defined in \eref{eqMatricBasis}. As the elements of the group algebra are expressed as a linear combination of permutation operators, one can apply the matric unit operator ${\hat{e}}_{ss}^\lambda$ to any primitive \generalizedspin{} to give either zero or a linear combination of permuted primitive \generalizedspin{s}. A key result of this construction is that there is an irreducible representation of the symmetric group associated with each $\lambda$ and that representations associated with different  $\lambda$  are inequivalent. \cite{rutherfordbook} Thus, the \generalizedspin{s} constructed in this way are in one-to-one correspondence with the irreducible representations of SU($n$) and with the SU($n$) multiplets occurring in the tensor product of $N$ fundamental multiplets. In other words, the operator ${\hat{e}}_{ss}^\lambda$ acts as a projector on the primitive \generalizedspin{s} to give a particular representation of SU($n$). 

A further key result is that the wave functions ${\hat{e}}_{rs}^\lambda  \vartheta_j (N, \lambda_z)$ are either zero or, independently of the choice of $r$ and $s$, they belong to the same symmetry type  $\lambda$. Due to the orthogonality of the matric unit operators, defined in \eref{eqMatrixUnitOrthogonality}, two such wave functions  ${\hat{e}}_{rs}^\lambda  \vartheta_j (N, \lambda_z)$ and  ${\hat{e}}_{r's}^\lambda  \vartheta_j (N, \lambda_z)$ are orthogonal if $r \ne r'$.  The operators ${\hat{e}}_{rs}^\lambda$ act as shift operators in the sense that
\[
{\hat{e}}_{rs}^\lambda  \left[ {{\hat{e}}_{ss}^\lambda  \vartheta_j (N, \lambda_z)} \right] = {\hat{e}}_{rs}^\lambda  \vartheta_j (N, \lambda_z),
\]
and so operating on the first projected function $\hat{e}_{ss}^\lambda  \vartheta_j$ yields a set of orthogonal functions belonging to the same symmetry type and, further, by acting with all such possible shift operators on all possible $\vartheta_j$, we obtain a complete basis of states having the same $\lambda$ and $\lambda_z$. 

For the SU(2) case it is possible to prove further that $\hat S^2 {\hat{e}}_{ss}^\lambda  \vartheta_j (N,S_z)= S(S+1){\hat{e}}_{ss}^\lambda  \vartheta_j (N,S_z)$ with $\lambda$ the representation corresponding to spin $S$ (the proof is given in Ref.~\onlinecite{paunczbook}). In other words, for the two-component case, the  \generalizedspin{s} are precisely spin eigenfunctions. 

Thus, we may decompose the complete space of \generalizedspin{s} $X_{{\rm{\gssubscript{}}}}$  in terms of the multiplet label $\lambda$ and $r$, and the multiplet state label $\lambda_z$.  Most generally, we write
\begin{equation}
X_{{\rm{\gssubscript{}}}} \left( N, \lambda, \lambda_z ,r \right) = \sum\limits_{s=1}^{f^{\lambda}}\sum\limits_{j=1}^{d \left( \lambda_z \right)} { b_{s,j} \, \hat e_{rs}^{\lambda} \vartheta_j (N, \lambda_z)},
\label{eqGeneralGeneralizedSpinWaveFunction}
\end{equation}
where $b_{s,j}$ are arbitrary coefficients. Since these wave functions are constructed in correspondence with the multiplets of SU($n$), the number of linearly independent values of $b_{s,j}$ is precisely the multiplicity, $M(\lambda,\lambda_z)$, of the SU($n$) multiplet state (this can be explicitly checked). It can be shown that these wave functions are mutually orthogonal in $\lambda$, $\lambda_z$, and $r$ due to the orthogonality relations satisfied by the matric unit operators (see Ref.~\onlinecite{paunczbook} for the proof in the SU(2) case).

The mapping to representations of the symmetric group manifests itself in the permutation symmetry of the wave function, for example, $X_{{\rm{\gssubscript{}}}}\left( {N,\lambda,\lambda_z ,1} \right)$, independent of $\lambda_z$, must be symmetric in labels 1 to $N_1$, in labels $N_1+1$ to $N_2$ and so on where $N_i$ are the integers describing the partition $\lambda$.  We shall describe the explicit forms of the \generalizedspin{s} in more detail in \aref{appendixGeneralizedSpinWaveFunctions}.

\subsection{Combination of Generalized Spin wave functions and and Spatial wave functions}
\label{subCombinationSpinSpatial}

In this section of the appendix we shall now describe the procedure for sewing together the spatial and \generalizedspin{s} to give a combined wave function with the correct symmetry property. We shall then focus on the explicit form of the spatial part of the wave function. We aim to determine the dimension of the vector space of spatial wave functions.

So far we have shown that  the most general expression for the basis of combined \generalizedspin{s} is given in \eref{eqAlternativeGeneralConstruction}. It is insightful to split up the antisymmetrization operator into parts acting separately on either the spatial or spin functions---a permutation ${\rm{\pi}}_i$ is equivalent to identical permutations ${\rm{\pi}}^{\Phi}_i$ acting on the spatial wave function and ${\rm{\pi}}^{X}_i$ acting on the \generalizedspin{}. Thus, \eref{eqAlternativeGeneralConstruction} can be written in the equivalent form:
\[
\begin{array}{l}
\psi   \left(N,\lambda,\lambda_z\right) \\ =  \frac{1}{\sqrt{N!}} \sum\limits_{r=1}^{f^\lambda} c_r \sum\limits_{i=1}^{N!} {{\mathop{\rm sgn}} \left( {{\rm{\pi }}_i}  \right)}  {\rm{\pi}}^{\Phi}_i \Phi _{{\rm{spatial}}} {\rm{\pi}}^{X}_i X_{{\rm{\gssubscript{}}}}\left( {N,\lambda,\lambda_z ,r} \right).\\
\end{array}
\]
Now, the most general expression for the \generalizedspin{s} is given in \eref{eqGeneralGeneralizedSpinWaveFunction}. An important result, which follows from the definition of the matric unit operators, is that the action of an arbitrary permutation operator ${\rm{\pi}}^{X}_i$ on a \generalizedspin{} is
\begin{equation}
{\rm{\pi}}^{X}_i X_{{\rm{\gssubscript{}}}} \left( N, \lambda, \lambda_z ,r \right) = \sum\limits_{s=1}^{f^{\lambda}} X_{{\rm{\gssubscript{}}}} \left( N, \lambda, \lambda_z ,s \right) U({\rm{\pi}}_i)_{sr}^{\lambda}
\label{eqActionOfPermutationOnX}
\end{equation}
Using this result, we can write \eref{eqAlternativeGeneralConstruction} without the antisymmetrization operator in the form
\[
\psi \left( N, \lambda, \lambda_z\right)  =  \frac{1}{\sqrt{f^\lambda}} \sum\limits_{r=1}^{f^\lambda} X_{{\rm{\gssubscript{}}}} \left( {N,\lambda,\lambda_z ,r} \right) \sum\limits_{s=1}^{f^\lambda} { c_s \Phi _{rs} },
\]
where we have introduced the spatial wave function $\Phi_{rs}$, which is defined as
\[
\Phi _{rs}  = \left(\frac{f^\lambda}{N!}\right)^{1/2} \sum\limits_{i=1}^{N!} U({\rm{\pi}}_i)_{rs}^{\lambda} {{\mathop{\rm sgn}} \left( {{\rm{\pi }}_i}  \right)}  {\rm{\pi}}^{\Phi}_i \Phi _{{\rm{spatial}}}.
\]
Finally, using the definition of the contragradient matric unit operators, we can write the spatial wave function in the form
\[
\Phi _{rs}  = \left( {\frac{{N!}}{{f^{\lambda} }}} \right)^{1/2} {\hat{e}}_{rs}^{\tilde \lambda } \Phi _{{\rm{spatial}}} .
\]
In the boson case, where we require the wave function to be fully symmetric [thus removing the ${{\mathop{\rm sgn}} \left( {{\rm{\pi }}_i}  \right)}$ factor], it clearly follows that $\tilde \lambda$ is replaced by $\lambda$. Note that $\Phi _{{\rm{spatial}}}$ is still given by \eref{eqPrimitiveSpatial}.

Now, in order to simplify our presentation, we shall define the $r$th part of spatial wave function to be the spatial wave function associated with the $r$th \generalizedspin{}, in other words,
\begin{equation}
\Phi \left( {N,\lambda,r} \right) = \sum\limits_{s = 1}^{f^\lambda} {c_s \hat e_{rs}^{\tilde \lambda } \Phi_{{\rm{spatial}}}  } ,
\label{eqRthSpatialWaveFunction}
\end{equation}
or with $\tilde \lambda$ replaced with $\lambda$ for the boson case. These spatial wave functions are mutually orthogonal in $\lambda$ and in $r$ due to the othogonality relations satisfied by the matric unit operators. Note that the explicit form of $\Phi \left( {N,\lambda,r} \right)$ will be
\begin{equation}
\Phi \left( {N,\lambda,r} \right) = \sum\limits_{i_1,\ldots, i_N = 1}^{N_{\mbox{\tiny orbitals}}} {a'_{i_1,\ldots,i_N} \left[ \phi_{i_1} (r_1) \ldots \phi_{i_N}(r_N)\right] } .
\label{eqExplicitRthSpatialWaveFunction}
\end{equation}
where now the coefficients $a'$ depend on the combinations of the $a$ coefficients (from \eref{eqPrimitiveSpatial}) and the $c$ coefficients [introduced in \eref{eqAlternativeGeneralConstruction}] and, in particular, they must contain linear dependencies. 

Thus, the most general form of the basis wave functions is
\begin{equation}
\psi \left( N, \lambda, \lambda_z\right)  \propto \sum\limits_{r=1}^{f^{\lambda}} {X_{{\rm{\gssubscript{}}}}\left( {N,\lambda,\lambda_z ,r} \right) \Phi \left( {N,\lambda ,r} \right)}.
\label{eqMostGeneralBasisWaveFunction}
\end{equation}
Since both $X_{{\rm{\gssubscript{}}}}\left( {N,\lambda,\lambda_z ,r} \right)$ and $\psi   \left( N, \lambda, r \right)$ are mutually orthogonal in $\lambda$, $\lambda_z$ and in $r$, the wave function $\psi \left( N, \lambda, \lambda_z\right) $ is itself othogonal to wave functions with different $\lambda$ and $\lambda_z$ values.  We now propose to simplify this result using the following theorem.

\vspace{1em}

\noindent
{\bf Theorem 1:} \emph{The wave function $\psi (N, \lambda, \lambda_z)$ in \eref{eqMostGeneralBasisWaveFunction} can be written in the form $\mathcal{\hat {A}} \Phi \left( {N,\lambda,1} \right) X_{{\rm{\gssubscript{}}}}(N,\lambda,\lambda_z,1)$ for fermions or  $\mathcal{\hat {S}} \Phi \left( {N,\lambda,1} \right) X_{{\rm{\gssubscript{}}}}(N,\lambda,\lambda_z,1)$ for bosons.}

\vspace{1em}

\noindent
{\bf Proof:} Let us demonstrate the proof in the fermion case only, since the boson case follows along very similar lines. 
The proof uses the following additional results: first, from \eref{eqGeneralGeneralizedSpinWaveFunction}, using the orthogonality condition for the matric unit operators, we have:
\[
\begin{array}{l}
X_{{\rm{\gssubscript{}}}}(N,\lambda,\lambda_z,r)  = \sum\limits_{s=1,j=1}^{f^{\lambda}, \, d \left( \lambda_z \right)} { b_{s,j} \, \hat e_{rs}^{\lambda} \vartheta_j (N, \lambda_z)} \\
={\hat{e}}_{r1}^{\lambda}\sum\limits_{s=1,j=1}^{f^{\lambda}, \, d \left( \lambda_z \right)} { b_{s,j} \, \hat e_{1s}^{\lambda} \vartheta_j (N, \lambda_z)}= {\hat{e}}_{r1}^{\lambda} X_{{\rm{\gssubscript{}}}}(N,\lambda,\lambda_z,1).
\end{array}
\]
Thus, all coefficients in $X_{{\rm{\gssubscript{}}}}(N,\lambda,\lambda_z,r) $ are linearly dependent on the set of coefficients in $X_{{\rm{\gssubscript{}}}}(N,\lambda,\lambda_z,1)$.

Similarly, the second result is that
\[
\sum\limits_{s = 1}^{f^\lambda} {c_s \hat e_{rs}^{\tilde \lambda } \Phi_{{\rm{spatial}}}  } = \hat e_{r1}^{\tilde \lambda} \sum\limits_{s = 1}^{f^\lambda} {c_s \hat e_{1s}^{\tilde \lambda } \Phi_{{\rm{spatial}}}  },
\]
or, alternatively,
\begin{equation}
\Phi \left( {N,\lambda,r} \right) = \hat e_{r1}^{\tilde \lambda} \Phi \left( {N,\lambda,1} \right),
\label{eqResultNeededInProof}
\end{equation}
and so the coefficients in $\Phi \left( {N,\lambda,r} \right)$ are linearly dependent on the coefficients in $\Phi \left( {N,\lambda,1} \right)$. 

Now consider the following construction
\begin{equation}
\psi (N, \lambda, \lambda_z) \propto \mathcal{\hat {A}} \Phi \left( {N,\lambda,1} \right) X_{{\rm{\gssubscript{}}}}(N,\lambda,\lambda_z,1).
\label{eqResultOfTheorem}
\end{equation}
Writing the sum over permutations in the antisymmetrization explicitly, this becomes 
\[
\psi (N, \lambda, \lambda_z) \propto \sum\limits_{i=1}^{N!} {{\mathop{\rm sgn}} \left( {{\rm{\pi }}_i}  \right)}  {\rm{\pi}}_i \left\{ \Phi \left( {N,\lambda,1} \right) X_{{\rm{\gssubscript{}}}}(N,\lambda,\lambda_z,1) \right\}.
\]
Then we can use that the permutation ${\rm{\pi}}_i$ is equivalent to identical permutations ${\rm{\pi}}^{\Phi}_i$ acting on the spatial wave function and ${\rm{\pi}}^{X}_i$ acting on the \generalizedspin{} to write instead
\begin{align*}
&\psi (N, \lambda, \lambda_z) \\&\propto \sum\limits_{i=1}^{N!} {{\mathop{\rm sgn}} \left( {{\rm{\pi }}_i}  \right)}  {\rm{\pi}}^\Phi_i  \Phi \left( {N,\lambda,1} \right)  {\rm{\pi}}^X_i X_{{\rm{\gssubscript{}}}}(N,\lambda,\lambda_z,1).
\end{align*}
Now, using \eref{eqActionOfPermutationOnX}, we have, equivalently,
\begin{align*}
&\psi (N, \lambda, \lambda_z) \\&\propto \sum\limits_{s=1}^{f^{\lambda}} \sum\limits_{i=1}^{N!} {{\mathop{\rm sgn}} \left( {{\rm{\pi }}_i}  \right)}  {\rm{\pi}}^\Phi_i  \Phi \left( {N,\lambda,1} \right)   U({\rm{\pi}}_i)_{s1}^{\lambda} X_{{\rm{\gssubscript{}}}} \left( N, \lambda, \lambda_z ,s \right) .
\end{align*}
Then, using the definition of the contragradient matric unit operator in \eref{eqContragradientMatricBasis}, we have
\[
\psi (N, \lambda, \lambda_z) \propto \sum\limits_{s=1}^{f^{\lambda}} {\hat{e}}_{s1}^{\tilde \lambda} \Phi \left( {N,\lambda,1} \right) X_{{\rm{\gssubscript{}}}} \left( N, \lambda, \lambda_z ,s \right) .
\]
Finally, using \eref{eqResultNeededInProof}, we obtain
\[
\psi (N, \lambda, \lambda_z) \propto \sum\limits_{s=1}^{f^{\lambda}} \Phi \left( {N,\lambda,s} \right) X_{{\rm{\gssubscript{}}}} \left( N, \lambda, \lambda_z ,s \right),
\]
which is proportional to \eref{eqMostGeneralBasisWaveFunction}, as required. Thus, we have proved theorem 1. 

For the boson case, we should start with a symmetrizer $ \mathcal{\hat {S}} $ instead of the antisymmetrizer in the expression for $\psi (N, \lambda, \lambda_z) $ and we must, consequently, replace the contragradient representations $\tilde \lambda$ by $\lambda$.

\vspace{1em}

\noindent
{\bf Corollary:} We have expressed all of the linearly independent terms in the basis wave functions $\psi (N, \lambda, \lambda_z)$ in terms of the product $\Phi \left( {N,\lambda,1} \right) X_{{\rm{\gssubscript{}}}}(N,\lambda,\lambda_z,1)$; the entire basis wave function can then be recovered from this single term by simply antisymmetrizing or symmetrizing in all the particle coordinates, and, further, this antisymmetrization procedure does not reduce the number of linearly independent terms. In other words, the dimension of the space of multicomponent basis wave functions $\psi  \left( N, \lambda, \lambda_z\right)$ is given by the dimension of the space of \generalizedspin{s} $X_{{\rm{\gssubscript{}}}} \left( {N,\lambda,\lambda_z ,1} \right)$ multiplied by the dimension of the space of spatial wave functions $\Phi \left( {N,\lambda ,1} \right)$. Our task now is to determine the dimensions of these vector spaces. We shall discuss the \generalizedspin{s} $X_{{\rm{\gssubscript{}}}}(N,\lambda,\lambda_z,1)$ in \aref{appendixGeneralizedSpinWaveFunctions} and we shall discuss the spatial wave functions $\Phi \left( {N,\lambda,1} \right)$  in \aref{appendixSpatialWaveFunctions}.

%%%%%%%%%%%%%%%%%%%%%%%%%%%%%%%%%%
%%%%%%%%%%%%%%%%%%%%%%%%%%%%%%%%%%

\section{Vector Space of Generalized Spin wave functions}
\label{appendixGeneralizedSpinWaveFunctions}

In this appendix we shall explain how to derive the space of \generalizedspin{s} $X_{{\rm{\gssubscript{}}}}\left( {N,\lambda,\lambda_z ,r} \right)$ in a variety of cases. First, we shall explain the simplifications that arise in the description of the two-component case (spin-1/2 eigenfunctions); we shall then explain the most general $n$-component case. In the final two sections we shall consider two special cases; the description of three-component systems with spin-1 and the description of four-component systems with spin-1/2 combined with a two-valley or bilayer degree of freedom. In each case, we shall describe how the procedure described in \aref{appendixConstructionOfMulticomponentWaveFunctions} is applied to generate the vector space of \generalizedspin{s}. 

\subsection{Two-Component Systems}
\label{appendixSpinEigenfunctions}

An important observation is that any two-component multiparticle system can be mapped onto a system with a spin-1/2 degree of freedom regardless of whether it contains fermionic or bosonic particles. The procedure of decomposing into SU(2) representations is then identical to the problem of the construction of spin eigenfunctions---a well studied problem in theoretical chemistry. \cite{paunczbook} Note that, in general, a wave function need only be an eigenfunction of spin if the system is invariant under spin rotations.  Nonetheless, since a decomposition in terms of spin eigenfunctions is akin to a decomposition in terms of SU(2), the spin eigenfunctions will always provide a complete basis in which to describe any two-component system. 

Spin eigenfunctions for a multiparticle system are defined to be eigenfunctions of the total spin operator $\hat{S}^2$ and total $z$-component of spin operator $\hat S_z$.  Such functions can be expressed as linear combinations of \emph{primitive spin wave functions}: to repeat the definition given in \sref{subPseudoTwoComponentResults}, a primitive spin wave function $\vartheta_i$ of a many particle system is an eigenfunction of the $S_z$ operator of every particle in the system. (They are effectively two-component  \generalizedspin{s} where $\alpha \equiv \left| { \uparrow  } \right\rangle$ and $\beta \equiv \left| { \downarrow } \right\rangle$.)

Spin eigenfunctions $X\left( {N,S,{S_z},r} \right)$ are constructed from linear combinations of these primitive spin wave functions. For a system of $N$ particles with spin eigenvalue $S$, there is a one-to-one correspondence between the spin eigenfunctions and the subset of orthogonal irreducible representations of the symmetric group denoted by Young tableaux of the shape \mbox{ $\lambda =\left[ {{\textstyle{1 \over 2}}N + S,{\textstyle{1 \over 2}}N - S} \right] $; see Ref.~\onlinecite{paunczbook}.} 

The following results apply specifically to the two-component case. For any primitive spin wave function $\vartheta_j$ it can be shown that
\[
\hat{e}_{rr}^\lambda  \vartheta_j \propto {\hat{e}}_{r1}^\lambda  \vartheta_1, \,\,\, \mbox{or} \,\,\, 0.
\]
A corollary is that one can generate the entire space of spin eigenfunctions from only the first primitive spin wave function $\vartheta _1$. The key expression for the construction of spin eigenfunctions is then
\begin{equation}
\label{eigenfunctionintermsofmatricunit}
X\left( {N,S,S_z ,r} \right) \propto {\hat{e}}_{r1}^\lambda  \vartheta _1 .
\end{equation}

Let us demonstrate these ideas with a simple example: the three particle state with eigenvalues $S=1/2$ and $S_z=1/2$. The associated Young tableaux are of the shape $\lambda =
\left[ 2 , 1 \right] $~:
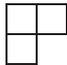
\[
\yng(2,1)
\]
There are two possible arrangements of standard tableau:
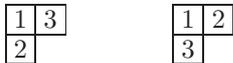
\[
\young(13,2) \qquad \qquad \young(12,3)
\]
and there are correspondingly two eigenfunctions with the same spin eigenvalue, each of which is generated by one of the group algebra basis operators 
\begin{align*}
 X \left( {3,{\textstyle{1 \over 2}},{\textstyle{1 \over 2}},1} \right)& =  \hat e^{[2,1]}_{11} \vartheta_1 \left(3, 1/2 \right) \\& \propto {\textstyle{1 \over {\sqrt 6 }}}\left( {2\left| { \uparrow  \uparrow  \downarrow } \right\rangle  - \left| { \uparrow  \downarrow  \uparrow } \right\rangle  - \left| { \downarrow  \uparrow  \uparrow } \right\rangle } \right) \\
 X\left( {3,{\textstyle{1 \over 2}},{\textstyle{1 \over 2}},2} \right) &= \hat e^{[2,1]}_{21} \vartheta_1 \left(3, 1/2 \right) \\& \propto {\textstyle{1 \over {\sqrt 2 }}}\left( {\left| { \uparrow  \downarrow  \uparrow } \right\rangle  - \left| { \downarrow  \uparrow  \uparrow } \right\rangle } \right). 
 \end{align*}
On inspection we see that $X\left( {3,{\textstyle{1 \over 2}},{\textstyle{1 \over 2}},1} \right)$ is symmetric under exchange of indices 1 and 2 with no particular symmetry conditions for index 3; moreover $X\left( {3,{\textstyle{1 \over 2}},{\textstyle{1 \over 2}},2} \right)$ is antisymmetric under exchange of the same indices. 

Recall the most general expression for the wave function given in \eref{eqResultOfTheorem}. In the two-component case, since $X (N,\lambda,\lambda_z,1) \equiv X (N,S,S_z,1) = \hat e^{\lambda}_{11} \vartheta_1 (N, S_z)$, we need only specify the first primitive spin wave function in order to describe the full wave function. Thus, in the two-component case the wave function (for fermions) can be succinctly expressed as
\begin{equation}
\psi (N, S, S_z) \propto \mathcal {\hat {A}} \Phi \left( {N,S,1} \right)  \vartheta_{1} (N, S_z).
\end{equation}

\subsection{Non Spin-Rotationally Invariant $n$-Component Systems}
\label{nonrotationallyinvariant}

In the most general $n$-component case, the system may not be invariant under spin rotations, or, if there is no spin degree of freedom present, rotations within the generalized spin space. The most general set of \generalizedspin{s} in this case are simply those specified by the decomposition into tensor products of $N$ fundamental  SU($n$) multiplets. The irreducible representations occurring in the decomposition are labeled by $\lambda$ and $r$, with no additional constraints---these are the functions described by \eref{eqGeneralGeneralizedSpinWaveFunction}. We note that $n$-component systems of this type have previously been conceived in Ref.~\onlinecite{pauncz1977}; however, the explicit wave functions have not been calculated.

Since primitive \generalizedspin{s} at different positions in the ordered set are related by permutations, it is possible to write \eref{eqGeneralGeneralizedSpinWaveFunction} [i.e., $X_{{\rm{\gssubscript{}}}}\left( {N,\lambda,\lambda_z ,r} \right)$] in the form:
\begin{equation}
\sum\limits_{s=1,j=1}^{f^{\lambda}, \, d \left( \lambda_z \right)} { b_{s,j} \, \hat e_{rs}^{\lambda} \vartheta_j (N, \lambda_z)}\equiv \sum\limits_{p=1}^{M(\lambda, \lambda_z)} b_{p} \hat e_{r1}^{\lambda} \vartheta_{p} (N, \lambda_z),
\label{eqMultipletFormGeneralizedSpin}
\end{equation}
where $p$ runs from 1 up to the multiplicity, $M(\lambda$, $\lambda_z)$, of the SU($n$) multiplet state; recall that this is given by the number of semistandard tableaux of shape $\lambda$ and component content $\lambda_z$. Note that since $p$ can be greater than 1 but is always less than or equal to $d(\lambda_z)$ (that is, the total number of primitive \generalizedspin{s} including, all possible orderings of components), we include only a certain subset of the lexicographically ordered primitive \generalizedspin{s} in the basis; these correspond precisely to the semistandard tableau shapes. For example, in the $\lambda=[3,1]$ state of SU(3) the semistandard tableaux are
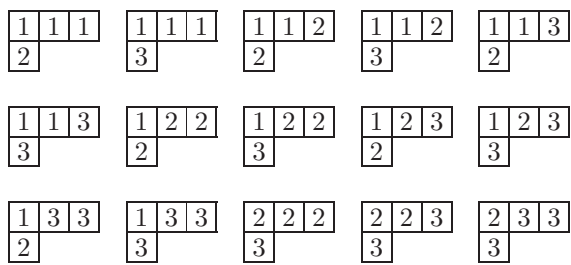
\[
\begin{array}{l}
\young(111,2) \quad \young(111,3) \quad \young(112,2) \quad \young(112,3) \quad \young(113,2) \\ \\
\young(113,3) \quad \young(122,2) \quad \young(122,3) \quad \young(123,2) \quad \young(123,3) \\ \\
\young(133,2) \quad \young(133,3) \quad \young(222,3) \quad \young(223,3) \quad \young(233,3) \\
\end{array}
\]
The corresponding primitive \generalizedspin{s} are given by interpreting the placement of the numbers 1 to $n$ in the semistandard tableau as the placement of the components $\alpha,\beta,\gamma,\ldots$ in the primitive \generalizedspin{s}.  The corresponding set of primitive \generalizedspin{s} that can occur in this state are then:
\[
\begin{array}{l} \left| {\alpha  \alpha  \alpha  \beta }\right\rangle , \left| {\alpha  \alpha  \alpha  \gamma }\right\rangle , \left| {\alpha  \alpha  \beta  \beta }\right\rangle , \left| {\alpha  \alpha  \beta  \gamma }\right\rangle , \left| {\alpha  \alpha  \gamma  \beta }\right\rangle ,\\ \left| {\alpha  \alpha  \gamma  \gamma }\right\rangle , \left| {\alpha  \beta  \beta  \beta }\right\rangle , \left| {\alpha  \beta  \beta  \gamma }\right\rangle , \left| {\alpha  \beta  \gamma  \beta }\right\rangle , \left| {\alpha  \beta  \gamma  \gamma }\right\rangle ,\\ \left| {\alpha  \gamma  \gamma  \beta }\right\rangle , \left| {\alpha  \gamma  \gamma  \gamma }\right\rangle , \left| {\beta  \beta  \beta  \gamma }\right\rangle , \left| {\beta  \beta  \gamma  \gamma }\right\rangle , \left| {\beta  \gamma  \gamma  \gamma }\right\rangle. \end{array}
\]
The set of $\vartheta_{p} (N, \lambda_z)$ for $N$ up to 4 and for SU(3) and SU(4) are listed in \tref{tableGeneralizedSpinWaveFunctions}. The results presented in \tref{tableGeneralizedSpinWaveFunctions} have been checked by explicitly applying the matric unit operators to all possible primitive \generalizedspin{s} and then determining the linearly independent combinations algebraically. The results of this brute-force approach for the numbers of linearly independent state precisely match with the multiplicities and multiplet dimensions deduced by the decomposition into irreducible representations of the Lie algebra of SU($n$) according to \aref{appendixMathPreliminaries}. 

Recall that in the tensor product of $N$ fundamental SU($n$) multiplets there are $f^{\lambda}$ irreducible representations associated with each symmetry type $\lambda$, for example with $\lambda=[2,1]$ we have $f^{\lambda}=2$, and, thus, there are  $f^{\lambda}$ \generalizedspin{s} of the form $X_{{\rm{\gssubscript{}}}}\left( {N,\lambda,\lambda_z ,r} \right)$. When we construct the fully symmetric or antisymmetric basis wave function by sewing together spin and spatial parts [\eref{eqAlternativeGeneralConstruction}], any orthogonal  \generalizedspin{s} of the same symmetry type and generated by the same root primitive \generalizedspin{} will appear in the same basis wave function. Thus, when counting the number of  \generalizedspin{s} we include this degeneracy (see \tref{tableGeneralizedSpinWaveFunctions}), but when counting the size of the vector space of \generalizedspin{s} we do not .

Recalling \eref{eqResultOfTheorem}, we aim to determine the most general form for $X_{{\rm{\gssubscript{}}}} (N,\lambda,\lambda_z,1)$.  We can use \eref{eqMultipletFormGeneralizedSpin} to write the basis of $X_{{\rm{\gssubscript{}}}} (N,\lambda,\lambda_z,1)$ in terms of $\hat e^{\lambda}_{11}$ applied to the set of primitive \generalizedspin{s} given in \tref{tableGeneralizedSpinWaveFunctions}. Thus, the minimal information to reconstruct the wave function is contained within that set of primitive \generalizedspin{s}. In the multicomponent case the wave function (for fermions) can then be succinctly expressed as
\begin{equation}
\psi (N, \lambda, \lambda_z, p) \propto \mathcal {\hat {A}} \Phi \left( {N,\lambda,1} \right)  \vartheta_{p} (N, \lambda_z),
\end{equation}
where the set of functions $\vartheta_{p} (N, \lambda_z)$ are taken from \tref{tableGeneralizedSpinWaveFunctions}.

\subsection{Spin-Rotationally Invariant $n$-Component Systems}
\label{subSpinRotationallyInvariant}

If it transpires that the system we are describing is spin-rotationally invariant, then it is convenient to construct a basis of spin eigenfunctions; the Hamiltonian will be diagonal in such a basis. We propose to construct appropriate spin eigenfunctions by forming particular linear combinations of the \generalizedspin{s} given in \eref{eqGeneralGeneralizedSpinWaveFunction}. The resulting wave functions will be labeled by spin eigenvalues $S$ and $S_z$. Let us begin by examining the origin of the basis in more detail.

A single particle that is symmetric under three-dimensional spin rotation is invariant under the three-dimensional rotation group O(3). \cite{hamermeshbook} These irreducible representations of the rotation group are characterized by the standard set of spin angular-momentum quantum numbers, $j$ and $j_z$. There exists a general mapping between such irreducible representations of the rotation group and irreducible representations of the group SU$(2j+1)$, \cite{hamermeshbook} and, in particular, there is a one-to-one mapping between a system of particles that individually have spin angular momentum $j=1/2$ and irreducible representations of SU(2) [since SU(2) is isomorphic to O(3) ]. In addition, there exists a further one-to-one mapping between irreducible representations of SU$(2j+1)$ and irreducible representations of the symmetric group, specifically the irreducible representations of SU$(2j+1)$ visualized by Young tableaux containing $N$ boxes in no more than $2j+1$ rows: $\lambda=[N_1,N_2,\ldots,N_{2j+1}]$ with $N_1 \ge N_2 \ge \ldots\ge N_{2j+1} \ge 0$.

Our general approach has been to decompose \generalizedspin{s} into a basis built from irreducible representations of the group SU($n$). Due to the isomorphism with the rotation group for the case of $n=2$, we find that our general approach in this case corresponds precisely to the construction of spin eigenfunctions. For $n>2$, however, the mapping becomes a little more complicated. In order to write our \generalizedspin{s} in a basis of spin eigenfunctions what we are really doing is decomposing representations of SU($n$) into representations of O(3). The procedure for this decomposition is explained in Ref.~\onlinecite{hamermeshbook}, chapter 11: The principle feature of this result is that this decomposition is not a one-to-one mapping for a general SU($n$) decomposition. In other words, due to this mapping, it follows that there are multiple spin eigenvalues corresponding to each $\lambda$. We must, therefore, label the basis by both $\lambda$ and by the spin eigenvalues $S$ and $S_z$, that is, our set of spin eigenfunctions are $X \left( N, \lambda, S, S_z, r \right)$, i.e., we have now replaced $\lambda_z$ with $S$ and $S_z$. These functions remain orthogonal in $\lambda$ and in $r$.

We shall now explain our procedure for how to construct spin eigenfunctions that describe spin-1 particles (the general case can be deduced along similar lines and checked using the information contained in Ref.~\onlinecite{hamermeshbook}). The explicit forms of the spin eigenfunctions can be determined as follows: First associate a spin value with our notation such that, for each particle, $\alpha$ represents the $s_z=1$ state, $\beta$ represents the $s_z =0$, and $\gamma$ represents the $s_z =-1$ state; we note that there could be several $\lambda_z$ values that could correspond to the same $S_z$ value, and so we form a linear combination of these states, for example $\lambda_z = \left\{1,1,1\right\}$ and $\lambda_z = \left\{0,3,0\right\}$ have the same $S_z$ value 0, and so we must work with the linear combination
\begin{align*}
&X \left( N, \lambda, S, S_z=0, r \right) \\ &= A X_{{\rm{\gssubscript{}}}} \left( N, \lambda, \left\{1,1,1\right\}, r \right) + B X_{{\rm{\gssubscript{}}}} \left( N, \lambda,  \left\{0,3,0\right\}], r \right),
\end{align*}
where $A$ and $B$ are arbitrary coefficients. In general, we would form a linear combination of all possible $\lambda_z$ for a given $S_z$. We then apply the $\hat S^2$ operator and solve the self-consistency condition
\begin{equation}
\hat S^2 X \left( N, \lambda, S, S_z, r \right)  = X \left( N, \lambda, S, S_z, r \right),
\label{eqApplySsquaredOperator}
\end{equation}
for the coefficients $A,B,\ldots$. 

We shall demonstrate our technique with some simple examples. Let us, first, consider $N=3$, $\lambda=[3]$, $S_z=1$, and $r=1$. In this case the valid primitive \generalizedspin{s} are:
\[
\left| \alpha \alpha \gamma \right\rangle ,  \left| \alpha \beta \beta \right\rangle.
\]
Applying $\hat S^2$ to this vector space gives
\[{\hat S^2}\hat e_{11}^{\left[ 3 \right]}\left( {\begin{array}{*{20}{c}}
   {\left| { \alpha \alpha \gamma } \right\rangle }  \\
   {\left| { \alpha \beta \beta} \right\rangle }  \\
\end{array}} \right) = \left( {\begin{array}{*{20}{c}}
   4 & 4  \\
   4 & {10}   \\
\end{array}} \right)\hat e_{11}^{\left[ 3 \right]}\left( {\begin{array}{*{20}{c}}
   {\left| { \alpha \alpha \gamma} \right\rangle }  \\
   {\left| {  \alpha \beta \beta} \right\rangle }  \\
\end{array}} \right).\]
Using this information we now apply \eref{eqApplySsquaredOperatorn} and we arrive at
\[
 \begin{array}{l}  \hat S^2 \hat e^{[3]}_{11} (A \left| \alpha \alpha \gamma \right\rangle +B \left| \alpha \beta \beta \right\rangle) \\
 =e^{[3]}_{11}  ( (4A+4B) \left|  \alpha \alpha \gamma \right\rangle + (4A+10B) \left|  \alpha \beta \beta \right\rangle) \\ = S(S+1) \hat e^{[3]}_{11} (A \left|  \alpha \alpha \gamma \right\rangle +B \left|  \alpha \beta \beta \right\rangle). \end{array}
\]
Hence, we must solve the eigenvalue equation
\[\left( {\begin{array}{*{20}{c}}
  4 & 4  \\
   4 &  {10}   \\
\end{array}} \right)\left( {\begin{array}{*{20}{c}}
   A  \\
   B  \\
\end{array}} \right) = S\left( {S + 1} \right)\left( {\begin{array}{*{20}{c}}
   A  \\
   B  \\
\end{array}} \right).\]
(Notice that the matrix appearing in the eigenvalue equation is the transpose of the matrix written in the first step.)
The solutions for $X \left( N, \lambda, S, S_z, 1 \right)$ in this case correspond to $S=1,3$, and the associated eigenvectors are
\[
\begin{array}{l}
 \hat e^{[3]}_{11} \left( \left| { \alpha \alpha \gamma} \right\rangle + 2 \left| { \alpha \beta \beta} \right\rangle \right), \\
 \hat e^{[3]}_{11} \left( \left| { \alpha \beta \beta} \right\rangle - 2  \left| { \alpha \alpha \gamma} \right\rangle \right),
\end{array}
\]
which are now precisely spin eigenfunctions. Given these results once can obtain the set of orthogonal spin eigenfunctions $X \left( N, \lambda, S, S_z, r \right)$ by applying the $\hat e^{[3]}_{r1}$ operator. 

The solutions obtained here ($S=1,3$) are precisely the spin eigenvalues corresponding to the decomposition of SU(3) into O(3) as described in Ref.~\onlinecite{hamermeshbook}. We have deliberately constructed the eigenfunctions in such a way that they can be specified by a linear combination of primitive \generalizedspin{s} $X_{\rm{lc}}$ (for example, $X_{\rm{lc}} =  \left| { \alpha \beta \beta} \right\rangle -2 \left| { \alpha \alpha \gamma} \right\rangle$) with a projection operator applied to them:
\[
X \left( N, \lambda, S, S_z, 1 \right) = \hat e^{\lambda}_{11}X_{\rm{lc}}.
\]
The projection operator does not depend on the set of spin eigenvalues, only on the symmetry type. Appropriate forms for $ X_{\rm{lc}} $ are given in \tref{tableSpinOneEigenfunctions}. Note that \tref{tableSpinOneEigenfunctions} only lists cases where $S=S_z$: spin eigenfunctions for $S_z<S$ can be deduced by explicitly applying a spin lowering operator $\hat S^-$. We have derived the results presented in \tref{tableSpinOneEigenfunctions} by precisely the procedure outlined in this appendix. 

Recall \eref{eqResultOfTheorem}. In the case described in the section, the minimal information required to construct the full wave function is just $X \left( N, \lambda, S, S_z, 1 \right)$. Since these functions can be expressed in the form $\hat e^{\lambda}_{11}X_{\rm{lc}} $, we have deduced that minimal information to reconstruct the wave function is in fact contained within these linear combinations $X_{\rm{lc}} $. Thus, the full wave function (for fermions) can be expressed as
\[
\psi (N, \lambda, S,S_z) \propto \mathcal {\hat {A}} \Phi \left( {N,\lambda,1} \right) X_{\rm{lc}} ,
\]

Let us give one further example briefly. Consider the $N=4$, $\lambda=[3,1]$, $S_z=1$ state. In this case, we must include the following set of primitive \generalizedspin{s}:
\[
\left| \alpha \alpha \beta \gamma \right\rangle , \left| \alpha  \alpha \gamma \beta \right\rangle , \left| \alpha  \beta  \beta  \beta \right\rangle
\]

Performing the same analysis we find:
\[{{\hat S}^2}\hat e_{11}^{\left[ {3,1} \right]}\left( {\begin{array}{*{20}{c}}
   {\left| \alpha \alpha \beta \gamma \right\rangle}  \\
   {\left| \alpha  \alpha \gamma \beta \right\rangle }  \\
   {\left| \alpha  \beta  \beta  \beta \right\rangle}  \\
\end{array}} \right) = \left( {\begin{array}{*{20}{c}}
   {10} & 0 & 2  \\
   2 & 4 & 2 \\
   4 & 4 & 6  \\
\end{array}} \right)\hat e_{11}^{\left[ {3,1} \right]}\left( {\begin{array}{*{20}{c}}
   {\left| \alpha \alpha \beta \gamma \right\rangle}  \\
   { \left| \alpha  \alpha \gamma \beta \right\rangle }  \\
   {\left| \alpha  \beta  \beta  \beta \right\rangle}  \\
\end{array}} \right)\]
In this case, we find that the possible $S$ values are 3, 2, and 1, precisely in correspondence with the underlying decomposition of SU(3) representations into O(3) representations. The resulting spin eigenfunctions obtained for this case are given in \tref{tableSpinOneEigenfunctions}. 

More generally, for a system of spin-$j$ particles (an example of an $n=2j+1$ component system) we can construct a basis of spin eigenfunctions by decomposing representations of SU$(2j+1)$ into representations of O(3), using a similar enumeration procedure to the three-component case discussed above. Tables of such decompositions are given in Ref.~\onlinecite{hamermeshbook},  for states up to $j=7/2$.

\subsection{Partially Spin-Rotationally Invariant $n$-Component Systems (Spin+Valley or Spin+Layer)}
\label{subGrapheneEigenfunctions}

A system such as graphene is a four-component system: there are two types of electrons corresponding to two valleys labeled by 1 and 2 or A and B. The same situation could arise in a bilayer system, where there could be two layers (1 and 2) of spin-1/2 particles. Although we are dealing with four-component wave functions here,  the spin component itself corresponds only to spin-1/2 (as opposed to the four-component spin-3/2 system  say). If the system is spin-rotationally invariant, but not invariant under valley symmetry, then it is fruitful to construct a basis in terms of spin eigenfunctions for the two types of valley electrons. As before, we propose to construct appropriate spin eigenfunctions by writing linear combinations of the \generalizedspin{s} that have the same $S_z$ eigenvalue and then solving a self-consistency equation as in \eref{eqApplySsquaredOperator}.

A four-component system has components labeled by $\alpha$, $\beta$, $\gamma$, and $\delta$. In order to write down spin eigenfunctions we specify the mapping: in valley 1, $\alpha$ corresponds to the $s_z=1/2$ state and $\beta$ corresponds to the $s_z =-1/2$ state and, in valley 2, $\gamma$ corresponds to the $s_z =+1/2$ state and $\delta$ corresponds to the $s_z =-1/2$ state. We can then proceed to construct spin eigenfunctions along the same lines as in \aref{subSpinRotationallyInvariant}: First, we enumerate all possible linearly independent four-component \generalizedspin{s}; we then apply the $\hat S^2$ operator to each. Finally we determine linear combinations of the four-component \generalizedspin{s} that result in spin eigenfunctions. A more rigorous mathematical underpinning of our method in terms of representation theory is presented in \aref{subSpinAndValleyEigenfunctions}.
 
As an example, consider the three-particle state with symmetry type [2,1] and $S_z =1/2$. Applying the $\hat S^2$ operator to the vector space of \generalizedspin{s} we find the following:
\begin{widetext}
\[{\hat S^2}\hat e_{11}^{\left[ 2,1 \right]}
\left(
\begin{array}{l}
{\left|  \alpha   \alpha   \beta \right\rangle} \\
{\left|  \alpha   \alpha   \delta  \right\rangle}\\
{\left|  \alpha   \beta   \gamma \right\rangle} \\
{\left|  \alpha   \gamma   \beta  \right\rangle}\\
{\left|  \alpha   \gamma   \delta  \right\rangle}\\
{\left|  \alpha   \delta   \gamma  \right\rangle}\\
 {\left| \beta   \gamma   \gamma  \right\rangle}\\
{ \left| \gamma   \gamma   \delta \right\rangle}
\end{array}
\right)
=
\left(
\begin{array}{cccccccc}
 \frac{3}{4} & 0 & 0 & 0 & 0 & 0 & 0 & 0 \\
 0 & \frac{7}{4} & 1 & -\frac{1}{2} & 0 & 0 & 0 & 0 \\
 0 & 2 & \frac{11}{4} & -1 & 0 & 0 & 0 & 0 \\
 0 & 0 & 0 & \frac{3}{4} & 0 & 0 & 0 & 0 \\
 0 & 0 & 0 & 0 & \frac{7}{4} & 1 & 1 & 0 \\
 0 & 0 & 0 & 0 & 1 & \frac{7}{4} & 1 & 0 \\
 0 & 0 & 0 & 0 & 1 & 1 & \frac{7}{4} & 0 \\
 0 & 0 & 0 & 0 & 0 & 0 & 0 & \frac{3}{4}
\end{array}
\right)
\hat e_{11}^{\left[ 2,1 \right]}
\left(
\begin{array}{l}
{\left|  \alpha   \alpha   \beta \right\rangle} \\
{\left|  \alpha   \alpha   \delta  \right\rangle}\\
{\left|  \alpha   \beta   \gamma \right\rangle} \\
{\left|  \alpha   \gamma   \beta  \right\rangle}\\
{\left|  \alpha   \gamma   \delta  \right\rangle}\\
{\left|  \alpha   \delta   \gamma  \right\rangle}\\
 {\left| \beta   \gamma   \gamma  \right\rangle}\\
{ \left| \gamma   \gamma   \delta \right\rangle}
\end{array}
\right)
\]
\end{widetext}
Using this information we implement \eref{eqApplySsquaredOperator}: The solutions for the spin eigenfunctions are given by determining the eigenvalues and eigenvectors of the matrix appearing here. We find in this example that $S=3/2$ occurs twice, and $S=1/2$ occurs 6 times. Due to this multiplicity we are forced to introduce an extra index, $k$, to distinguish between these linearly independent and orthogonal states with the same $S$ eigenvalue. Our spin eigenfunctions are now written as $X \left( N, \lambda, S, k, S_z, r \right)$. As in \aref{subSpinRotationallyInvariant}, the solutions for the spin eigenfunctions (given by the eigenvectors of the above matrix)  can be written as a projection operator applied to a given linear combination of primitive \generalizedspin{s}:
\[
X \left( N, \lambda, S, k, S_z, 1 \right) = \hat e^{\lambda}_{11}X_{\rm{lc}}.
\]
The appropriate linear combinations $X_{\rm{lc}}$ of primitive  \generalizedspin{s} are given in \tref{tableGrapheneBasis}. Note that \tref{tableGrapheneBasis} only lists cases where $S=S_z$: spin eigenfunctions for $S_z<S$ can be deduced by explicitly applying a spin lowering operator $\hat S^-$. We have derived the results presented in \tref{tableGrapheneBasis} by precisely the procedure outlined in this appendix. Also, following the same reasoning given in \aref{subSpinRotationallyInvariant}, the minimal information to construct the wave function is contained within the expressions $X_{\rm{lc}}$, and, thus, the full wave function can be reconstructued from 
\[
\psi (N, \lambda, S, k, S_z) \propto \mathcal {\hat {A}} \Phi \left( {N,\lambda,1} \right) X_{\rm{lc}}.
\]

%%%%%%%%%%%%%%%%%%%%%%%%%%%%%%%%%%
%%%%%%%%%%%%%%%%%%%%%%%%%%%%%%%%%%

\section{Spatial wave functions for Quantum Hall Systems}
\label{appendixSpatialWaveFunctions}

In this appendix we will apply the general techniques for the construction of spatial wave functions derived in \aref{appendixConstructionOfMulticomponentWaveFunctions} to the specific case of translationally invariant lowest Landau level (LLL) wave functions of fixed degree $L$. Specifically, our task is to determine the functions $\Phi \left( N, \lambda, 1\right)$ which will provide the minimal information required to reconstruct the complete wave function. The enumeration of these functions will lead us to a derivation of the dimension of the space of spatial wave functions. In \sref{subProcedureForSpatial} we shall describe how to construct appropriate LLL wave functions in general, and we shall explain how the vector space of LLL spatial wave functions can be enumerated algebraically using a set of projection operators. In \sref{subCalculationOfDimensions} we shall explain a more general procedure to calculate the dimensions of the vector space of LLL spatial wave functions. 

\subsection{Procedure for Construction of Spatial wave functions in the LLL}
\label{subProcedureForSpatial}

Spatial wave functions describing multiparticle LLL states are built from sums of products of single-particle LLL wave functions and, as we argued previously, at the beginning of \sref{secPseudoMultiPseudoReview}, it follows that the spatial wave functions must be (translationally invariant) analytic polynomials of a fixed degree in the complex relative coordinates $\tilde z_i$. For a state with fixed relative angular-momentum eigenvalue $L$ the most general form of spatial wave function is given by
\[\Phi_{{\rm{spatial}}}   = \sum\limits_{{i_1},\ldots,{i_N}} {{a_{{i_1}\ldots{i_N}}} \tilde z_1^{{i_1}}\ldots\tilde z_N^{{i_N}}} \,\,\,\quad {\rm{with}}\quad \sum\limits_{m = 1}^N {{i_m} = L}. \]
with arbitrary coefficients $a_{{i_1}\ldots{i_N}}$ [cf. \eref{eqPrimitiveSpatial}]. For example, for $N=3$ and $L=2$ the most general spatial wave function is
\[
\begin{array}{l}
\Phi_{{\rm{spatial}}} = a_{200} \tilde z_1^2  + a_{020} \tilde z_2^2 \\ + a_{002} \tilde z_3^2  + a_{110} \tilde z_1 \tilde z_2  + a_{011} \tilde z_2 \tilde z_3  + a_{101} \tilde z_1 \tilde z_3 .
\end{array}
\]
Using this definition, we could, in principle, now use \eref{eqRthSpatialWaveFunction} to produce an expression for $\Phi \left( {N,\lambda,1} \right)$ in the form given in \eref{eqExplicitRthSpatialWaveFunction}. In practice, we can employ a further short cut to this procedure that reproduces exactly the same result. Recalling the proof of theorem 1, we showed that
\[
\Phi \left( {N,\lambda,r} \right) = \hat e_{r1}^{\tilde \lambda} \Phi \left( {N,\lambda,1} \right).
\]
This result implies the following idempotency relation: 
\[
\Phi \left( {N,\lambda,1} \right) = \hat e_{11}^{\tilde \lambda} \Phi \left( {N,\lambda,1} \right).
\]
Thus, we see that if we start by writing $ \Phi \left( {N,S,1} \right)$ as the most general form of spatial wave function $\Phi_{{\rm{spatial}}} $ given above, then the action of applying $\hat e_{11}^{\tilde \lambda}$ will automatically project out the most general form of the first spatial wave function, given in \eref{eqExplicitRthSpatialWaveFunction}. We can interpret this process of applying the symmetric group algebra basis operator as a process of introducing many linear dependencies between the original set of linearly independent coefficients $a_{i_1 \ldots i_N}$. The result will be a polynomial of a particular symmetry imposed by the representation of the symmetric group $\lambda$, which may still contain some linearly independent coefficients $a'_{i_1 \ldots i_N}$, the number of which is the dimension of the space of spatial wave functions. 

We have now deduced that the vector space of LLL spatial wave functions can be projected out from the space of most general LLL wave functions. In principle, this method can be employed to determine the dimensions of the basis (listed in \tref{tableDimensionsOfPolynomialSpaces}) and the explicit forms of the basis functions. However, due to the algebraic complexity of constructing the most general form of the spatial wave functions, and the fact that the number of terms in $\hat e_{11}^{\lambda}$ grows as $N!$ with the number of particles $N$, this brute-force approach is not viable in general. In the next section we shall describe a method to calculate the dimensions of the space of LLL spatial wave functions in general. 

\subsection{Calculating the Dimension of the Vector Space of Spatial wave functions}
\label{subCalculationOfDimensions}

In this section we shall explain the method used to calculate the values listed in \tref{tableDimensionsOfPolynomialSpaces}. Using these results we shall then reinterpret the form of the spatial wave functions $\Phi \left( {N,\lambda,1} \right) $ and in doing so we shall complete our explanation of the results presented in \tref{tablePrimitivePolynomialsFermions}.

The spatial wave function $\Phi \left( {N,\lambda,1} \right) $ takes the form of a translationally invariant, analytic polynomial in the coordinates $z_i$, satisfying a particular set of permutation symmetries imposed by the underlying symmetric group representation. In general, it contains many linearly independent coefficients, the number of which depends on the degree of the polynomial. This is called the dimension of the polynomial space.  In order to determine how many independent parameters occur at a given degree we construct generating functions for the polynomial space dimension, and to do this we first need to construct a linearly independent basis in which to describe the polynomials.

Recall from \sref{secPseudoMultiPseudoReview} that any antisymmetric polynomial in $N$ variables can always be written as a symmetric polynomial by factoring out a Jastrow factor, \eref{eqJastrowFactor}. There are many ways to construct a basis of symmetric polynomials; the one we utilize is the basis of elementary symmetric polynomials, which we defined in \eref{eqElementarySymmetricPolyDefinition}. If we impose the condition that the polynomials must be translationally invariant then we must write elementary symmetric polynomials in terms of relative coordinates ${\tilde z}_i$, which we defined in \eref{eqTildeCoordinates}. Recall also that translational invariance results in \eref{eqTranslationallyInvariantSymmPoly}. These modified elementary symmetric polynomials form a basis of translationally invariant symmetric polynomials, in other words we can form any translationally invariant symmetric polynomial from a sum of products of the modified elementary symmetric polynomials. (More precisely, these polynomials form what is called a ring of translationally invariant symmetric polynomials.)\cite{liptrap2010}

The polynomials such as $\Phi \left( {N,\lambda,1} \right) $ are not fully symmetric but, instead, they have symmetries in subsets of particle indices as dictated by the shape $\lambda$ of their associated Young tableaux.  As a starting point we shall try to use the basis of elementary symmetric polynomials to span the space of polynomials with subsets of symmetries. We require, in addition that the polynomials are translationally invariant and this places a restriction on the basis,
\begin{equation}
e_{1,1\ldots N } =  0.
\label{eqTranlationalInvarianceConstraint}
\end{equation}

The dimension $d_{{\rm{sym}}} \left( {L,N} \right)$ of the space of translationally invariant symmetric polynomials in $N$ variables and of degree $L$ is defined in terms of the generating function
\[
Z_N \left( q \right) = \prod\limits_{m = 2}^N {\frac{1}{{1 - q^m }}}  = \sum\limits_{L = 0}^\infty  {q^L } d_{{\rm{sym}}} \left( {L,N} \right),
\]
so
\begin{equation}
d_{{\rm{sym}}} \left( {L,N} \right) = \left[ {\frac{1}{{L!}}\left( {\frac{d}{{dq}}} \right)^L Z_N \left( q \right)} \right]_{q = 0}.
\label{eqDimensionFullSymmetricPoly}
\end{equation}

We can write down a generating function for the space of polynomials with subsets of symmetries by multiplying the generating functions of the subsets (for proof, see \aref{subAppendixGeneratingFunctions}). 

Our task now is to construct a basis of partially symmetric polynomials that reflects the symmetric group representations. It happens that the two component case is the simplest to explain, and so we shall discuss that case first before generalizing our result to the multicomponent case. 

\subsubsection{Two-Component Case}

The basis of two-component wave functions is identical to the basis of spin eigenfunctions. We shall, thus, consider the spatial functions associated with an $N$-particle state of spin $S$.  

For bosons the particular case of interest is a translationally invariant polynomial that is symmetric in arguments 1 to  ${\textstyle{1 \over 2}}N + S$ and in the arguments ${\textstyle{1 \over 2}}N + S +1$ to $N$. Due to \eref{eqTranlationalInvarianceConstraint} we must eliminate elementary symmetric polynomials of degree 1 from the generating function of one subset. The generating function for translationally invariant, symmetric polynomials with two symmetry subsets is then
\begin{equation}
Z^{\rm{boson}}_{N,S} \left( q \right) =\prod\limits_{m = 1}^{{\textstyle{N \over 2}} + S} {\frac{1}{{1 - q^m }}} \prod\limits_{n = 2}^{{\textstyle{N \over 2}} - S} {\frac{1}{{1 - q^n }}} .
\label{eqBosonGeneratingFunctionTwoBody}
\end{equation}

For example, we might consider polynomials of order $L=3$ in $N=5$ indices which are symmetric in indices 1,2 and 3, and separately symmetric in indices 4 and 5. In terms of elementary symmetric polynomials we can construct:
\[\begin{array}{l}
 {e_{1,123}}{e_{2,123}} \,\,, \quad {e_{1,123}}{e_{2,45}} \,\,, \quad {e_{3,123}} \,\,, \quad e_{1,123}^3. \\
 \end{array}\]
This number, 4, is precisely given by appropriately differentiating the generating function given in \eref{eqBosonGeneratingFunctionTwoBody}, for the case of $N=5$ and $S=1/2$.

For fermions we require that the polynomial is separately antisymmetric in the two symmetry subsets instead of separately symmetric in the two subsets, and so the space of symmetric polynomials only comes into play at polynomial orders higher than the order of the Jastrow factor. To start the generating function at a particular order, $J$, we must multiply by $q^J$ (for a proof see \aref{subAppendixGeneratingFunctions}). In order to take into account the Jastrow factors associated with each of the symmetry subsets we must have
\begin{equation}
\begin{array}{l}
 J = {\textstyle{1 \over 2}}\left( {{\textstyle{N \over 2}} - S} \right)\left( {{\textstyle{N \over 2}} - S - 1} \right) + {\textstyle{1 \over 2}}\left( {{\textstyle{N \over 2}} + S} \right)\left( {{\textstyle{N \over 2}} + S - 1} \right) \\ \\
 \,\,\,\,\,\, = {\textstyle{{N^2 } \over 4}} - {N \over 2} + S^2.  \\
 \end{array}
 \label{eqDegreeOfJastrowFactor}
\end{equation}

The generating function for translationally invariant, antisymmetric polynomials with two symmetry subsets is then
\begin{equation}
Z^{\rm{fermion}}_{N,S} \left( q \right) = q^J \prod\limits_{m = 1}^{{\textstyle{N \over 2}} + S} {\frac{1}{{1 - q^m }}} \prod\limits_{n = 2}^{{\textstyle{N \over 2}} - S} {\frac{1}{{1 - q^n }}} .
\label{eqFermionGeneratingFunctionTwoBody}
\end{equation}

More precisely these generating functions actually account for polynomials with this type of symmetry \emph{or greater symmetry}.  For example, a polynomial which is fully symmetric is also, by definition, symmetric in any subset of its arguments. A polynomial that is symmetric in two subsets of its arguments is also symmetric in any further subdivision of those subsets. What we really require is a generating function for polynomials of one particular symmetry only, which we shall denote by $\tilde Z_{N,S}$, and we emphasize that Eqs.~\ref{eqBosonGeneratingFunctionTwoBody} and \ref{eqFermionGeneratingFunctionTwoBody} fail to do this. In \aref{subAppendixPermutationModules} we make the statement ``greater symmetry'' more precise; in particular, we show that for tableaux containing the same number of boxes, a tableau $\lambda$ is of ``greater symmetry'' than a tableau $\mu$ if and only if $S_{\lambda} \ge S_{\mu}$, where $S$ is the corresponding spin eigenvalue of the tableau.

What we call $\tilde Z_{N,S}$ is the generating function for polynomials that are eigenfunctions of $S^2$  (note that this is independent of $S_z$ eigenvalue). Below, $Z_{N,S} \left( q \right)$ represents either $Z^{\rm{fermion}}_{N,S} \left( q \right)$ defined in \eref{eqFermionGeneratingFunctionTwoBody} or $Z^{\rm{boson}}_{N,S} \left( q \right)$ defined in \eref{eqBosonGeneratingFunctionTwoBody}. Our derivation of the generating function $\tilde Z_{N,S}$ is as follows. First, it is easy to see that
\[
 Z_{N,{\textstyle{N \over 2}}} \left( q \right) = \tilde Z_{N,{\textstyle{N \over 2}}} \left( q \right),
 \]
which is simply the statement that for the state where $S=N/2$, the wave function is fully symmetric (bosons) or fully antisymmetric (fermions). For the next highest state we have
 \[
 Z_{N,{\textstyle{N \over 2}} - 1} \left( q \right) = \tilde Z_{N,{\textstyle{N \over 2}}} \left( q \right) + \tilde Z_{N,{\textstyle{N \over 2}} - 1} \left( q \right),
\]
that is, the polynomial space whose dimension is generated by either \eref{eqBosonGeneratingFunctionTwoBody} or  \eref{eqFermionGeneratingFunctionTwoBody} with $S=N/2-1$, is composed of both fully symmetric polynomials and polynomials that are symmetric in $N-1$ indices. We can rearrange this equation to give an expression for the generating function that we want to get at:
\[
\tilde Z_{N,{\textstyle{N \over 2}} - 1} \left( q \right) =  Z_{N,{\textstyle{N \over 2}} - 1} \left( q \right) - Z_{N,{\textstyle{N \over 2}}} \left( q \right).
\]
In general, we have
\[
Z_{N,S} \left( q \right) = \sum\limits_{i = 0}^{{\textstyle{N \over 2}} - S} {\tilde Z_{N,{\textstyle{N \over 2}} - i} \left( q \right)} ,
\]
and so
\begin{equation}
\label{eqTwoBodyFullGeneratingFunction}
\tilde Z_{N,S} \left( q \right) = Z_{N,S} \left( q \right) - Z_{N,S + 1} \left( q \right) .
\end{equation}
These generating functions are used to calculate the results displayed in \tref{tableDimensionsOfPolynomialSpaces}. A more rigorous proof of this result is given in \aref{subAppendixPermutationModules}.

Although the generating functions provide us with a means to calculate the dimension of the space of polynomials at a given degree, they shed no light on the explicit form of the polynomials. An insight into the forms of the polynomials can be obtained by first rewriting the generating function using the following result (see \aref{subAppendixGeneratingFunctions}):
\begin{equation}
\tilde Z_{N,\,S} \left( q \right) = q^J \prod\limits_{n = 2}^N {\frac{1}{{1 - q^n }}} \sum\limits_{k = 0}^{k_{\max} } {b'_k q^k },
\label{eqFactoredGeneratingFunction}
\end{equation}
where ${b'}_k$ are positive integer coefficients and
\[
k_{\max}  = \left( {{\textstyle{N \over 2}} - S} \right)\left( {{\textstyle{N \over 2}} + S} \right).
\]

We interpret the expression for the generating function in \eref{eqFactoredGeneratingFunction} as describing a polynomial basis comprising \emph{primitive polynomials} at degrees $J$ up to $J+k_{\max}$, each of which has ${b'}_k$ linearly independent coefficients. In order to indicate the polynomial degrees at which the primitive polynomials occur, we include the following functions, $Y_{N,S}$, in \tref{tableYfunctions}:
\begin{equation}
Y_{N,S}=\frac{{\tilde Z_{N,S} \left( q \right)}}{{\prod\limits_{l = 2}^N {{\textstyle{1 \over {1 - q^l }}}} }} = q^J \sum\limits_{k = 0}^{k_{\max} } {b'_k q^k } .
\label{eqYfunction}
\end{equation}
Some examples of the  ${b'}_k$ thus can be read off from \tref{tableYfunctions}.

The conclusion of this section is the following key statement: the most general polynomial of a given symmetry type is a linear combination of the primitive polynomials multiplied by any valid translationally invariant fully symmetric polynomial.

%%%%%%%%%%%%%%%%%%%%%%%%%%%%%%%%%%

\begin{table}[t]
%\begin{center}
\begin{tabular}{c|c}
\hline
\hline
\multicolumn{1}{c|}{$\lambda$}   & {$Y_{N,S}$}\\
\hline
\multicolumn{1}{c|}{$[2]$}   & {$1$}\\
\multicolumn{1}{c|}{$[1,1]$}   & {$q$}\\
\hline
\multicolumn{1}{c|}{$[3]$}   & {$1$}\\
\multicolumn{1}{c|}{$[2,1]$}   & {$q+q^2$}\\
\multicolumn{1}{c|}{$[1,1,1]$}   & {$q^3$}\\
\hline
\multicolumn{1}{c|}{$[4]$}   & {$1$}\\
\multicolumn{1}{c|}{$[3,1]$}   & {$q+q^2+q^3$}\\
\multicolumn{1}{c|}{$[2,2]$}   & {$q^2+q^4$}\\
\multicolumn{1}{c|}{$[2,1,1]$}   & {$q^3+q^4+q^5$}\\
\multicolumn{1}{c|}{$[1,1,1,1]$}   & {$q^5$}\\
\hline
\multicolumn{1}{c|}{$[5]$}   & {$1$}\\
\multicolumn{1}{c|}{$[4,1]$}   & {$q+q^2+q^3+q^4$}\\
\multicolumn{1}{c|}{$[3,2]$}   & {$q^2+q^3+q^4+q^5+q^6$}\\
\multicolumn{1}{c|}{$[3,1,1]$}   & {$q^3+q^4+2q^5+q^6+q^7$}\\
\multicolumn{1}{c|}{$[2,2,1]$}   & {$q^4+q^5+q^6+q^7+q^8$}\\
\multicolumn{1}{c|}{$[2,1,1,1]$}   & {$q^6+q^7+q^8+q^9$}\\
\multicolumn{1}{c|}{$[1,1,1,1,1]$}   & {$q^{10}$}\\
\hline
\hline
\end{tabular}
\caption{These functions, $Y_{N,S}$, are calculated as in \eref{eqYfunction}, by taking the ratios of the generating functions for polynomials of a particular symmetry type $\lambda$ in \eref{eqKostkaDecomposition} [or \eref{eqTwoBodyFullGeneratingFunction} for the two-component case], to the generating function of a fully symmetric polynomial for the same number of particles. As explained in the text, the $q$ polynomial degree of each term indicates the presence of a primitive polynomial at that degree in the spatial wave function and the value of the coefficient in the $q$ polynomial indicates the number of linearly independent parameters constituting the primitive polynomial at that degree (cf. Tables~\ref{tablePrimitivePolynomialsFermions} and \ref{tablePrimitivePolynomialsBosons}).}
\label{tableYfunctions}
%\end{center}
\end{table}

%%%%%%%%%%%%%%%%%%%%%%%%%%%%%%%%%%

\subsubsection{Generlization to the Multicomponent Case}

In the multicomponent case, we have already shown that the full spatial wave function can be reconstructed from just the first part, $\Phi (N,\lambda,1)$.  In this section we aim to describe the form of this first part of the spatial wave function, a translationally invariant analytic polynomial of a particular symmetry, in terms of the elementary symmetric polynomials. We would like to derive an expression for the dimensions of the spaces of such polynomials, as we did for the two-component case, by constructing a set of generating functions.

We start by writing down a generating function for the space of polynomials with an arbitrary number of symmetry subsets, and we do this by simply multiplying together the generating functions corresponding to each of the different subsets (previously we only needed to consider two subsets). To take into account  the translational invariance of the polynomial, we use the relative coordinates also defined in \sref{secPseudoMultiPseudoReview}. This introduces the constraint given in \eref{eqTranlationalInvarianceConstraint}. To enforce this constraint, we remove one of the sets of $e_1$ terms by starting one of the generating functions from a lower bound of 2 rather than 1 (similar to the argument given in for the two-component case),
\begin{equation}
Z_{\lambda}^{{\rm{boson}}} = \prod\limits_{{m_1} = 1}^{{N_1}} {\frac{1}{{1 - {q^{{m_1}}}}}\prod\limits_{{m_2} = 1}^{{N_2}} {\frac{1}{{1 - {q^{{m_2}}}}}} } \ldots \prod\limits_{{m_{n}} = 2}^{{N_{n}}} {\frac{1}{{1 - {q^{{m_n}}}}}}
\label{eqBosonGeneratingFunctionManyBody}
\end{equation}
or for conjugate tableaux we would have to take out an appropriate Jastrow factor of degree
\[J = \sum\limits_{i = 1}^{n} {\frac{1}{2}{N_i}\left( {{N_i} - 1} \right)}.\]
The generating function in this case is
\begin{equation}
Z_{\lambda}^{{\rm{fermion}}} = {q^J}\prod\limits_{{m_1} = 1}^{{N_1}} {\frac{1}{{1 - {q^{{m_1}}}}}\prod\limits_{{m_2} = 1}^{{N_2}} {\frac{1}{{1 - {q^{{m_2}}}}}} } \ldots \prod\limits_{{m_{n}} = 2}^{{N_{n}}} {\frac{1}{{1 - {q^{{m_n}}}}}}.
\label{eqFermionGeneratingFunctionManyBody}
\end{equation}

In fact, the generating functions written down here describe a space of polynomials that has a certain symmetry type or a greater symmetry. For example, the generating function $Z_{\lambda=[2,1]}^{{\rm{boson}}} $ gives the dimension of the vector space of polynomials of symmetry type $\lambda=[2,1]$ added to the dimension of the vector space of polynomials of symmetry type $\lambda=[3]$. More precisely, we say that the vector space generated by $Z_{\lambda}^{{\rm{boson}}} $ contains two subvector spaces, each of which corresponds to a particular irreducible representation of the symmetric group. It follows that, for our purposes, determining the vector spaces of polynomials associated with irreducible representations of the symmetric group,  we are looking for the dimensions of these subvector spaces. If we wish to perform this decomposition of vector spaces for any arbitrary shape of tableau then an important question to address is which subvector spaces are included, in general, and how many times is each subvector space included (as there is no reason, in principle, why a subvector space cannot be included multiple times in the decomposition). The answer to this question can be found in the underlying mathematical structure of the construction of polynomial spaces corresponding to representations of the symmetric group (objects known as modules in the mathematical nomenclature); we shall simply state the result here, leaving the details of the underlying mathematics to \aref{subAppendixPermutationModules}. In order to state the result we must, first, introduce a notation that makes precise the statement ``greater symmetry'' \cite{saganbook}: when comparing two Young tableaux shapes $\mu=[\mu_1,\mu_2,\ldots\mu_n]$ and $\lambda=[\lambda_1,\lambda_2,\ldots\lambda_l]$, we say that $\mu$ dominates  (has ``greater symmetry'' than) $\lambda$, written $ \mu \, \underbar{$\triangleright$} \, \lambda $  if
\[
{\mu _1} + {\mu _2} + \ldots+ {\mu _i} \ge {\lambda _1} + {\lambda _2} + \ldots + {\lambda _i}\,\,\,\,\,\,\forall \,\,i \ge 1, 
\]
with ${\mu _{i > l}} = {\lambda _{i > n}} = 0. $
For example, $[3,3] \, \underbar{$\triangleright$} \,  [2,2,1,1]$. Note that there are cases when tableaux are incomparable, such as $[3,3]$ and $[4,1,1]$. One way to avoid such incomparable tableau shapes by restricting possible shapes to tableaux having no more than two rows, which is precisely the two-component case. For proof of this result, see \aref{subAppendixPermutationModules}. Below, $Z_{\lambda} \left( q \right)$ represents either $Z^{\rm{fermion}}_{\lambda} \left( q \right)$ defined in \eref{eqFermionGeneratingFunctionManyBody} or $Z^{\rm{boson}}_{\lambda} \left( q \right)$ defined in \eref{eqBosonGeneratingFunctionManyBody}. Using this notation, the general relation between the $Z_{\lambda}$ generating functions and the generating functions corresponding to irreducible representations of the symmetric group, $\tilde Z_{\lambda}$, is
\begin{equation}
{Z_{\lambda}} = \sum\limits_{\mu \, \underbar{$\triangleright$} \, \lambda } {K_{\mu \lambda }}{{\tilde Z}_\mu } .
\label{eqGeneratingFunctionKostkaForm}
\end{equation}
In this equation the coefficients $K_{\mu \lambda }$ are called Kostka numbers. \cite{kostka1882} It is clear that \eref{eqGeneratingFunctionKostkaForm} is a matrix equation. Since the matrix of Kostka numbers is nonsingular, the relation can be inverted, which gives us our key result:
\begin{equation}
{{\tilde Z}_\lambda } = \sum\limits_{\mu \, \underbar{$\triangleright$} \, \lambda } \left( K \right)_{\mu \lambda }^{ - 1}{Z_{\mu}}.
\label{eqKostkaDecomposition}
\end{equation}
A formal derivation of this result is given in \aref{subAppendixPermutationModules}. In \tref{tableKostkaNumbers} we have listed selected inverse matrices of Kostka numbers. Using these numbers, we can construct, for example,
\[
{{\tilde Z}_{[2,1,1]}^{{\rm{boson}}}} = Z_{[2,1,1]}^{{\rm{boson}}} - Z_{[3,1]}^{{\rm{boson}}} - Z_{[2,2]}^{{\rm{boson}}} + Z_{[2,1,1]}^{{\rm{boson}}}\]
Notice that if we restrict the allowed tableau shapes to have no more than two, rows then we recover the special case of the result given in \eref{eqTwoBodyFullGeneratingFunction}. An identical argument applies for the description of conjugate tableaux $Z_{\lambda}^{{\rm{fermion}} }$.

We have used these results to calculate the lists of the dimensions of vector spaces associated with irreducible representations of the symmetric group listed in \tref{tableDimensionsOfPolynomialSpaces} for the symmetry types not already included in the SU(2) case.

%%%%%%%%%%%%%%%%%%%%%%%%%%%%%%%%%%

\begin{table}[t]
\centering
\subfloat[$(K)^{-1}_{\mu \lambda}$ for $N=3$]{
\begin{tabular}{*{5}{c}}
\hline \hline
 & \multicolumn{1}{c}{$\lambda=$} & [3] & [2,1] & [1,1,1] \\[2pt]
\hline
\multicolumn{1}{c}{$\mu=$} & \multicolumn{1}{c|}{[3]} & 1 & $-$1 & 1 \\
\multicolumn{1}{c}{} & \multicolumn{1}{c|}{[2,1]} & 0 & 1 & $-$2  \\
\multicolumn{1}{c}{} & \multicolumn{1}{c|}{[1,1,1]} & 0 & 0 & 1 \\[2pt]
\hline \hline
\end{tabular}
}
\\
\subfloat[$(K)^{-1}_{\mu \lambda}$ for $N=4$]{
\begin{tabular}{*{7}{c}}
\hline \hline
 & \multicolumn{1}{c}{$\lambda=$} & [4] & [3,1] & [2,2] & [2,1,1] & [1,1,1,1] \\[2pt]
\hline
\multicolumn{1}{c}{$\mu=$} & \multicolumn{1}{c|}{[4]} & 1 & $-$1 & 0  & 1 & $-$1 \\
\multicolumn{1}{c}{} & \multicolumn{1}{c|}{[3,1]} & 0 & 1 & $-$1 & $-$1 & 2  \\
\multicolumn{1}{c}{} & \multicolumn{1}{c|}{[2,2]} & 0 & 0 & 1 & $-$1 & 1\\
\multicolumn{1}{c}{} & \multicolumn{1}{c|}{[2,1,1]} & 0 & 0 & 0 & 1 & $-$3 \\
\multicolumn{1}{c}{} & \multicolumn{1}{c|}{[1,1,1,1]} & 0 & 0 & 0 & 0 & 1 \\[2pt]
\hline \hline
\end{tabular}
}
\\
\subfloat[$(K)^{-1}_{\mu \lambda}$ for $N=5$]{
\begin{tabular}{*{9}{c}}
\hline \hline
\multicolumn{1}
{c}{} & \multicolumn{1}{c}{$\lambda=$} & [5] & [4,1] & [3,2] & [3,1,1] & [2,2,1] & [2,1,1,1] & [1,1,1,1,1] \\[2pt]
\hline
\multicolumn{1}{c}{$\mu=$} & \multicolumn{1}{c|}{[5]} & 1 & $-$1 & 0 & 1 & 0 & $-$1 & 1 \\
\multicolumn{1}{c}{} & \multicolumn{1}{c|}{[4,1]} & 0 & 1 & $-$1 & $-$1 & 1 & 1 & $-$2 \\
\multicolumn{1}{c}{} & \multicolumn{1}{c|}{[3,2]} & 0 & 0 & 1 & $-$1 & $-$1 & 2 & $-$2 \\
\multicolumn{1}{c}{} & \multicolumn{1}{c|}{[3,1,1]} & 0 & 0 & 0 & 1 & $-$1 & $-$1 & 3 \\
\multicolumn{1}{c}{} & \multicolumn{1}{c|}{[2,2,1]} & 0 & 0 & 0 & 0 & 1 & $-$2 & 3 \\
\multicolumn{1}{c}{} & \multicolumn{1}{c|}{[2,1,1,1]} & 0 & 0 & 0 & 0 & 0 & 1 & $-$4 \\
\multicolumn{1}{c}{} & \multicolumn{1}{c|}{[1,1,1,1,1]} & 0 & 0 & 0 & 0 & 0 & 0 & 1 \\[2pt]
\hline \hline
\end{tabular}
}
\caption{A list of inverse Kostka matrices for different Young tablaux sizes. Lists of the Kostka matrices $K_{\mu \lambda}$ up to $N=8$ are given in Ref.~\onlinecite{kostka1882}.}
\label{tableKostkaNumbers}
\end{table}

%%%%%%%%%%%%%%%%%%%%%%%%%%%%%%%%%%

\subsubsection{Calculation of Primitive Polynomials}

We shall conclude our argument by describing the procedure by which we have calculated the primitive polynomials presented in Tables~\ref{tablePrimitivePolynomialsFermions} and \ref{tablePrimitivePolynomialsBosons}. We start with the most general fixed degree translationally invariant polynomial as the spatial wave function:
\[
\Phi_{{\rm{spatial}}}   = \sum\limits_{{i_1},\ldots,{i_n}} {{a_{{i_1}\ldots{i_N}}} \tilde z_1^{{i_1}}\ldots\tilde z_N^{{i_N}}} \quad\,\,{\rm{with}}\quad \sum\limits_{m = 1}^N {{i_m} = L}. 
\]
We then apply the appropriately constructed symmetric group algebra projection operator: $\Phi \left( {N,\lambda,1} \right) = \hat e_{11}^{\lambda} \Phi_{{\rm{spatial}}}$ (with $\lambda = [N_1,N_2,\ldots,N_n]$). This will leave a sum of terms that are of the correct permutation symmetry, and, hence, they can be written in terms of a basis of elementary symmetric polynomials (multiplied by a Jastrow factor, for antisymmetry types):
\begin{align*}
&\Phi \left( {N,\lambda,1} \right) =\\& \sum\limits_{{i_1},\ldots,{i_N}} {b_{{i_1}\ldots{i_N}}}
e_{1,{N_1}}^{{i_1}} \left( {{{\tilde z}_1}\ldots{{\tilde z}_{N_1}}} \right)\ldots e_{{N_1},{N_1}}^{{i_{N_1}}}\left( {{{\tilde z}_1}\ldots{{\tilde z}_{N_1}}} \right)
\\& \ldots e_{1,{N_n}}^{{i_{N_{n-1}+1}}} \left( {{{\tilde z}_{N_{n-1}+1}}\ldots{{\tilde z}_{N_n}}} \right)
\ldots e_{{N_n},{N_n}}^{{i_{N_n}}}\left( {{{\tilde z}_{N_{n-1}+1}}\ldots{{\tilde z}_{N_n}}} \right)
 \\& {\rm{with}} \qquad \sum\limits_{m = 1}^{N_1} m \,{i_m} + \ldots +\sum\limits_{m = {N_n+1}}^{N_n} {(m-N_n)} \,{i_m}= L.  
\end{align*}
%\begin{align*}
%&\Phi \left( {N,S,1} \right) \\& = \sum\limits_{{i_1},\ldots,{i_n}} {{b_{{i_1}\ldots{i_N}}}e_{2,N}^{{i_1}}\left( {{{\tilde z}_1}\ldots{{\tilde z}_N}} \right)\ldots e_{N,N}^{{i_N}}\left( {{{\tilde z}_1}\ldots{{\tilde z}_N}} \right)}  \\&
% {\rm{with}} \quad \sum\limits_{m = 1}^N {m{i_m} = L}.  
%\end{align*}
This is equivalent to a basis involving primitive polynomials and fully symmetric polynomials only. We start with linearly independent coefficients $a_{{i_1}\ldots{i_n}}$, but once we apply the projection operator we have a smaller set of coefficients, $b_{{i_1}\ldots{i_n}}$. These coefficients are not linearly independent in general. We can express the $b$ coefficients in terms of the $a$ coefficients, and, in principle, it is possible to solve these systems of equations to determine all possible linear dependences between the $b$ coefficients. Indeed, using this procedure, we have independently verified the number of linearly independent coefficients predicted by the generating function method up to polynomial degree ten.

If we perform this construction at a polynomial degree exactly corresponding to a primitive polynomial, then we can be sure that one of the basis polynomials (i.e., the terms multiplying one of the $b_{{i_1}\ldots{i_n}}$ coefficients) can be considered as the primitive polynomial. We can therefore extract the primitive polynomial by setting an appropriate number of the $b_{{i_1}\ldots{i_n}}$ coefficients to zero, such that the remaining expression contains the correct number of linearly independent coefficients required for the primitive polynomial. For $N \le 5$, almost all of the primitive polynomials  contain only one arbitrary coefficient and so this procedure can be reduced to setting all but one of the $b_{{i_1}\ldots{i_n}}$ coefficients to zero, such that different primitive polynomials are not related to each other by multiplication by any translationally invariant fully symmetric polynomial in $N$ variables. The polynomials given in \tref{tablePrimitivePolynomialsFermions} are all derived using  this technique. We have elucidated the details of the technique here in order to point out that the primitive polynomials listed in \tref{tablePrimitivePolynomialsFermions} are defined in a somewhat arbitrary way. Despite this, once we reconstruct the spatial wave functions from primitive polynomials multiplied by translationally invariant symmetric polynomials with a fixed combined degree, then these functions will fully span the basis.

We began this construction procedure with the most general possible $\Phi_{{\rm{spatial}}}$. In practice, we are free to choose a less general starting spatial wave function if we know what properties to expect in the final spatial wave function after projection (in other words, we can effectively pre-empt the linear dependencies that will be introduced between the coefficients and this saves  a great deal of computational time). The first part of the spatial wave function will be associated with the first standard tableau of the shape $\lambda = [N_1,N_2,\ldots,N_n]$. The corresponding polynomial will be translationally invariant and symmetric or antisymmetric in indices 1 to $N_1$, in indices $N_1+1$ to $N_2$, and so on.  Using this information, we can choose the starting spatial wave function to be the most general translationally invariant analytic polynomial with these symmetries at a given degree, $ \Phi _{\rm reduced} $. These symmetries are, in fact, uniquely associated with the $r=1$ standard tableau and, as a result, it follows that
\begin{equation}
\Phi \left( {N,\lambda,1} \right) ={e_{11}^{\tilde \lambda }{\Phi _{{\rm{reduced}}{\rm{}}}}}.
\end{equation}

We shall now discuss a simple example illustrating the whole construction procedure in the case of a two-component system. For simplicity, there are no undetermined parameters in this example. Let us consider the three-particle $L=1$, $S=1/2$ , $S_z=1/2$ fermion wave function. We begin by constructing the spatial wave function associated with the first spin eigenfunction. To do this we, first, construct the $r=1$ standard tableau associated with $N=3$ and $S=1/2$. We know that the spin eigenfunction $X(N=3,S=1/2,S_z=1/2,r=1)$ is constructed from
\[
\young(12,3)
\]
The fermion spatial wave function associated with the spin eigenfunction $X(3,1/2,1/2,1)$ is then constructed from the contragradient standard tableau, which is
\[
\young(13,2)
\]
For bosons we instead construct the spatial wave function from the uppermost tableau shown here.

We have already calculated $X(3,1/2,1/2,1)$ in \aref{appendixSpinEigenfunctions}, and it is clearly symmetric under exchange of the first two spins but has no other symmetry. If we require that the overall wave function is antisymmetric (for fermions) then we know immediately that the spatial wave function  $\Phi(N=3,S=1/2,r=1)$ will at the very least be antisymmetric under exchange of the first two coordinates. Using this information we can choose the primitive spatial wave function to be
\[
\Phi _{{\rm{reduced}}} = \left( {\tilde z_1  - \tilde z_2 } \right).
\]
We then apply a projection operator to generate $\Phi(3,1/2,1)$ (see \aref{appendixFurtherMaths} for the explicit form of this operator)
\[
\Phi(3,1/2,1) = \hat e_{1,1}^{\left[ {2,1} \right]} \Phi  = \hat e_{1,1}^{\left[ {2,1} \right]} \left( {\tilde z_1  - \tilde z_2 } \right) = \left( {\tilde z_1  - \tilde z_2 } \right)
\]
As we have explained, this term is associated with the first primitive spin wave function and we know that all of the physics is contained in the construction
\[
\left| { \uparrow  \uparrow  \downarrow } \right\rangle \left( {\tilde z_1  - \tilde z_2 } \right).
\]
Given this information we can effectively determine the remainder of the wave function by antisymmetrizing in all the particle indices [see \eref{eqUpUpDownExpandedForm}]. 

Although simply antisymmetrizing does indeed give the correct result, it may be necessary to write the result explicitly in terms of spin eigenfunctions. In this example we have $f^{\lambda} = 2$. We associate $\Phi(3,1/2,1)$ with the first spin eigenfunction:
\[
X \left( {3,1/2,1/2,1} \right) = {\textstyle{1 \over {\sqrt 6 }}}\left( {2\left| { \uparrow  \uparrow  \downarrow } \right\rangle  - \left| { \uparrow  \downarrow  \uparrow } \right\rangle  - \left| { \downarrow  \uparrow  \uparrow } \right\rangle } \right).
\]
The second spin eigenfunction is
\[
X \left( {3,1/2,1/2,2} \right) = {\textstyle{1 \over {\sqrt 2 }}}\left( {\left| { \uparrow  \downarrow  \uparrow } \right\rangle  - \left| { \downarrow  \uparrow  \uparrow } \right\rangle } \right).
\]
The associated spatial wave function is given by applying the operator (written in terms of permutation operators)
\[
\hat e_{21}^{\left[ {2,1} \right]}  = \frac{{\sqrt 3 }}{2}\left\{ {\left( {312} \right) - \left( {132} \right) - \left( {231} \right) + \left( {321} \right)} \right\}
\]
to the first part of the spatial wave function. This results in the polynomial
\[
\Phi \left( {3,1/2,2} \right) = {\textstyle{1 \over {\sqrt 3 }}} \left( {2 \tilde z_3  - \tilde z_1  - \tilde z_2 } \right).
\]
The full wave function must be a the sum of these two terms: for $L=1$ this is
\begin{align*}
 \left| {1,1/2,1/2 , 1} \right\rangle &\propto \left( {2\left| { \uparrow  \uparrow  \downarrow } \right\rangle  - \left| { \uparrow  \downarrow  \uparrow } \right\rangle  - \left| { \downarrow  \uparrow  \uparrow } \right\rangle } \right)\left( {\tilde z_1  - \tilde z_2 } \right) \\
    &- \left( {\left| { \uparrow  \downarrow  \uparrow } \right\rangle  + \left| { \downarrow  \uparrow  \uparrow } \right\rangle } \right)\left( {2 \tilde z_3 - \tilde z_1  - \tilde z_2  } \right). \\
 \end{align*}
Comparing with \eref{eqUpUpDownExpandedForm} we see that once the terms corresponding to each of the primitive spin wave functions are collected up then the results do indeed agree.

To obtain the equivalent boson wave function in this case we can simply interchange the associations between the spin and spatial parts. We can, of course, apply the same technique described in this example to construct the wave function at any degree, starting from the first part of the spatial wave function at that degree.

%%%%%%%%%%%%%%%%%%%%%%%%%%%%%%%%%%
%%%%%%%%%%%%%%%%%%%%%%%%%%%%%%%%%%

\section{Further Mathematical Details}
\label{appendixFurtherMaths}

\subsection{Young Operators}

The concept of Young operators may be familiar to many readers. The Young operators are idempotent projection operators that generate minimal left ideals in the regular representation of the symmetric group; in other words, the Young operators form a resolution of the identity that generates the irreducible representations of the symmetric group algebra. \cite{hamermeshbook} By definition this means that they form a basis spanning the group algebra.

The Young operators are constructed as follows: with each standard tableau, $T_r$, we associate an operator
\[
Y_r = Q_r P_r,
\]
where $P_r$ is product over all rows in the tableau of the sum of all permutations that permute the numbers in the same row and $Q_r$ is product over all columns in the tableau of the sum of all permutations that permute the numbers in the same column multiplied by the sign of those permutations. Essentially, $P_r$ symmetrizes in the row indices and $Q_r$ antisymmetrizes in the column indices. The Young operator is the product of these two operators.

The set of Young operators can be used to generate all of the results derived in \aref{appendixConstructionOfMulticomponentWaveFunctions}; however, the resulting projected functions do not form a fully orthogonal basis. \cite{paunczbook} It is more convenient, therefore, to use a set of operators that automatically project out an orthogonal basis of \generalizedspin{s}.

The advantage of the matric basis, defined in \aref{appendixMathPreliminaries}, is that it enables us to use the representation matrices of any irreducible representation to construct a corresponding basis of the group algebra, and, for convenience, we choose the orthogonal representation to ensure orthogonal basis functions. 

\vspace{1em}

\subsection{Matric Basis Operators}

In \tref{tableMatricUnitOperators} below we list the operators $\hat e^{\lambda}_{r1}$ that are applied to the first part of the spatial wave function in order to generate the spatial part of the wave function associated with the $r^{{\rm{th}}}$ spin eigenfunction. The operators are written in terms of permutation operators: we use the notation explained in Ref.~\onlinecite{hamermeshbook}, for example. In this notation the term $ \left( {p_1 ,p_2 ,\ldots,p_N } \right) $ denotes the permutation operation that relabels index 1 with index $p_1$, index 2 with $p_2$, and so on.

The operators here are calculated using a {MATHEMATICA}-based computer program. \cite{multicomponentWaveFunctionProgram} This program can be used to calculate expressions for operators up to $N=5$ and for any valid $\lambda$.

%%%%%%%%%%%%%%%%%%%%%%%%%%%%%%%%%%

\onecolumngrid

\begin{center}
\begin{table}[H]
\centering
\subfloat[Matric unit operators associated with the orthogonal representation of the symmetric group.]{
\begin{tabular}{l}
\hline \hline
\multicolumn{1}{l}{$\begin{array}{l} \\[-0.5em] \hat e_{11}^{\left[ {2,1} \right]}  = \frac{{1}}{2}\left\{ { 2 \left( {123} \right) + 2 \left( {213} \right) - \left( {312} \right) - \left( {132} \right) - \left( {231} \right) - \left( {321} \right)} \right\}\\ \\\hat e_{21}^{\left[ {2,1} \right]}  = \frac{{\sqrt 3 }}{2}\left\{ {\left( {312} \right) + \left( {132} \right) - \left( {231} \right) - \left( {321} \right)} \right\}\\  \end{array}$}\\
\multicolumn{1}{l}{$\begin{array}{l} \\ \hat e_{11}^{\left[ {3,1} \right]}  =
\frac{{\rm{1}}}{{\rm{3}}}\left\{ \begin{array}{l}
 3\left( {{\rm{1234}}} \right) + 3\left( {{\rm{2134}}} \right) + 3\left( {{\rm{3124}}} \right) + 3\left( {{\rm{1324}}} \right) + 3\left( {{\rm{2314}}} \right) + 3\left( {{\rm{3214}}} \right) - \left( {{\rm{4213}}} \right) -  \\
 \left( {{\rm{2413}}} \right) - \left( {{\rm{1423}}} \right) - \left( {{\rm{4123}}} \right) - \left( {{\rm{2143}}} \right) - \left( {{\rm{1243}}} \right) - \left( {{\rm{1342}}} \right) - \left( {{\rm{3142}}} \right) - \left( {{\rm{4132}}} \right) \\
  - \left( {{\rm{1432}}} \right) - \left( {{\rm{3412}}} \right) - \left( {{\rm{4312}}} \right) - \left( {{\rm{4321}}} \right) - \left( {{\rm{3421}}} \right) - \left( {{\rm{2431}}} \right) - \left( {{\rm{4231}}} \right) - \left( {{\rm{3241}}} \right) - \left( {{\rm{2341}}} \right) \\
 \end{array} \right\}
\\ \\
\hat e_{21}^{\left[ {3,1} \right]}  = \frac{{\sqrt 2 }}{3}\left\{
 {\begin{array}{l}
 2\left( {4213} \right) + 2\left( {2413} \right) + 2\left( {1423} \right) + 2\left( {4123} \right) + 2\left( {2143} \right) + 2\left( {1243} \right)  - \left( {1342} \right) - \left( {3142} \right) - \left( {4132} \right) \\
  - \left( {1432} \right) - \left( {3412} \right) - \left( {4312} \right) - \left( {4321} \right) - \left( {3421} \right) - \left( {2431} \right) - \left( {4231} \right) - \left( {3241} \right) - \left( {2341} \right) \\
 \end{array}} \right\} \\ \\
\hat e_{31}^{\left[ {3,1} \right]}  = \sqrt {\frac{2}{3}} \left\{  \begin{array}{l}  \left( {1342} \right) + \left( {3142} \right) + \left( {4132} \right) + \left( {1432} \right)  + \left( {3412} \right) + \left( {4312} \right) \\ - \left( {4321} \right) - \left( {3421} \right) - \left( {2431} \right) - \left( {4231} \right) - \left( {3241} \right) - \left( {2341} \right) \\ \end{array} \right\} \\  \end{array}$}\\
\multicolumn{1}{l}{$\begin{array}{l} \\ \hat e_{11}^{\left[ {2,2} \right]}  =
\frac{{\rm{1}}}{{\rm{2}}}\left\{ \begin{array}{l}
 2\left( {{\rm{1234}}} \right) + 2\left( {{\rm{2134}}} \right) - \left( {{\rm{3124}}} \right) - \left( {{\rm{1324}}} \right) - \left( {{\rm{2314}}} \right) - \left( {{\rm{3214}}} \right) - \left( {{\rm{4213}}} \right) - \left( {{\rm{2413}}} \right) \\
  - \left( {{\rm{1423}}} \right) - \left( {{\rm{4123}}} \right) + 2\left( {{\rm{2143}}} \right) + 2\left( {{\rm{1243}}} \right) - \left( {{\rm{1342}}} \right) - \left( {{\rm{3142}}} \right) - \left( {{\rm{4132}}} \right) - \left( {{\rm{1432}}} \right) \\
  + 2\left( {{\rm{3412}}} \right) + 2\left( {{\rm{4312}}} \right) + 2\left( {{\rm{4321}}} \right) + 2\left( {{\rm{3421}}} \right) - \left( {{\rm{2431}}} \right) - \left( {{\rm{4231}}} \right) - \left( {{\rm{3241}}} \right) - \left( {{\rm{2341}}} \right) \\
 \end{array} \right\}
\\ \\ \hat e_{21}^{\left[ {2,2} \right]}  = \frac{{\sqrt 3 }}{2}  \left\{  \begin{array}{l}
 \left( {3124} \right) + \left( {1324} \right) - \left( {2314} \right) - \left( {3214} \right) + \left( {4213} \right) + \left( {2413} \right) - \left( {1423} \right) - \left( {4123} \right) \\
  + \left( {1342} \right) + \left( {3142} \right) - \left( {4132} \right) - \left( {1432} \right) + \left( {2431} \right) + \left( {4231} \right) - \left( {3241} \right) - \left( {2341} \right) \\
 \end{array} \right\}\\ \\ \end{array}
$}\\
\hline \hline
\end{tabular}
} \\
\subfloat[Matric unit operators associated with the contragradient orthogonal representation of the symmetric group.]{
\begin{tabular}{l}
\hline \hline
\multicolumn{1}{l}{$\begin{array}{l} \\[-0.5em] \hat e_{11}^{\widetilde{\left[ {2,1} \right]}}  = \frac{{1}}{2}\left\{ { 2 \left( {123} \right) - 2 \left( {213} \right) - \left( {312} \right) + \left( {132} \right) - \left( {231} \right) + \left( {321} \right)} \right\}\\ \\\hat e_{21}^{\widetilde{\left[ {2,1} \right]}}  = \frac{{\sqrt 3 }}{2}\left\{ {\left( {312} \right) - \left( {132} \right) - \left( {231} \right) + \left( {321} \right)} \right\}\\  \end{array}$}\\
\multicolumn{1}{l}{$\begin{array}{l} \\  \hat e_{11}^{\widetilde{\left[ {3,1} \right]}}  =
\frac{{\rm{1}}}{{\rm{3}}}\left\{ \begin{array}{l}
 3\left( {{\rm{1234}}} \right) - 3\left( {{\rm{2134}}} \right) + 3\left( {{\rm{3124}}} \right) - 3\left( {{\rm{1324}}} \right) + 3\left( {{\rm{2314}}} \right) - 3\left( {{\rm{3214}}} \right) - \left( {{\rm{4213}}} \right) +  \\
 \left( {{\rm{2413}}} \right) - \left( {{\rm{1423}}} \right) + \left( {{\rm{4123}}} \right) - \left( {{\rm{2143}}} \right) + \left( {{\rm{1243}}} \right) - \left( {{\rm{1342}}} \right) + \left( {{\rm{3142}}} \right) - \left( {{\rm{4132}}} \right) \\
  + \left( {{\rm{1432}}} \right) - \left( {{\rm{3412}}} \right) + \left( {{\rm{4312}}} \right) - \left( {{\rm{4321}}} \right) + \left( {{\rm{3421}}} \right) - \left( {{\rm{2431}}} \right) + \left( {{\rm{4231}}} \right) - \left( {{\rm{3241}}} \right) + \left( {{\rm{2341}}} \right) \\
 \end{array} \right\}
\\ \\
 \hat e_{21}^{\widetilde{\left[ {3,1} \right]}}  = \frac{{\sqrt 2 }}{3}\left\{ \begin{array}{l}
 2 \left( {4213} \right) - 2\left( {2413} \right) + 2\left( {1423} \right) - 2\left( {4123} \right) + 2\left( {2143} \right) - 2\left( {1243} \right) - \left( {1342} \right) + \left( {3142} \right) - \left( {4132} \right)  \\
  + \left( {1432} \right) - \left( {3412} \right) + \left( {4312} \right) - \left( {4321} \right) + \left( {3421} \right) - \left( {2431} \right) + \left( {4231} \right) - \left( {3241} \right) + \left( {2341} \right) \\
 \end{array} \right\} \\ \\
 \hat e_{31}^{\widetilde{\left[ {3,1} \right]}}  = \sqrt {\frac{2}{3}} \left\{ \begin{array}{l} \left( {1342} \right) - \left( {3142} \right) + \left( {4132} \right) - \left( {1432} \right) + \left( {3412} \right) - \left( {4312} \right) \\- \left( {4321} \right) + \left( {3421} \right) - \left( {2431} \right) + \left( {4231} \right) - \left( {3241} \right) + \left( {2341} \right) \\ \end{array} \right\} \\ 
 \end{array}$}\\
\multicolumn{1}{l}{$ \begin{array}{l} \\ \hat e_{11}^{\widetilde{\left[ {2,2} \right]}}  =
\frac{{\rm{1}}}{{\rm{2}}}\left\{ \begin{array}{l}
 2\left( {{\rm{1234}}} \right) - 2\left( {{\rm{2134}}} \right) - \left( {{\rm{3124}}} \right) + \left( {{\rm{1324}}} \right) - \left( {{\rm{2314}}} \right) + \left( {{\rm{3214}}} \right) - \left( {{\rm{4213}}} \right) + \left( {{\rm{2413}}} \right) \\
  - \left( {{\rm{1423}}} \right) + \left( {{\rm{4123}}} \right) + 2\left( {{\rm{2143}}} \right) - 2\left( {{\rm{1243}}} \right) - \left( {{\rm{1342}}} \right) + \left( {{\rm{3142}}} \right) - \left( {{\rm{4132}}} \right) + \left( {{\rm{1432}}} \right) \\
  + 2\left( {{\rm{3412}}} \right) - 2\left( {{\rm{4312}}} \right) + 2\left( {{\rm{4321}}} \right) - 2\left( {{\rm{3421}}} \right) - \left( {{\rm{2431}}} \right) + \left( {{\rm{4231}}} \right) - \left( {{\rm{3241}}} \right) + \left( {{\rm{2341}}} \right) \\
 \end{array} \right\}
 \\ \\
\hat e_{21}^{\widetilde{\left[ {2,2} \right]}}  = \frac{{\sqrt 3 }}{2}\left\{ \begin{array}{l}
 \left( {3124} \right) - \left( {1324} \right) - \left( {2314} \right) + \left( {3214} \right) + \left( {4213} \right) - \left( {2413} \right) - \left( {1423} \right) + \left( {4123} \right) \\
  + \left( {1342} \right) - \left( {3142} \right) - \left( {4132} \right) + \left( {1432} \right) + \left( {2431} \right) - \left( {4231} \right) - \left( {3241} \right) + \left( {2341} \right) \\
 \end{array} \right\}\\ \\ \end{array}$}\\
\hline \hline
\end{tabular}
}
\caption{Some examples of the matric unit operators. These operators form an orthogonal basis of the symmetric group algebra in correspondence with the orthogonal representation of the symmetric group.}
\label{tableMatricUnitOperators}
\end{table}
\end{center}
\twocolumngrid

%%%%%%%%%%%%%%%%%%%%%%%%%%%%%%%%%%

\subsection{Construction of Spin Eigenfunctions for Systems with Valley or Layer Degrees of Freedom}
\label{subSpinAndValleyEigenfunctions}

In \aref{subGrapheneEigenfunctions} we discussed the construction of spin eigenfunctions describing systems containing two types of electrons corresponding to two valleys labeled by 1 and 2 or A and B.

The general mathematical procedure for the construction of spin eigenfunctions corresponding to multiple sets of spinful particles is explained in detail in Ref.~\onlinecite{katriel1991}. The procedure provides a group-theoretical explanation of the results derived by the brute-force approach presented in \aref{subGrapheneEigenfunctions}. What we are doing is starting with a decomposition of wave functions into representations of SU($n$) and then further decomposing into representations of SU($2$) $\times$ SU($2$) (for two sets). In general, we could, in principle, decompose into representations of SU($a$) $\times$ SU($b$) $\times$ \ldots . That would describe a system of $a$ component particles in set 1, a system of $b$ component particles in set 2, and so on. 

Let us illustrate the method with the following example: decomposing a four-component system into two two-component systems (which applies to the case of graphene when the system is spin-rotationally invariant). Given a number of electrons, $N$, in the system we partition this number into two parts (representing the number of electrons in each set). For example, with $N=3$ we can have the partitions [3,0], [2,1], [1,2], and [0,3]. Next we construct SU(2) spin eigenfunctions corresponding to each set of partitions. In this example we would construct spin eigenfunctions corresponding to the following SU(2) Young tableaux shapes:
\[\begin{array}{l}
[3,0]: \quad \yng(3) \quad ; \quad \mbox{-} \\ \\
\qquad   \qquad S=3/2 \quad ; \quad S=0 \\ \\
\qquad   \qquad \yng(2,1) \quad ;\quad \mbox{-} \\ \\
 \qquad   \qquad S=1/2 \quad ; \quad S=0 \\ \end{array} \]
  \[
  \begin{array}{l}
[2,1]: \quad \yng(2) \quad ; \quad \yng(1) \\  \\
 \qquad   \qquad S=1 \quad ; \quad S=1/2 \\ \\
 \qquad   \qquad \yng(1,1) \quad ; \quad \yng(1) \\ \\
  \qquad   \qquad S=0 \quad ; \quad S=1/2 \\
\end{array}\]
and vice versa for the [0,3] and [1,2] partitions. The corresponding SU(2) spin angular momentum eigenvalues are written below the Young tableaux. The next step is to construct SU(4) Young tableaux by forming tensor products between all combinations of the SU(2) Young tableaux for each partition. This is done according to the Littlewood--Richardson rule; see for example, Ref.~\onlinecite{hamermeshbook}. Simultaneously, we add the angular momenta of the two sets for each partition. For the [3,0] and [0,3] partitions this is a trivial procedure since there is only one set containing electrons. For the [2,1] we construct
\[
\yng(1,1) \otimes \yng(1) = \yng(2,1) \oplus \yng(1,1,1)
\]
and we add the angular momenta according to the standard procedure: $0+1/2 =1/2$. Also we have
\[
\yng(2) \otimes \yng(1) = \yng(3) \oplus \yng(2,1)
\]
and $1 + 1/2 = 3/2,1/2$. The final step is to associate all possible shapes of SU(4) tableaux deduced from combining each pair of cluster tableaux with all possible results of the corresponding addition of angular momenta. For example, the SU(4) tableaux shape $\lambda=[2,1]$ can occur with angular momenta $S=3/2$ in two places (once in cluster partition [2,1] and once in partition [1,2] ). The same shape can occur with angular momentum $S=1/2$ in six places (once in [3,0] and [0,3] and twice in [2,1] and [1,2]). This result matches our previous calculation: the shape $\lambda=[3]$ occurs with $S=3/2$ in four places and with $S=1/2$ in two places. Finally, the shape $\lambda=[1,1,1]$ occurs with only with $S=1/2$ in two places. We have checked that all of these results are consistent with the alternative brute-force method described in \aref{subGrapheneEigenfunctions}.

In general, one might want to consider a system in which there are more than two sets of particles. We might also wish to consider sets of spin-$j$ particles. In principle the procedure for constructing the basis of pseudopotentials is a simple generalization of the procedure described here for the case of graphene: we write down all valid symmetry types for each set of cluster partitions and all the corresponding angular-momentum eigenvalues. For example, if we have three sets of electrons, then we would write down the set of all SU(2) tableaux for each partition of $N$ into the three sets and then construct all possible combinations of tensor products between these tableaux. In the case where $j \ne 1/2$ some additional complications arise due to the multiple angular-momentum eigenvalues associated with each tableaux shape (the result of what is called fractional parentage). The resolution of these issues is described in Ref.~\onlinecite{katriel1991}.

\subsection{Generating Functions for Symmetric Polynomials}
\label{subAppendixGeneratingFunctions}

\subsubsection*{Properties of Generating Functions}

The generating function for a translationally invariant symmetric polynomial in $N$ variables is given by \cite{macdonaldbook}:
\[
Z_N \left( q \right) = \prod\limits_{m = 2}^N {\frac{1}{{1 - q^m }}}.
\]
Consider a product of two generating functions for symmetric polynomials
\[
Z_{N_1 } \left( q \right)Z_{N_2 } \left( q \right).
\]
The dimension of the space of polynomials at degree $L$ is given by
\[
d \left( {L,N_1,N_2} \right) = \left[ {\frac{1}{{L!}}\left( {\frac{d}{{dq}}} \right)^L Z_{N_1 } \left( q \right)Z_{N_2 } \left( q \right)} \right]_{q = 0} ,
\]
but we can expand this expression out using the Leibniz rule as follows:
\[
\left( {\frac{d}{{dq}}} \right)^L Z_{N_1 } Z_{N_2 }  = \sum\limits_{k = 0}^L {{}^LC_k \left( {\frac{d}{{dq}}} \right)^{L - k} Z_{N_1 } \left( {\frac{d}{{dq}}} \right)^k Z_{N_1 } } .
\]
So the dimension of the space of the combined symmetry is given by the sum of the dimensions of the spaces of polynomials of the same symmetry but with all possible subdivisions of the angular momentum, $L$. This is precisely the result we require for the generating function of a polynomial with two combined symmetries. This result clearly generalizes to a product of arbitrarily many generating functions.

Now consider a modified generating function given by multiplying by a factor of $q^J$. The dimension of the space of polynomials arising from this modified generating function is effectively a special case of the above result:
\[
\begin{array}{l}
 d\left( {L,N} \right) = \left[ {\frac{1}{{L!}}\left( {\frac{d}{{dq}}} \right)^L \left( {q^J \prod\limits_{m = 2}^N {\frac{1}{{1 - q^m }}} } \right)} \right]_{q = 0}  \\
= \left[ {\frac{1}{{L!}}\sum\limits_{k = 0}^L {{}^LC_k \left( {\frac{d}{{dq}}} \right)^{L - k} q^J \left( {\frac{d}{{dq}}} \right)^k \left( {\prod\limits_{m = 2}^N {\frac{1}{{1 - q^m }}} } \right)} } \right]_{q = 0}  \\
= \left[ {\frac{1}{{L!}}{}^LC_{L - J} \,J!\left( {\frac{d}{{dq}}} \right)^{L - J} \left( {\prod\limits_{m = 2}^N {\frac{1}{{1 - q^m }}} } \right)} \right]_{q = 0}  \\
= \left[ {\frac{1}{{\left( {L - J} \right)!}}\left( {\frac{d}{{dq}}} \right)^{L - J} \left( {\prod\limits_{m = 2}^N {\frac{1}{{1 - q^m }}} } \right)} \right]_{q = 0}  \\
= d_{{\rm{sym}}} \left( {L - J,N} \right) \\
 \end{array}
\]

\subsubsection*{Dividing Generating Functions}

In \aref{appendixSpatialWaveFunctions} we made use of the following results for the two-component case:
\begin{align*}
&\frac{{Z_{N,\,S} \left( q \right)}}{q^J {\prod\limits_{n = 2}^N {\frac{1}{{1 - q^n }}} }} = \frac{{\prod\limits_{l = 2}^{{\scriptscriptstyle{N \over 2}} + S} {\frac{1}{{1 - q^l }}} \prod\limits_{m = 1}^{{\scriptscriptstyle{N \over 2}} - S} {\frac{1}{{1 - q^m }}} }}{{\prod\limits_{n = 2}^N {\frac{1}{{1 - q^n }}} }} 
\\&= \prod\limits_{j = 1}^{{\scriptscriptstyle{N \over 2}} - S} {\frac{{\left( {1 - q^{N + 1 - j} } \right)}}{{\left( {1 - q^j } \right)}}}  =  
{N \brack \frac{N}{2} -S}_q.
\end{align*}

The final expression is a so-called $q$-binomial coefficient. \cite{andrewsbook} We can, thus, simplify the generating function to
\[
 Z_{N,\,S} \left( q \right) = q^J \prod\limits_{n = 2}^N {\frac{1}{{1 - q^n }}} \sum\limits_k {b_k q^k } ,
\]
where $b_k$ are positive integer coefficients. Using the definition of $q$-binomials we can show
\[
\sum\limits_k {b_k }  = {}^NC_{{\textstyle{N \over 2}} - S} .
\]
Using the result from \eref{eqTwoBodyFullGeneratingFunction} we have
\[
\frac{{\tilde Z_{N,\,S} \left( q \right)}}{{\prod\limits_{n = 2}^N {\frac{1}{{1 - q^n }}} }} = \,q^J {N \brack \frac{N}{2} -S}_q  - q^{1 + J + 2S} {N \brack \frac{N}{2} -S-1}_q,
\]
with $J$ given in \eref{eqDegreeOfJastrowFactor}. It follows that
\[
\tilde Z_{N,\,S} \left( q \right) = q^J \prod\limits_{n = 2}^N {\frac{1}{{1 - q^n }}} \sum\limits_{k = 1}^{k_{\max} } {b'_k q^k } ,
\]
where ${b'}_k$ are positive integer coefficients, which can be derived from the previous formula above in terms of the $q$-binomial coefficients, and
\[
k_{\max}  = \left( {{\textstyle{N \over 2}} - S} \right)\left( {{\textstyle{N \over 2}} + S} \right).
\]
A similar result also holds in the multicomponent case for $\tilde Z_{\lambda}$ defined in \eref{eqKostkaDecomposition}. 

\subsection{Vector Spaces Associated with Representations of the Symmetric Group}
\label{subAppendixPermutationModules}

In this final appendix we shall present an alternative derivation of the results presented in \aref{appendixSpatialWaveFunctions}. Our aim is to construct a basis of polynomials that are in a one-to-one correspondence with irreducible representations of the symmetric group, and, hence, are labeled by Young tableau shapes $\lambda=[N_1,N_2,\ldots, N_n]$. The starting point of our derivation is the set of symmetric polynomials that have subsets of symmetries. Such polynomials can also be enumerated in terms of integer partitions and, hence, Young tableau shapes. The question is: how are these polynomial constructions related? In this appendix we shall explain how this question can be posed in a more precise mathematical sense, in terms of the representation theory of the symmetric group. Having demonstrated an equivalence to a problem in mathematics, we shall describe the solution and then finally explain how this solution can be applied in the context of our problem.

To begin with, recall the simple argument given in \aref{appendixSpatialWaveFunctions}, which applied to the case of Young tableaux restricted to having no more than two rows. In that case we stated that the basis of polynomials of a given symmetry type actually contains polynomials of that particular symmetry type or of a greater symmetry type. We should make this statement more precise. Mathematically speaking the term ``\emph{greater symmetry}'' is equivalent to whether one tableau shape dominates another. \cite{saganbook} When comparing two Young tableau shapes $\mu=[\mu_1,\mu_2,\ldots, \mu_n]$ and $\lambda=[\lambda_1,\lambda_2,\ldots, \lambda_l]$, we say that $\mu$ dominates (has ``greater symmetry'' than) $\lambda$, written $ \mu \, \underbar{$\triangleright$} \, \lambda $  if
\[
{\mu _1} + {\mu _2} + \ldots+ {\mu _i} \ge {\lambda _1} + {\lambda _2} + \ldots + {\lambda _i}\,\,\,\,\,\,\forall \,\,i \ge 1,
\]
with
\[
{\mu _{i > l}} = {\lambda _{i > n}} = 0.
\]
In the special case where the Young tableau $\lambda$ contains $N$ boxes in two rows, with $N/2+S_{\lambda}$ boxes in the first row, this definition reduces to just
\[\begin{array}{l}
 {\mu _1} \ge {\lambda_1}\,\,\,\,\,\, \Rightarrow \,\,\,\,\,\,\frac{N}{2} + {S_\mu } \ge \frac{N}{2} + {S_\lambda }, \\
 {\mu _1} + {\mu _2} \ge {\lambda _1} + {\lambda_2}\,\,\,\,\,\, \Rightarrow \,\,\,\,\,\,N \ge N. \\
 \end{array}\]
The second statement is clearly always satisfied, and the first statement is true for all spin eigenvalues $S_\lambda \le S_\mu$, where $S_\lambda$ is the spin eigenvalues associated with symmetry type $\lambda$. This is precisely the way in which we defined ``greater symmetry'' in \aref{appendixSpatialWaveFunctions}. An important consequence of this result is that there are no possible cases of ambiguous dominance (that is, where two tableaux shapes do satisfy some but not all of the conditions for dominance, for example, [3,3] and [4,1,1]). We can represent the set of dominances diagrammatically in a Hasse diagram. \cite{saganbook} For the case of Young tableaux containing $N$ boxes in only two rows the Hasse diagram is shown in \fref{figHasseDiagrams} (a). The argument given in \aref{appendixSpatialWaveFunctions} to derive \eref{eqTwoBodyFullGeneratingFunction} follows directly form the ordered structure of this set of dominaces.

More generally, if we want to consider arbitrary shapes of Young tableaux, the set of dominances becomes more complicated. For example the Hasse diagram for all \mbox{$N=5$} Young tableaux is shown in \fref{figHasseDiagrams} (b). In general we must be more careful take into account the exact structure of the inclusions each of these symmetry subsets, and in particular we cannot rule out that in general one subset may be included multiple times within a subset of greater symmetry. At this juncture we shall introduce an equivalent mathematical formulation of this problem.

%%%%%%%%%%%%%%%%%%%%%%%%%%%%%%%%%%

\begin{figure}
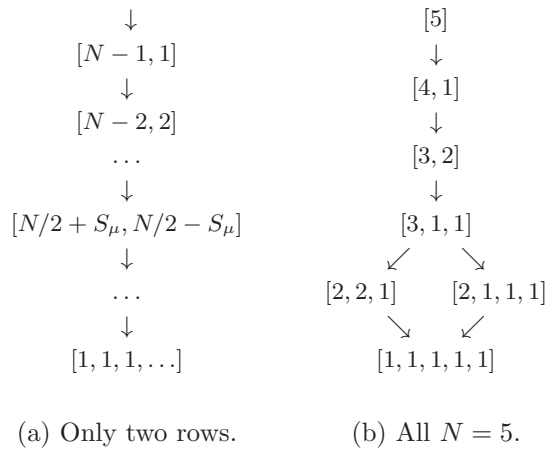

\begin{center}
\subfloat[Only two rows.]{
$
\begin{array}{c} \ \downarrow \\ \left[ {N-1,1} \right] \\ \downarrow \\ \left[ {N-2,2} \right] \\ \ldots \\ \downarrow \\ \left[ {N/2+S_\mu,N/2-S_\mu} \right] \\ \downarrow \\ \ldots \\ \downarrow \\ \left[ {1,1,1,\ldots} \right] \\ \\
\end{array}
$
}
\qquad
\subfloat[All $N=5$.]{
$
\begin{array}{c} [5] \\ \downarrow \\ \left[ {4,1} \right] \\ \downarrow \\ \left[ {3,2} \right] \\ \downarrow  \\ \left[ {3,1,1} \right] \\ \swarrow \qquad \searrow \\ \left[ {2,2,1} \right] \qquad \left[ {2,1,1,1} \right]\\  \searrow \qquad \swarrow \\ \left[ {1,1,1,1,1} \right] \\ \\
\end{array}
$
}
\end{center}
\caption{Examples of Hasse diagrams representing Young tableau shapes of ``greater symmetry''. \cite{saganbook}}
\label{figHasseDiagrams}
\end{figure}

%%%%%%%%%%%%%%%%%%%%%%%%%%%%%%%%%%

In the mathematical theory of representations of the symmetric group the symbol $M^{\lambda}$ denotes the polynomial space spanned by objects that correspond to the set of all Young tableaux of shape ${\lambda}$. \cite{saganbook} For example, $M^{[3,1]}$ has the basis
\[
\young(234,1) \qquad \young(134,2) \qquad \young(124,3) \qquad \young(123,4)
\]
We say that $M^{\lambda}$ forms a representation of the symmetric group called a permutation module (a more strict definition is that a module over a ring is an algebraic structure generalizing the notion of a vector space over a field). \cite{saganbook} In the context of polynomials, the permutation module forms a vector space for the ring of polynomials associated with a particular symmetry type. This is because the symmetry type of a polynomial should correspond to a Young tableau shape, as we have already explained, and because there should be no restriction on the order of the indices in a symmetric polynomial, so we must take into account all possible ways to place the numbers in a tableau of a given shape. The dimension of this polynomial vector space is, thus, given by the generating functions that we have denoted by $Z_{\lambda}$.

Our aim is to construct the space of polynomials corresponding to irreducible representations of the symmetric group. In the mathematical theory of representations of the symmetric group the symbol $S^{\lambda}$ denotes the polynomial spaces that correspond to irreducible representations of the symmetric group (the notation is not to be confused with our $S_\lambda$, which is the spin eigenvalue associated with symmetry type $\lambda$). These vector spaces are spanned by objects that correspond to the standard Young tableaux of shape ${\lambda}$. For example, $S^{[3,1]}$ has the basis
\[
\young(134,2) \qquad \young(124,3) \qquad \young(123,4)
\] 
$S^{\lambda}$ are called Specht modules. \cite{saganbook} In the context of polynomials, these Specht modules correspond exactly to the space of polynomials that becomes associated with irreducible representations of the symmetric group, in other words, precisely the space of polynomials that we require for our investigation. The dimensions of this space of polynomials are given by the generating functions that we have denoted by $\tilde{Z_{\lambda}}$.

A well-studied mathematical problem is to decompose the permutation module $M^{\lambda}$ in terms of irreducible representations of the symmetric group or, equivalently, in terms of Specht modules. It turns out that it is possible to write $M^{\lambda}$ as a direct sum of Specht modules, and the result is \cite{saganbook}
\[
{M^\lambda } \cong \mathop  \oplus \limits_{\lambda  \underbar{$\triangleright$} \mu } {K_{\mu \lambda }}{S^\mu }.
\]
The coefficients appearing here are precisely the Kostka numbers that we discussed in \aref{subCalculationOfDimensions}. \cite{kostka1882}

Finally, since this relation involves a direct sum of irreducible representations, it follows that the dimensions of these irreducible representations simply add together, and we find
\[
{d_{{M^\lambda }}} = \sum\limits_{\lambda  \triangleright \mu } {{K_{\mu \lambda }}{d_{{S^\mu }}}}. 
\]
It also follows that an equivalent relation must also be satisfied by the corresponding generating functions, and, hence, we find \eref{eqGeneratingFunctionKostkaForm}.

%%%%%%%%%%%%%%%%%%%%%%%%%%%%%%%%%%
%%%%%%%%%%%%%%%%%%%%%%%%%%%%%%%%%%

%\bibliographystyle{../dphilThesis/references/prsty}	%	Phys rev style references	
%\bibliography{../dphilThesis/references/myrefs} 		% Bibtex file

\begin{thebibliography}{10}

\bibitem{prangebook}
For classic reviews of quantum Hall physics, see R. Prange and S.M. Girvin,
  {\it The Quantum Hall Effect}, (Springer--Verlag, New York, 1987).

\bibitem{haldane1983}
F.~D.~M. Haldane, Phys. Rev. Lett. {\bf \textbf{51}},  605  (1983).

\bibitem{laughlin1983a}
R.~B. Laughlin, Phys. Rev. Lett. {\bf \textbf{50}},  1395  (1983).

\bibitem{trugman1985}
S.~A. Trugman and S. Kivelson, Phys. Rev. B {\bf 31},  5280  (1985).

\bibitem{nayak2008}
C. Nayak, S.~H. Simon, A. Stern, M. Freedman, and S.~D. Sarma, Rev. Mod. Phys.
  {\bf \textbf{80}},  1083  (2008).

\bibitem{bishara2009}
W. Bishara and C. Nayak, Phys. Rev. \textbf{B} {\bf \textbf{80}},  121302(R)
  (2009).

\bibitem{cooper2004}
N.~R. Cooper, Phys. Rev. Lett. {\bf 92},  220405  (2004).

\bibitem{moore1991}
G. Moore and N. Read, Nucl. Phys. \textbf{B} {\bf \textbf{360}},  363  (1991).

\bibitem{greiter1991}
M. Greiter, X.-G. Wen, and F. Wilczek, Phys. Rev. Lett. {\bf \textbf{66}},
  3205  (1991).

\bibitem{read1999}
N. Read and E. Rezayi, Phys. Rev. \textbf{B} {\bf \textbf{59}},  8084  (1999).

\bibitem{simon2007b}
S.~H. Simon, E.~H. Rezayi, N.~R. Cooper, and I. Berdnikov, Phys. Rev.
  \textbf{B} {\bf \textbf{75}},  075317  (2006).

\bibitem{green2002}
D. Green, ArXiv: {\bf 0202455},    (2002).

\bibitem{simon2007c}
S.~H. Simon, E.~H. Rezayi, and N.~R. Cooper, Phys. Rev. \textbf{B} {\bf
  \textbf{75}},  075318  (2007).

\bibitem{bernevig2008b}
B.~A. Bernevig and F.~D.~M. Haldane, Phys. Rev. Lett. {\bf 100},  246802
  (2008).

\bibitem{jackson2013}
T.~S. {Jackson}, N. {Read}, and S.~H. {Simon},  Phys. Rev. B {\bf 88},  075313
  (2013).

\bibitem{simon2010}
S.~H. Simon, E.~H. Rezayi, and N. Regnault, Phys. Rev. B {\bf 81},  121301
  (2010).

\bibitem{du1995}
R.~R. Du, A.~S. Yeh, H.~L. Stormer, D.~C. Tsui, L.~N. Pfeiffer, and K.~W. West,
  Phys. Rev. Lett. {\bf \textbf{75}},  3926  (1995).

\bibitem{cho1998}
H. Cho {\it et~al.}, Phys. Rev. Lett. {\bf \textbf{81}},  2522  (1998).

\bibitem{leadley1997}
D.~R. Leadley {\it et~al.}, Phys. Rev. Lett. {\bf 79},  4246  (1997).

\bibitem{bishop2007}
N.~C. Bishop, M. Padmanabhan, K. Vakili, Y.~P. Shkolnikov, E.~P. De~Poortere,
  and M. Shayegan, Phys. Rev. Lett. {\bf 98},  266404  (2007).

\bibitem{padmanabhan2010b}
M. Padmanabhan, T. Gokmen, and M. Shayegan, Phys. Rev. Lett. {\bf 104},  016805
   (2010).

\bibitem{dean2011}
C.~R. Dean {\it et~al.}, Nature Physics {\bf \textbf{ 7}},  693�696  (2011).

\bibitem{bolotin2009}
K. Bolotin, F. Ghahari, M.~D. Shulman, H.~L. Stormer, and P. Kim, Nature {\bf
  \textbf{462}},  196  (2009).

\bibitem{du2009}
X. Du, I. Skachko, F. Duerr, A. Luican, and E.~Y. Andrei, Nature {\bf
  \textbf{462}},  192  (2009).

\bibitem{lai2004}
K. Lai, W. Pan, D.~C. Tsui, S. Lyon, M. M\"uhlberger, and F. Sch\"affler, Phys.
  Rev. Lett. {\bf \textbf{93}},  156805  (2004).

\bibitem{arovas1999}
D.~P. Arovas, A. Karlhede, and D. Lillieh\"o\"ok, Phys. Rev. \textbf{B} {\bf
  \textbf{59}},  13147  (1999).

\bibitem{eng2007}
K. Eng, R.~N. McFarland, and B.~E. Kane, Phys. Rev. Lett. {\bf 99},  016801
  (2007).

\bibitem{eisensteinbook}
J.P. Eisenstein in ``Perspectives in Quantum Hall Effects", S. Das Sarma and A.
  Pinczuk, eds., Wiley, New York (1997); also S.M. Girvin and A.H. MacDonald,
  {\it ibid}.

\bibitem{eisenstein2004a}
J. Eisenstein, Science {\bf \textbf{305}},  950  (2004).

\bibitem{eisenstein2004b}
J.~P. Eisenstein and A.~H. MacDonald, Nature {\bf \textbf{432}},  691  (2004).

\bibitem{cooper2008}
N.~R. Cooper, Advances in Physics {\bf \textbf{57}},  539  (2008).

\bibitem{palmer2006}
R.~N. Palmer and D. Jaksch, Phys. Rev. Lett. {\bf 96},  180407  (2006).

\bibitem{hormozi2012}
L. Hormozi, G. M\"oller, and S.~H. Simon, Phys. Rev. Lett. {\bf 108},  256809
  (2012).

\bibitem{halperin1983}
B.~I. Halperin, Helv. Phys. Acta {\bf \textbf{56}},  75  (1983).

\bibitem{yoshioka1989}
D. Yoshioka, A.~H. MacDonald, and S.~M. Girvin, Phys. Rev. B {\bf 39},  1932
  (1989).

\bibitem{macdonald1989}
A.~H. MacDonald, D. Yoshioka, and S.~M. Girvin, Phys. Rev. B {\bf 39},  8044
  (1989).

\bibitem{haldane1988}
F.~D.~M. Haldane and E.~H. Rezayi, Phys. Rev. Lett. {\bf 60},  956  (1988).

\bibitem{rezayi1987a}
E.~H. Rezayi, Phys. Rev. B {\bf 36},  5454  (1987).

\bibitem{rezayi1987b}
E. Rezayi and F.~D.~M. Haldane, Bull. Am. Phys. Soc {\bf 32},  892  (1987).

\bibitem{ardonne1999}
E. Ardonne and K. Schoutens, Phys. Rev. Lett. {\bf \textbf{82}},  5096  (1999).

\bibitem{read2000}
N. Read and D. Green, Phys. Rev. \textbf{B} {\bf \textbf{61}},  10267  (2000).

\bibitem{ardonne2001}
E. Ardonne, N. Read, E. Rezayi, and K. Schoutens, Nucl. Phys. \textbf{B} {\bf
  \textbf{607}},  549  (2001).

\bibitem{ardonne2002}
E. Ardonne, F.~J.~M. van Lankvelt, A.~W.~W. Ludwig, and K. Schoutens, Phys.
  Rev. \textbf{B} {\bf \textbf{65}},  041305(R)  (2002).

\bibitem{reijnders2002}
J.~W. Reijnders, F.~J.~M. van Lankvelt, K. Schoutens, and N. Read, Phys. Rev.
  Lett. {\bf \textbf{89}},  120401  (2002).

\bibitem{reijnders2004}
J.~W. Reijnders, F.~J.~M. van Lankvelt, K. Schoutens, and N. Read, Phys. Rev. A
  {\bf \textbf{69}},  023612  (2004).

\bibitem{barkeshli2010}
M. Barkeshli and X.-G. Wen, Phys. Rev. B {\bf 82},  233301  (2010).

\bibitem{estienne2012}
B. Estienne and B.~A. Bernevig, Nucl. Phys. \textbf{B} {\bf \textbf{857}},  185
   (2012).

\bibitem{ardonne2011}
E. Ardonne and N. Regnault, Phys. Rev. \textbf{B} {\bf \textbf{84}},  205134
  (2011).

\bibitem{yang2008}
K. Yang and E.~H. Rezayi, Phys. Rev. Lett. {\bf \textbf{101}},  216808  (2008).

\bibitem{simon2007a}
S.~H. Simon, E.~H. Rezayi, and N.~R. Cooper, Phys. Rev. \textbf{B} {\bf
  \textbf{75}},  195306  (2007).

\bibitem{liptrap2010}
J. Liptrap, ArXiv: {\bf 1004.0364v1},    (2010).

\bibitem{paunczbook}
R. Pauncz, {\it Spin Eigenfunctions}, (Plenum Press, New York, 1979).

\bibitem{hamermeshbook}
M. Hamermesh,{\it Group Theory and its Application to Physical Problems},
  (Addison-Wesley, Reading, MA, 1964).

\bibitem{hund1927}
F. Hund, Z. Physik {\bf \textbf{43}},  788  (1927).

\bibitem{halzenbook}
F. Halzen, and A.D. Martin, {\it Quarks and Leptons: An Introductory Course in
  Modern Particle Physics}, (John Wiley \& Sons, New York, 1984).

\bibitem{toke2007}
C. T{\"o}ke and J.~K. Jain, Phys. Rev. \textbf{B} {\bf \textbf{75}},  245440
  (2007).

\bibitem{goerbig2007}
M.~O. Goerbig and N. Regnault, Phys. Rev. \textbf{B} {\bf \textbf{75}},
  241405(R)  (2007).

\bibitem{lim2011}
L.-K. Lim, M.~O. Goerbig, and C. Bena, Phys. Rev. \textbf{B} {\bf \textbf{84}},
   115404  (2011).

\bibitem{multicomponentWaveFunctionProgram}
S. C. Davenport. [http://www2.physics.ox.ac.uk/contacts /people/davenport]
  (2011).

\bibitem{bergholtz2007}
E.~J. Bergholtz, T.~H. Hansson, M. Hermanns, and A. Karlhede, Phys. Rev. Lett.
  {\bf 99},  256803  (2007).

\bibitem{bergholtz2008}
E.~J. Bergholtz, T.~H. Hansson, M. Hermanns, A. Karlhede, and S. Viefers, Phys.
  Rev. B {\bf 77},  165325  (2008).

\bibitem{ardonne2008}
E. Ardonne, E.~J. Bergholtz, J. Kailasvuori, and E. Wikberg, J. Stat. Mech.
  {\bf P04016},    (2008).

\bibitem{seidel2011}
A. Seidel and K. Yang, Phys. Rev. B {\bf 84},  085122  (2011).

\bibitem{wen2008b}
X.-G. Wen and Z. Wang, Phys. Rev. B {\bf 78},  155109  (2008).

\bibitem{davenport2013a}
S.~C. Davenport, E. Ardonne, N. Regnault, and S.~H. Simon, Phys. Rev. B {\bf
  87},  045310  (2013).

\bibitem{estienne2009}
B. Estienne and R. Santachiara, J. Phys. A: Math. Theor. {\bf \textbf{42}},
  245209  (2009).

\bibitem{bernevig2009}
B.~A. Bernevig, V. Gurarie, and S.~H. Simon, J. Phys. A: Math. Theor. {\bf
  \textbf{42}},  245206  (2009).

\bibitem{Read2009}
N. Read, Phys. Rev. \textbf{B} {\bf \textbf{79}},  245304  (2009).

\bibitem{hermanns2011a}
M. Hermanns, N. Regnault, B.~A. Bernevig, and E. Ardonne, Phys. Rev. B {\bf
  83},  241302  (2011).

\bibitem{andrewsbook}
G.E. Andrews, {\it The Theory of Partitions (Encyclopedia of Mathematics and
  its Applications)}, (Cambridge University Press, Cambridge , 1984).

\bibitem{rutherfordbook}
D.E. Rutherford, {\it Substitutional Analysis}, (Edinburgh University Press,
  Edinburgh, 1948).

\bibitem{davenport2012a}
S.~C. Davenport and S.~H. Simon, Phys. Rev. \textbf{B} {\bf \textbf{85}},
  075430  (2012).

\bibitem{pfeiferbook}
W. Pfeifer, {\it The Lie Algebras of su(N)}, (Birkh\"{a}user, Basal, 2003).

\bibitem{pauncz1977}
R. Pauncz and J. Katriel, Chemical Physics Letters {\bf \textbf{46}},  319
  (1977).

\bibitem{saganbook}
B.E. Sagan, {\it The Symmetric Group: Representations, Combinatorical
  Algorithms, and Symmetric Functions}, (Springer, Berlin, 2001).

\bibitem{kostka1882}
C. Kostka, Crelle's J. {\bf 93},  89  (1882).

\bibitem{katriel1991}
J. Katriel, Annals of Physics {\bf \textbf{211}},  1  (1991).

\bibitem{macdonaldbook}
I. G. Macdonald, {\it Symmetric Functions and Hall Polynomials}, (Oxford
  University Press, Oxford, 1979).

\end{thebibliography}

%\begin{comment}

%\end{comment}

\end{document}